\documentclass[a4paper,12pt, fleqn]{article}
\usepackage[a4paper, total={6.5in, 10in}]{geometry}
\usepackage{graphicx}
\usepackage{algorithmicx}
\usepackage{amsmath}
\usepackage{amssymb}
\usepackage{upgreek}
\usepackage{multicol}
\usepackage{color}
\usepackage{changepage}
\usepackage{booktabs}
\usepackage{multirow}
\usepackage{siunitx}
\usepackage{rotating}
\usepackage{float}
\usepackage{Tabbing}
\usepackage{dsfont}
\usepackage{amsthm}
\usepackage{amstext}
\usepackage{hyperref}
\usepackage{cleveref}
\usepackage[normalem]{ulem}
\usepackage{enumerate}
\usepackage{array}
\usepackage{caption}
\usepackage{graphicx}
\graphicspath{ {./images/} }
\usepackage[english]{babel}
\usepackage[round]{natbib}
\usepackage[title]{appendix}
\usepackage{authblk}
\usepackage[ruled,vlined]{algorithm2e}
\usepackage{marvosym}
\usepackage{setspace}

\hypersetup{
    colorlinks=true,
    linkcolor=blue,
    filecolor=magenta,      
    urlcolor=cyan,
    citecolor=blue
    }

\usepackage{enumitem}
\usepackage[dvipsnames,svgnames]{xcolor}

\DeclareMathOperator*{\argmin}{arg\,min}

\newcommand{\iid}{\overset{\text{iid}}{\sim}}

\newcommand{\abs}[1]{\left\lvert#1\right\rvert}
\newcommand{\norm}[1]{\left\lVert#1\right\rVert}
\newcommand{\E}{\mathbb{E}}
\renewcommand{\P}{\mathbb{P}}

\newcommand{\setEPl}[1]{\setEP{T}^l}

\newcommand{\bzero}{\boldsymbol{0}}
\newcommand{\by}{\boldsymbol{y}}
\newcommand{\bx}{\boldsymbol{x}}

\newcommand{\bu}{\boldsymbol{u}}

\newcommand{\bb}{\boldsymbol{b}}
\newcommand{\bB}{\boldsymbol{B}}

\newcommand{\bz}{\boldsymbol{z}}

\newcommand{\bX}{\boldsymbol{X}}

\newcommand{\bR}{\boldsymbol{R}}

\newcommand{\bS}{\boldsymbol{S}}
\newcommand{\bD}{\boldsymbol{D}}

\newcommand{\bI}{\boldsymbol{I}}
\newcommand{\bJ}{\boldsymbol{J}}

\newcommand{\bbeta}{\boldsymbol{\beta}}
\newcommand{\bxi}{\boldsymbol{\xi}}

\newcommand{\bchi}{\boldsymbol{\chi}}

\newcommand{\bSigma}{\boldsymbol{\Sigma}}
\newcommand{\bOmega}{\boldsymbol{\Omega}}

\newcommand{\diag}{\text{diag}}
\newcommand{\ba}{\boldsymbol{a}}

\newcommand{\bC}{\boldsymbol{C}}
\newcommand{\bA}{\boldsymbol{A}}
\newcommand{\mA}{\mathds{A}}

\newcommand{\cB}{\mathcal{B}}
\newcommand{\bc}{\boldsymbol{c}}
\newcommand{\bbf}{\boldsymbol{f}}

\newcommand{\bPsi}{\boldsymbol{\Psi}}

\newcommand{\bepsilon}{\boldsymbol{\epsilon}}
\newcommand{\blambda}{\boldsymbol{\lambda}}

\newcommand{\cX}{\mathcal{X}}

\newcommand{\ind}[1]{\mathds{1}_{\{#1\}}}
\title{Sparse High-Dimensional Vector Autoregressive Bootstrap\thanks{The first and second author were financially supported by the Dutch Research Council (NWO) under grant number 452-17-010. The first author is also affiliated with the Center for Research in Energy:~Economics and Markets, CoRE, funded by InCommodities.}}
\author[a]{Robert Adamek}
\author[b]{Stephan Smeekes}
\author[b]{Ines Wilms}
\affil[a]{Department of Economics and Business Economics, Aarhus University}
\affil[b]{Department of Quantitative Economics,
Maastricht University}
\affil[ ]{\texttt{r.adamek@econ.au.dk,\{s.smeekes,i.wilms\}@maastrichtuniversity.nl}}
\date{\today}

\newtheorem{theorem}{Theorem}
\newtheorem{corollary}{Corollary}
\newtheorem{lemma}{Lemma}
\Crefname{lemma}{Lemma}{Lemmas}
\Crefname{corollary}{Corollary}{Corollaries}
\Crefname{theorem}{Theorem}{Theorems}
\Crefname{definition}{Definition}{Definitions}
\Crefname{example}{Example}{Examples}
\Crefname{remark}{Remark}{Remarks}
\Crefname{assumption}{Assumption}{Assumptions}
\Crefname{algorithm}{Algorithm}{Algorithms}

\theoremstyle{definition}
\newtheorem{assumption}{Assumption}
\newtheorem*{assumption*}{Assumption}
\newtheorem{remark}{Remark}
\newtheoremstyle{example}{3pt}{3pt}{}{\parindent}{}{.}{}{}
\newtheorem{example}{Example}

\usepackage[]{setspace}
\begin{document}
\maketitle
\onehalfspacing
\doublespacing
\begin{abstract}
We introduce a high-dimensional multiplier bootstrap for time series data based on capturing dependence through a sparsely estimated vector autoregressive model. 
We prove its consistency for inference on high-dimensional means
under two different moment assumptions on the errors, namely sub-gaussian moments 
and a finite number of absolute moments. In establishing these results, we derive a Gaussian approximation for the maximum mean of a linear process, which may be of independent interest.
%Our simulation studies demonstrate the potential of the method in finite samples, even if the data are not generated by a finite-order vector autoregression.

\end{abstract}
\doublespacing
    
\section{Introduction}\label{sec:intro}
We introduce theory for bootstrapping the distribution of high-dimensional means of sparse, finite order, stable vector autoregressive (VAR) processes. For an $N$-dimensional vector of time series $\bx_t=(x_{1,t},\dots,x_{N,t})^\prime$, we provide an approximation for the distribution of $\max\limits_{1\leq j\leq  N}\abs{\frac{1}{\sqrt{T}}\sum\limits_{t=1}^T x_{j,t}}$, where the number of variables $N$ is potentially much larger than the sample size $T$, and can asymptotically grow faster than $T$. This prototypical statistic is commonly considered in high-dimensional settings, 
see e.g.~the closely related work of \cite{ChangZhengZhouZhou2017}, \cite{CCK13}, \cite{CCK17}, \cite{chernozhukov2020nearly}, \cite{chernozhukov2022high}, \cite{giessing2020bootstrapping}, \cite{zhang2017gaussian}, 
or the review by \cite{chernozhukov2022high} who investigate the properties of this estimator for independent data. In this paper, we extend these results to high-dimensional linear processes, including stable VARs. 
Related work in time series settings include \cite{ChangChenWu2024}, \cite{CCK2019}, \cite{ChiangKatoSasaki2021}, \cite{kurisu2024}, \cite{ZhangCheng2014bootstrapping}, and \cite{zhang2018gaussian},
who provide Gaussian approximations under various forms of dependence. 

The VAR sieve bootstrap is well-known in the low-dimensional time series bootstrapping literature, see e.g.~\cite{ChangPark2003}, \cite{MeyerKreiss15}, \cite{paparoditis1996}, \cite{Park2002},
and Section 12.2 of \cite{kilian2017structural}. 
It fits a VAR to the time series data, resamples the residuals 
of the estimated VAR, and 
re-applies the VAR recursively
to place the dependence back into the bootstrap sample. Under appropriate conditions, 
the VAR sieve bootstrap allows for valid inference. 
We extend this 
approach to high dimensions 
where the VAR  is estimated by the lasso (see \cite{Tibshirani96}) or another sparse estimation method, and use a multiplier (or wild) bootstrap to resample the residuals. 
Our work is related to that of \cite{bi2021ar}, \cite{Trapani2013} and \cite{Krampe19}. 
The two former papers assume a dense structure on the data, and apply the VAR sieve bootstrap to a low-dimensional set of factors. The latter consider a sparse setting, providing bootstrap inference for desparsified estimators of VAR coefficients. We assume a data-generating process (DGP) similar to the one considered in \cite{Krampe19}.

All theoretical results in this paper are established under two different sets of assumptions on the errors. First, we assume
the errors have sub-gaussian moments, which generally allows $N$ to grow at an exponential rate of $T$. 
Second, we assume 
that the errors have some finite number of absolute moments, which effectively restricts the growth of $N$ to some polynomial rate of $T$.  
In Section \ref{sec:bootexplained}, we introduce the multiplier bootstrap for sparsely estimated high-dimensional VARs.
In Section \ref{sec:HDCLTlinear}, we 
start by providing a high-dimensional central limit theorem (HDCLT) for linear processes in Theorem \ref{thm:HDCLTforLP}, which may be of independent interest.  In Section \ref{sec:HDCLTVAR}, we introduce the stable VAR model, and show that under consistent estimation, the long run covariance structure is recovered with high probability. Theorem \ref{thm:CovarianceCloseness} provides a consistency result for the covariance matrix. 
In Section \ref{sec:bootstrapworld}, we show that the bootstrap's  behaviour is asymptotically similar to that of the original sample. In particular, Theorem \ref{thm:HDCLTforboot} provides a HDCLT for the bootstrap process which mirrors that of Theorem \ref{thm:HDCLTforLP}, and Theorem \ref{thm:bootstrapconsistency} shows consistency of the bootstrap. Section \ref{sec:lassoexample} then shows how these results can be 
used to establish validity of inference in VARs estimated by the lasso. Section \ref{sec:simulations} demonstrates the finite sample performance of our proposed method in a simulation exercise, Section \ref{sec:empirical} demonstrates the method in an empirical application, and Section \ref{sec:conclusion} concludes.

\textit{Notation.} 
For a random variable $x$, $\norm{x}_{L_p}=\left(\E\abs{x}^p\right)^{1/p}$ denotes the $L_p$ norm, and $\norm{x}_{\psi_2}=\inf\left\lbrace c>0:\E\exp(\abs{x}^2/c^2)\leq 2\right\rbrace$ denotes the Orlicz norm. For any $N$ dimensional vector $\boldsymbol{x}$, $\left\Vert \boldsymbol{x}\right\Vert_p=\left(\sum\limits_{j=1}^{N}\left\vert x_j\right\vert^p\right)^{1/p}$ 
denotes the $p$-norm, with the familiar convention that $\norm{\boldsymbol{x}}_0 = \sum_{i} \ind{\abs{x_i}>0}$ and $\left\Vert \boldsymbol{x}\right\Vert_{\infty}=\max\limits_{i}\left\vert
x_i\right\vert$. 
For a matrix $\bA$, we let $\norm{\bA}_p = \max_{\norm{\bx}_p = 1} \norm{\bA \bx}_p$ for any $p \in [0, \infty]$ and $\norm{\bA}_{\max}=\max\limits_{i,j}\left\vert a_{i,j}\right\vert$. 
$\Lambda_{\min}(\bA)$ and $\Lambda_{\max}(\bA)$ denote the smallest and largest eigenvalues of $\bA$, and $\rho(\bA)$ the spectral radius of $\bA$, i.e. the largest absolute eigenvalue of $\bA$, or equivalently $\rho(\bA)=\lim\limits_{k\to\infty}\norm{\bA^k}^{1/k}$ for any induced norm $\norm{\cdot}$. For $\bA$ a square matrix, we let its zero-th power $\bA^0=\bI$.
We use $\overset{p}{\to}$ and $\overset{d}{\to}$ to denote convergence in probability and distribution respectively. Depending on the context, $\sim$ denotes equivalence in order of magnitude of sequences, or equivalence in distribution. 
We frequently make use of arbitrary positive finite constants $C$ (or its sub-indexed version $C_i$) whose values may change from line to line throughout the paper, but they are always independent of the time and cross-sectional dimension.
Similarly, generic sequences converging to zero as $T\to\infty$ are denoted by  $\eta_T$  (or its sub-indexed version $\eta_{i,t}$). When they are used, it should be understood that there exists some constant $C$ or sequence $\eta_T\to0$ such that the given statement holds.

\section{Vector Autoregressive Bootstrap}\label{sec:bootexplained}
We introduce our proposed bootstrap procedure for sparsely estimated high-dimensional VARs and subsequently discuss how it can be used to perform inference on high-dimensional time series.

\subsection{Bootstrap for High-Dimensional VARs}
Let $\bx_t$ be an $N$-dimensional time series process. We assume the data is generated by a stable, finite order, high-dimensional VAR($K$) model
\begin{equation}\label{eq:DGPVAR}
    \bx_t=\sum\limits_{k=1}^{K}\bA_k\bx_{t-k}+\bepsilon_t 
    ,~t=1,\dots,T,
\end{equation}
with autoregressive parameter matrices $\bA_k \ (k=1,\ldots,K)$, independent errors $\bepsilon_t $ with $\E\bepsilon_t=\bzero$ and covariance matrix $\bSigma_{\epsilon} :=\frac{1}{T}\sum\limits_{t=1}^T\E\bepsilon_t\bepsilon_t^\prime$, and $\bx_t=\bepsilon_t=\bzero$ for $t<1$.
We can re-write 
\Cref{eq:DGPVAR} as a collection of linear equations
\begin{equation*}
    x_{j,t}=\sum\limits_{k=1}^K\ba_{j,k}\bx_{t-k}+\epsilon_{j,t}=\underset{1\times KN}{\bbeta_j^\prime}\underset{KN\times 1}{\cX_{t}}+\epsilon_{j,t},~ j=1,\dots,N,~t=1,\dots,T,
\end{equation*}
where $\ba_{j,k}$ is the $j$th row of $\bA_k$, $\bbeta_{j}=(\ba_{j,1},\dots,\ba_{j,K})^\prime$, and $\cX_t=(\bx_{t-1}^\prime,\dots,\bx_{t-K}^\prime)^\prime$.
We denote data stacked into a matrix as $\underset{T\times N}{\bX}=(\bx_{1}^\prime, \dots, \bx_{T}^\prime)^\prime$.
The lasso estimator of equation $j$ is defined as
\begin{equation}\label{eq:problem}
     \hat\bbeta_j=\argmin\limits_{\bbeta_j^*\in\mathds{R}^{KN}}\frac{1}{T}\sum\limits_{t=1}^T\left(x_{j,t}-\bbeta_{j}^{*\prime}\cX_t\right)^2+2\lambda_j\norm{\bbeta_j^*}_1, 
\end{equation}
where $\lambda_j$ is a tuning parameter that determines the degree of penalization in equation $j$, and can be selected independently in each equation.
For tuning parameter selection, one could use e.g.~the theoretically founded method of \cite{kock2024data}, the iterative plug-in procedure described in Section 5.1 of \cite{adamek2021lasso}, or information criteria. 

Once all equations $j=1,\ldots, N$ are estimated by the lasso, we 
collect the VAR coefficient estimates as follows
\begin{equation*}
    \left[\begin{array}{ccc}
         \hat\bA_1&\cdots &\hat\bA_k 
    \end{array}\right]=\left[\begin{array}{c}
         \hat\bbeta_1^\prime\\
         \vdots\\
         \hat\bbeta_{N}^\prime
    \end{array}\right].
\end{equation*}
 
Our object of interest is the scaled high-dimensional mean $$ Q= \max\limits_{1\leq j\leq N}\abs{\frac{1}{\sqrt{T}}\sum\limits_{t=1}^{T} x_{j,t}}$$
of the sparse VAR.
To approximate its  
distribution, 
we apply the VAR multiplier bootstrap summarized in 
Algorithm \ref{alg:boot}.
When $B$ is sufficiently large, the CDF of $Q$ can be approximated by the quantiles of the ordered statistics $Q^{*(1)}, \dots, Q^{*(B)}$. Note that while we derive results for the maximum absolute mean, this bootstrap procedure is equally valid for statistics such as  $\max\limits_{1\leq j\leq N}\frac{1}{\sqrt{T}}\sum\limits_{t=1}^Tx_{j,t}$ or  $\min\limits_{1\leq j\leq N}\frac{1}{\sqrt{T}}\sum\limits_{t=1}^Tx_{j,t}$, which would allow for one-sided tests, or tests with an asymmetric rejection region.

\begin{algorithm}[t]
\nl Given the sample $\{\bx_t\}_{t=1}^{T}$, compute the statistic $Q=\max\limits_{1\leq j\leq N}\abs{\frac{1}{\sqrt{T}}\sum\limits_{t=1}^{T} x_{j,t}}$\;
\nl Demean the data to obtain $\tilde\bx_t=\bx_t-\bar{\bx}$, where $\bar\bx=\frac{1}{T}\sum\limits_{t=1}^{T}\bx_{t}$\;
\nl Let  $\hat\bA_1,\dots,\hat\bA_K$ be the lasso estimates in the \Cref{eq:DGPVAR} model for the demeaned data, where unobserved values of the lags are padded with zeroes, i.e.~we let $\tilde\bx_t=\bzero$ for $t<1$\; 
\nl Set $\hat\bepsilon_t=\tilde\bx_t-\sum\limits_{k=1}^K\hat\bA_k\tilde\bx_{t-k}$ for $t=1,\dots, T$\;
\medskip
\nl \For{$b \in \{1, \ldots, B\}$}{
\medskip
\nl Generate $\gamma_1, \dots, \gamma_T$ from a $N(0,1)$ distribution\;
\nl Set $\bepsilon_{t}^* =\hat\bepsilon_t\gamma_t$ for $t = 1, \dots, T$\;
\nl  Build $\bx_t^*$ recursively from $\bx_t^*=\sum\limits_{k=1}^K\hat\bA_k\bx_{t-k}^*+\bepsilon_{t}^*$ for $t=1,\dots T$, letting $\bx_t^*=\bzero$ for $t<1$\;
\nl Compute and store the statistic $Q^{*b}=\max\limits_{1\leq j\leq N}\abs{\frac{1}{\sqrt{T}}\sum\limits_{t=1}^Tx_{j,t}^*}$\;
}
\caption{VAR Multiplier Bootstrap}\label{alg:boot}
\end{algorithm}
\begin{remark} \label{remark:Klags}
So far, we treated the number of lags $K$  in the VAR as known, which is typically not the case in practice.
Indeed, Algorithm \ref{alg:boot} requires one to choose $K$. One of the lasso's advantages is that it performs well when the number of regressors is large, provided the parameters are sparse. This means it is less harmful to include many redundant lags, compared to low-dimensional estimation  methods which suffer in terms of efficiency. Therefore, if the practitioner believes the true VAR order is some $K\leq K_{\max}$, one may simply take $K=K_{\max}$, and let the lasso penalize any redundant lags to 0. 
For example, the informative upper bound in Section 5 of \cite{HecqMargaritellaSmeekes2023} appears to work well for this purpose, see Algorithm \ref{alg:lag_selection} in Appendix C. Alternatively, one could use the hierarchical lag structure approach of \cite{Wilms2020Hlag} that embeds lag selection into the estimation procedure. 
\end{remark}
 
\begin{remark} \label{remark:nonstat}
It may happen that the estimated VAR is not 
stable, even if the true underlying process is. Proper functioning of our method requires, however, that the bootstrap process is stable. 
In low-dimensional settings, this can be dealt with by using an estimation method that guarantees 
stable estimates, such as Yule-Walker estimation. However, to our knowledge, a similar method has not yet been proposed for high-dimensional settings. 
In case of non-stability, we  suggest  to manually correct the estimates 
by uniformly shrinking all entries of $\hat{\bA}_1,\dots,\hat{\bA}_K$ towards 0 to ensure stability of the bootstrap process. Specifically, we compute the roots $z_1,\dots z_k$ of the lag polynomial $\abs{\bI-\hat{\bA}_1 z - \hat{\bA}_2 z^2 \dots -\hat{\bA}_K z^k}=0$, and if $z_{\min}:=\min\limits_{i}\abs{z_i}$ is smaller than $1/0.999$, we use a stabilized version of the VAR with coefficient matrices $\tilde{\bA}_k=\hat{\bA}_k\left(0.999z_{\min}\right)^k$, which has all roots outside the unit circle by construction, and is therefore stable. In \Cref{sec:HDCLTVAR}, we justify that this correction is asymptotically negligible.

\end{remark}
 
\subsection{Bootstrap Inference on (Approximate) Means}
Statistics such as the scaled mean $Q$ are useful in high-dimensional settings, since they allow us to simultaneously test a high-dimensional set of hypotheses. For example, let $\mu_{j}=\E x_{j,t}$ be the means of a high-dimensional stable autoregressive process, and 
assume we are interested in testing 
the hypothesis 
\begin{equation*}
    H_0:\mu_{1}=\dots=\mu_{N}=0\text{ vs. }H_1: \mu_j\neq 0\text{ for at least one }j.
\end{equation*}
Under the null hypothesis, this process follows \Cref{eq:DGPVAR}, which allows us to directly test the null 
using the quantiles of $Q^{*(1)}, \dots, Q^{*(B)}$. Specifically, 
one would reject the null 
at significance level $\alpha$ if $Q>Q^{*(B[1-\alpha])}$.
To know for \textit{which} means 
the null can be rejected, 
one can use the stepdown procedure of \cite{romano2005exact}, as detailed in Section 5 of \cite{CCK13} or Section 4.5 of \cite{chernozhukov2022high}. Importantly, this procedure is asymptotically exact -- non-conservative -- as it takes into account the possible correlations between statistics, instead of using the conservative worst case of independence.

More generally, this bootstrap procedure can be used to test any high-dimensional set of hypotheses, provided its test statistic can be expressed as an approximate mean, that is, $\frac{1}{\sqrt{T}}\sum\limits_{t=1}^Tx_{j,t}+o_p(1)$. While we do not formally consider this extension here, we can adapt the arguments in Section 5 of \cite{CCK13} (which do not rely on independent data) to establish this result in our context as well. This opens up the way for applications to statistics that are much more general than just sample means, as many statistics of practical interest, such as (high-dimensional) regression estimates, can be written in this form. Our results therefore form a first step towards a more general bootstrap theory for high-dimensional inference using VAR models on statistics that can be well-approximated by the mean of a linear process.

\section{HDCLT for Linear Processes}\label{sec:HDCLTlinear}
In this section, we establish a high-dimensional CLT for linear processes, which is a useful result in its own right, but also a vital building block to establish theoretical results for the bootstrap.
We therefore give it a 
self-contained treatment in this section, before applying it to the VAR process in \Cref{eq:DGPVAR} and covering the theory for the bootstrap in the following sections.

Under appropriate invertibility
conditions, it is well-known that the VAR process in \Cref{eq:DGPVAR} can be written in the following infinite order vector moving average (VMA) form
\begin{equation}\label{eq:DGPVMA}
    \bx_t=\sum\limits_{k=0}^{\infty}\bB_k\bepsilon_{t-k}=\cB(L)\bepsilon_t,
        ~t=1,\dots,T,
\end{equation}
where $\cB(z)=\sum\limits_{k=0}^{\infty}\bB_k z^k=\left(\bI-\sum\limits_{k=1}^K\bA_k z^k\right)^{-1}$, and $L$ is the lag operator. We derive a Gaussian approximation for linear processes of the form in \Cref{eq:DGPVMA}, which builds on and extends similar approximations for  independent and identically distributed (i.i.d.)~processes by \cite{chernozhukov2020nearly} and others (see \Cref{sec:intro}). 
   
Specifically, we show that the distribution of $\max\limits_{1\leq j\leq N}\abs{\frac{1}{\sqrt{T}}\sum\limits_{t=1}^T x_{j,t}}=\norm{\frac{1}{\sqrt{T}}\sum\limits_{t=1}^T\bx_t}_{\infty}$ can be asymptotically approximated by $\norm{\bz}_{\infty}$, with $\bz\sim N(\bzero,\bSigma)$ and $\bSigma$ an appropriate covariance matrix. This result parallels well-known results in low-dimensional settings, where scaled means of linear processes converge in distribution to a Gaussian random variable as $T\to\infty$. However, in our high-dimensional setting, we consider the case where $N$ and $T$ diverge simultaneously, and $\norm{\frac{1}{\sqrt{T}}\sum\limits_{t=1}^T\bx_t}_{\infty}$ does not converge to a well defined limit; the maximum over a growing number of elements generally also grows. As such, we instead show that their distributions grow closer together asymptotically, in the sense that the Kolmogorov distance between between $\norm{\frac{1}{\sqrt{T}}\sum\limits_{t=1}^T\bx_t}_{\infty}$ and $\norm{\bz}_\infty$ converges to 0.
Even though to our knowledge, there does not exist a closed-form expression for the CDF of $\norm{\bz}_{\infty}$, it can be approximated for any $N$ by Monte Carlo simulation, making it a useful asymptotic approximation in practice. 
    
The broad sketch of our proof is as follows. We use the Beveridge-Nelson decomposition to write
\begin{equation}\label{eq:BN}
    \frac{1}{\sqrt{T}}\sum_{t=1}^T\bx_t=\frac{1}{\sqrt{T}}\sum_{t=1}^T\cB(1)\bepsilon_t-\frac{1}{\sqrt{T}}\tilde{\cB}(L)\left(\bepsilon_{T}-\bepsilon_{0}\right),
\end{equation}
where $\tilde{\cB}(z)=\sum\limits_{j=0}^\infty\sum\limits_{k=j+1}^\infty\bB_k z^j$. The first term is a scaled sum of independent errors with covariance matrix $\bSigma:=\cB(1)\bSigma_{\bepsilon}\cB(1)^\prime$, $\sigma^2_{j}:=\bSigma_{(j,j)}$, and can therefore be approximated by a Gaussian maximum thanks to \cite{chernozhukov2020nearly} when $\bSigma$ is non-degenerate and the $\bepsilon_t$'s satisfy certain moment conditions (see \Cref{lma:ChernozhukovHDCLT}). The second term 
is an asymptotically negligible leftover under certain summability conditions on the VMA coefficient matrices $\bB_k$ (see \Cref{lma:leftover}). 
Formally, we make the following assumptions:
\begin{assumption}\label{ass:covariance}
     Let $\Lambda_{\min}\left(\bSigma\right)\geq 1/C$ and $\max\limits_{1\leq j\leq N}\sigma_j\leq C$.
\end{assumption}
\begin{assumption}\label{ass:subgaussian}
    Let the vector $\bepsilon_t$ satisfy \textit{one} of  the following moment conditions 
    \begin{enumerate}
        \item\label{ass:subgaussian1} $\max\limits_{j,t}\norm{\epsilon_{j,t}}_{\psi_2}\leq C$.
        \item\label{ass:subgaussian2}  $\max\limits_{j,t}\norm{\epsilon_{j,t}}_{L_m}\leq C$, for some constant $m\geq 4$.
    \end{enumerate}
\end{assumption}
We derive our results 
under two different moment assumptions. In \Cref{ass:subgaussian}.\ref{ass:subgaussian1} we require that the errors are uniformly sub-gaussian over $j$ and $t$; or in \Cref{ass:subgaussian}.\ref{ass:subgaussian2} that the moments possess some number ($m$) of finite absolute moments.
By equation (2.15) in \cite{vershynin2019high}, \Cref{ass:subgaussian}.\ref{ass:subgaussian2} follows automatically for all $m$ from \Cref{ass:subgaussian}.\ref{ass:subgaussian1}, making the latter a considerably less stringent assumption. Under these assumptions, \Cref{thm:HDCLTforLP} provides an upper bound on the Kolmogorov distance between our statistic of interest and a Gaussian maximum:

\begin{theorem}[Gaussian approximation for linear processes]\label{thm:HDCLTforLP} Consider a linear process $\bx_t$ as in \Cref{eq:DGPVMA}, let \Cref{ass:covariance} hold, and define $\tilde{S}:=\sum\limits_{j=0}^{\infty} \norm{\bB_{j}}_\infty$, $S_q:=\sum\limits_{j=0}^{\infty}\left(\sum\limits_{k=j+1}^\infty \norm{\bB_{k}}_\infty\right)^{q}$, and
\begin{equation*}
     {J_{N,T}}:=\sup\limits_{y\in\mathbb{R}}\abs{\P\left(\norm{\frac{1}{\sqrt{T}}\sum_{t=1}^T\bx_t}_{\infty}\leq y\right)-\P\left(\norm{ \bz}_\infty\leq y\right)},
\end{equation*}
where $\bz\sim N(\bzero,\bSigma)$. 
\begin{enumerate}
\item Under \Cref{ass:subgaussian}.\ref{ass:subgaussian1},
\begin{equation*}
    {J_{N,T}}
    \leq C\left( \frac{(\tilde{S} d_{N})^2\log(N)^{3/2}\log(T)}{\sqrt{T}}+\frac{(\tilde{S} d_{N})^2\log(N)^2}{\sqrt{T}}+\frac{\log(N)d_{N}\sqrt{S_2}}{\sqrt{T}}+\frac{1}{\log(N)}\right),
\end{equation*}
where $d_{N}=C\sqrt{\log(N)}$.
    \item Under \Cref{ass:subgaussian}.\ref{ass:subgaussian2}, 
   \begin{equation*}\begin{split}
    {J_{N,T}}
    \leq&  C\Bigg(\frac{ (\tilde{S} d_{N})^2(\log N)^{3/2}\log (T)}{\sqrt{T}}+\frac{(\tilde{S} d_{N})^4\log(N)^2\log (T)}{T^{1-2/m}}\\
    &+\left[\frac{(\tilde{S} d_{N})^{2m} \log(N)^{3m/2-4}\log(T)\log (NT)}{T^{m/2-1}}\right]^{\frac{1}{m-2}}+\left( Nd_{N}^mS_1^m \right)^{\frac{1}{m+1}}\left[\frac{\sqrt{\log(N)}}{\sqrt{T}}\right]^{\frac{m}{m+1}}\Bigg),
\end{split}\end{equation*}
where $d_{N}=C N^{1/m}\eta_{T}^{-1}$.
\end{enumerate}
\end{theorem}
Under \Cref{ass:subgaussian}.\ref{ass:subgaussian1}, convergence of 
this upper bound 
to 0 depends on the size of the terms $\tilde{S}$ and $S_2$, and the relative growth rates of $N$ and $T$. As $N$ only enters 
in logs compared to $\sqrt{T}$ in the denominator, it is possible to have $N$ 
grow at some exponential rate of $T$. Under \Cref{ass:subgaussian}.\ref{ass:subgaussian2}, $N$ enters the numerator at a polynomial rate through the sequence $d_{N}$; this effectively restricts the growth rate of $N$ to some polynomial of $T$, though it can still grow faster than $T$ when $m$ is sufficiently large. 
Our results under these two sets of assumptions therefore mainly differ
 (apart from the different proof strategies required for each case), 
in this regard:
if exponential growth of $N$ is desirable, we need finite exponential moments of $\bepsilon_t$; whereas if polynomial growth of $N$ is sufficient, we only need finite polynomial moments of $\bepsilon_t$.

\begin{remark} \label{remark:extensions}
This Gaussian approximation can be extended to hold over sets of rectangles, i.e.\ for  $\sup\limits_{A\in\mathcal{R}}\abs{\P\left(\frac{1}{\sqrt{T}}\sum_{t=1}^T\bx_t\in A\right)-\P\left( \bz\in A\right)}$. For the details of this extension, see e.g.\ \cite{CCK13} or \cite{CCK17}. Similarly, Theorem \ref{thm:HDCLTforLP} can be extended to  approximate means, i.e.\ $\sup\limits_{A\in\mathcal{R}}\abs{\P\left(\hat{\bS}\in A\right)-\P\left( \bz\in A\right)}$, where $\hat{\bS}=\frac{1}{\sqrt{T}}\sum_{t=1}^T\bx_t + \bR_T$, with $\norm{\bR_T}_{\infty}=o_p(1/\sqrt{\log N})$ along the lines of  Lemma 1 in \cite{chernozhukov2022high}.
\end{remark}

\section{Application to VAR Models}\label{sec:HDCLTVAR}
\Cref{thm:HDCLTforLP} is a key building block in our derivations for the bootstrap, as it can be applied to our VAR in \Cref{eq:DGPVAR} under appropriate conditions. In this section, we explain our assumptions on the VAR process, and on the consistency properties of lasso estimation. 
While the lasso is our 
running example, the following theoretical results do not rely on the lasso specifically, and are equally valid for any other estimation method which satisfies our consistency conditions. We return to the lasso in \Cref{sec:lassoexample}, where we show examples of it satisfying these conditions. 

For the following exposition, it is useful to define the companion matrix 
\begin{equation*}
    \mA=\left(\begin{array}{cccc}
         \bA_1& \bA_2 &\hdots & \bA_{K}\\
         \bI & \bzero &\hdots&\bzero\\
         \vdots& \ddots &&\vdots\\
         \bzero &\hdots&\bI&\bzero
    \end{array}\right).
\end{equation*}
of the VAR in \Cref{eq:DGPVAR}. This matrix allows us to re-write the VAR($K$) as a VAR(1) with 
\begin{equation*}
    \cX_t=\mA\cX_{t-1}+\left[\begin{array}{c}
         \bepsilon_t \\
         \bzero
    \end{array}\right], 
\end{equation*}
and allows for a simple expression for the corresponding VMA coefficients in \Cref{eq:DGPVMA}:  $\bB_k=\bJ\mA^k\bJ^\prime$, where $\underset{N\times KN}{\bJ}=\left(\bI,\bzero,\dots,\bzero\right)$.\footnote{See page 279 of \cite{paparoditis1996}.} This inversion is only possible if the VAR is invertible. 
\begin{assumption}\label{ass:VARsummable}
    Let $\norm{\mA^j}_{\infty}\leq \psi_N \theta^j,$
    for some $0<\theta\leq C<1$, all $j\in \mathds{N}_0$, and $1\leq \psi_N<\infty$ a sequence potentially growing as $N\to\infty$.
\end{assumption}
\Cref{ass:VARsummable} is based on Assumption 1(ii) of \cite{Krampe19}, and its purpose is twofold. First, it allows us to derive summability properties for the quantities $\tilde{S}$ and $S_q$ in \Cref{sec:HDCLTlinear}, since $\norm{\bB_j}_{\infty}\leq \norm{\mA^j}_{\infty}\leq \psi_N\theta^j$. Second, it implies that the VAR process in \Cref{eq:DGPVAR} is stable, since $\rho(\mA)=\lim\limits_{k\to\infty}\norm{\mA^k}_{\infty}^{1/k}\leq\lim\limits_{k\to\infty}\left(\psi_N\theta^k\right)^{1/k}=\theta$, and it can therefore be inverted into a VMA. Based on this inequality, it is also clear that when $k$ is large, $\norm{\mA^k}_{\infty}\approx \rho(\mA)^k\leq \theta^k$, i.e., the powers of $\mA$ will eventually converge at an approximately exponential rate. The \textit{magnitude} of $\psi_N$ controls the magnitude of $\norm{\mA}_\infty$, which may be substantially larger than 1 even in VAR models with low persistence. The \textit{growth rate} of $\psi_N$ controls how quickly $\norm{\mA^k}_\infty$ approaches $\theta^k$, as the dimension of $\mA$ increases. 
Sequences of VAR models which require $\psi_N$ to grow were (to our knowledge) first highlighted 
in \cite{liu2021bernstein}, who relate the growth of $\psi_N$ to spatial dependence, as opposed to temporal dependence tied to $\theta$. 

While our results allow for DGPs with $\psi_N$ growing, it should be noted that such DGPs suffer in terms of convergence rates required for bootstrap validity, and many are already implicitly excluded by \Cref{ass:covariance}. To illustrate this, consider a VAR(1) where $\norm{\mA^j}_{\infty}$ grows with $N$. In many cases this leads to  $\cB(1)=\sum\limits_{j=0}^{\infty} \mA^j$ growing with $N$ as well, resulting in $\sigma_j^2$ growing. However, this is not always the case, and \Cref{ex:off_diagonal_VAR} in Appendix \ref{app:misc} shows a DPG which satisfies \Cref{ass:covariance} while requiring $\psi_N$ to grow exponentially with $N$.

Next, we make the following assumptions about consistency of the estimators $\hat\mA$, and the residuals $\hat\bepsilon_t$:
 \begin{assumption}\label{ass:VARconsistency}
 For a sequence $\xi_{N,T}$, define the set $\mathcal{P}:=\left\lbrace\norm{\hat\mA-\mA}_{\infty}\leq \xi_{N,T}\right\rbrace$. Assume that $\psi_N\xi_{N,T}\leq \bar{C}(1-\theta)^2$ for some $0<\bar{C}<1$, and $\lim\limits_{N,T\to\infty}\P(\mathcal{P})= 1$.
\end{assumption}
\begin{assumption}\label{ass:resconsistency}
For a sequence $\phi_{N,T}$, define the set $\mathcal{Q}:=\left\lbrace\max\limits_{1\leq j\leq N}\frac{1}{T}\norm{\hat\bepsilon_j-\bepsilon_j}_2^2\leq  \phi_{N,T}\right\rbrace$, where $\bepsilon_{j}=(\epsilon_{j,1},\dots,\epsilon_{j,T})^\prime$ and similarly for $\hat\bepsilon_{j}$. Assume that $\lim\limits_{N,T\to\infty}\P(\mathcal{Q})=1$.
\end{assumption}

While we leave the sequences $\xi_{N,T}$ and $\phi_{N,T}$ unspecified and derive later results in terms of these sequences, the reader may think of them as $\xi_{N,T}$ converging at a rate close to $\frac{1}{\sqrt{T}}$ and $\phi_{N,T}$ close to $\frac{1}{T}$ for reasonable estimators. Regarding the assumption that $\psi_N\xi_{N,T}\leq \bar{C}(1-\theta)^2$, a sufficient condition to satisfy this is that $\psi_N\xi_{N,T}\to0$ and $N,T$ are sufficiently large. However, this formulation highlights that our requirements on $\xi_{N,T}$ -- and therefore on the estimation error $\norm{\hat{\mA}-\mA}_{\infty}$ -- are stricter for VARs with large temporal and/or spatial dependence. We elaborate more on these rates when using the lasso in \Cref{sec:lassoexample}. 

\begin{remark} \label{remark:K}

The lag length $K$ is an important feature of the assumed data-generating process, though we do not address its role 
separately
in our assumptions or theoretical results. For many estimation methods, including the lasso, $K$ implicitly affects $\xi_{N,T}$ and $\phi_{N,T}$, because the number of parameters which need to be estimated is $NK$, and the dimension of $\mA$ is $NK\times NK$.
\end{remark}

In our proof strategy, we make use of the probabilistic sets denoted by calligraphic letters $\mathcal{P}$ to $\mathcal{U}$. They describe events involving functions of the random variables $\bx_t$ and $\bepsilon_t$, and can therefore only hold with a certain probability. For the sets $\mathcal{P}$ and $\mathcal{Q}$, we assume that they hold with probability converging to 1 as $N,T\to \infty$. For the other sets, they are chosen in such a way that we can show they hold with probability converging to 1 under our assumptions. For example, relevant to this section are the sets 
 \begin{equation*}
        \mathcal{R}_{1}:=\left\lbrace\max\limits_{1\leq j\leq N}\abs{\frac{1}{T}\sum\limits_{t=1}^T\epsilon_{j,t}^2}\leq C\log(N)\right\rbrace,~ \mathcal{R}_{2}:=\left\lbrace\max\limits_{1\leq j\leq N}\abs{\frac{1}{T}\sum\limits_{t=1}^T\epsilon_{j,t}^2}\leq CN^{2/m}\eta_T^{-1}\right\rbrace,
\end{equation*}
and 
\begin{equation*}
    \mathcal{S}_1:=\left\lbrace\norm{\frac{1}{T}\sum\limits_{t=1}^T\bepsilon_{t}\bepsilon_{t}^\prime-\bSigma_\epsilon}_{\max}\leq C\frac{\sqrt{\log(N)}}{\sqrt{T}}\right\rbrace,~\mathcal{S}_2:=\left\lbrace\norm{\frac{1}{T}\sum\limits_{t=1}^T\bepsilon_{t}\bepsilon_{t}^\prime-\bSigma_\epsilon}_{\max}\leq \frac{N^{4/m}}{T^{3/4}}\eta_T^{-1}\right\rbrace.
\end{equation*}
The different subscripts of these sets indicate for which version of \Cref{ass:subgaussian} they are intended. We show they hold with high probability in \Cref{lma:epsilonConsistent,lma:nohatstoexp}. 
Note that many of our intermediate results are phrased as non-random bounds on random quantities, which hold on these sets, i.e., these bounds hold with probability 1 conditionally on these random events occurring. For the main result in \Cref{thm:bootstrapconsistency}, we then show that the probability of all these random events occurring jointly converges to 1, such that these non-random bounds hold asymptotically.  

The main result of this section 
concerns the consistency of our estimate of $\bSigma$, namely $\hat\bSigma:=\hat\cB(1)\hat\bSigma_{\bepsilon}\hat\cB(1)^\prime$, with $\hat\bSigma_{\bepsilon}:=\frac{1}{T}\sum\limits_{t=1}^T\hat\bepsilon_t\hat\bepsilon_{t}^\prime$, $\hat\cB(z)=\bI+\sum\limits_{k=1}^{\infty}\hat\bB_k z^{k}$, $\hat\cB(z)=\bI+\sum\limits_{k=1}^{\infty}\hat\bB_k z^{k}$.
Unsurprisingly, the form of $\hat\bSigma$ mirrors that of $\bSigma$, since we 
apply the same Beveridge-Nelson decomposition in \Cref{eq:BN} to the bootstrap process. 
To do so, 
the estimated VAR is required to be invertible, i.e. $\rho(\hat\mA)<1$; we show that this is the case with probability converging to 1
in \Cref{lma:trueVMAsummable}.4. This justifies our suggested invertibility 
correction in \Cref{remark:nonstat},
since it is asymptotically negligible.
In
\Cref{thm:CovarianceCloseness} we establish a covariance closeness result which plays a crucial role in showing consistency of our proposed bootstrap method in the next section. 

\begin{theorem}\label{thm:CovarianceCloseness} Let
\Cref{ass:VARsummable,ass:VARconsistency} hold and define the set 
\begin{equation*}
    \mathcal{T}_{1}:=\left\lbrace\norm{\hat\bSigma-\bSigma}_{\max}\leq C\psi_N^2\left[ \phi_{N,T}+d_{N}\sqrt{\phi_{N,T}}+\frac{d_N}{\sqrt{T}}+\xi_{N,T}\psi_N\right]\right\rbrace.
\end{equation*}
Under \Cref{ass:subgaussian}.\ref{ass:subgaussian1}, on $\mathcal{P}\bigcap\mathcal{Q}\bigcap\mathcal{R}_{1}\bigcap\mathcal{S}_{1}$,  $\mathcal{T}_1$ holds. 

Furthermore, define the set 
\begin{equation*}
    \mathcal{T}_{2}:=\left\lbrace\norm{\hat\bSigma-\bSigma}_{\max}\leq C\psi_N^2\left[ \phi_{N,T}+d_{N}\sqrt{\phi_{N,T}}+d_N^4+\xi_{N,T}\psi_N\right]\right\rbrace.
\end{equation*}
Under \Cref{ass:subgaussian}.\ref{ass:subgaussian2}, on $\mathcal{P}\bigcap\mathcal{Q}\bigcap\mathcal{R}_{2}\bigcap\mathcal{S}_{2}$,  $\mathcal{T}_2$ holds. $d_N$ is defined as in \Cref{thm:HDCLTforLP} respectively.
\end{theorem}

\section{Bootstrap Consistency}\label{sec:bootstrapworld}
In this section, we introduce some of the bootstrap-related notation, and flesh out the exact properties of the processes $\bx_t^*$ and $\bepsilon_t^*$. In \Cref{thm:HDCLTforboot}, we then give a Gaussian approximation for the bootstrap process, mirroring \Cref{thm:HDCLTforLP}. 
Finally,  \Cref{thm:bootstrapconsistency} provides the main result of bootstrap consistency.

As is customary in the bootstrap literature, we define the following bootstrap conditional notation: Let $\P^*\left(\cdot\right)$ denote the bootstrap probability 
conditional on the sample $\bX$, and $\E^*\left(\cdot\right)$ the expectation with respect to $\P^*$,
and similarly ley $\norm{x}_{\psi_2}^*:=\inf\left\lbrace c>0:\E^*\exp(\abs{x}^2/c^2)\leq 2\right\rbrace$ and  $\norm{x}_{L_p}^*:=\left(\E^*\abs{x}^p\right)^{1/p}$ denote the corresponding conditional norms.
We let
\begin{equation*}\begin{split}
    \bepsilon_t^*:=\left\lbrace\begin{array}{cc}
    \hat\bepsilon_t\gamma_t &  t=1,\dots,T\\
     \bzero& t< 1
\end{array}\right.,~\gamma_t\iid N(0,1),
\end{split}\end{equation*}
and $\bx_t^*$ built from $\bepsilon_t^*$   
\begin{equation}\label{eq:DGPbootstrap}\begin{split}
\bx_{t}^*:=\left\lbrace\begin{array}{cc}
    \sum\limits_{k=1}^{K}\bA_k^*\bx_{t-k}^*+\bepsilon_t^* &  t=1,\dots,T\\
     \bzero& t<1
\end{array}\right.
\end{split}\end{equation}
where $\bA_{k}^*:=\hat\bA_k$. 
By construction, the bootstrap processes $\bx_t^*$ and $\bepsilon_t^*$ then follow a VAR process mirroring \Cref{eq:DGPVAR}, and can be inverted under appropriate conditions to a VMA process mirroring \Cref{eq:DGPVMA}: $\hat\bB_k=\bJ\hat\mA^k\bJ^\prime$, where $\underset{N\times KN}{\bJ}=\left(\bI,\bzero,\dots,\bzero\right)$.
This then also leads to the bootstrap versions of $\tilde{S}$ and $S_q$,
and the following bootstrap equivalent of \Cref{thm:HDCLTforLP}.
\begin{theorem}[Gaussian approximation for the bootstrap process]\label{thm:HDCLTforboot} Let $\bx_t^*$ be a linear process as in \Cref{eq:DGPbootstrap}, let \Cref{ass:covariance,ass:VARsummable,ass:VARconsistency} hold.
Define the sets
\begin{equation*}
\mathcal{U}_1:=\left\lbrace\max\limits_{j,t}\abs{\epsilon_{j,t}}\leq \sqrt{\log(N)}\log(T)\right\rbrace,~\mathcal{U}_2:=\left\lbrace\max\limits_{j,t}\abs{\epsilon_{j,t}}\leq (NT)^{1/m}\eta_{T}^{-1}\right\rbrace,
\end{equation*}
the bootstrap VMA coefficient sums $\tilde{S}^*:=\sum\limits_{j=0}^\infty \norm{\hat\bB_j}_{\infty}$, $S_q^*:=\sum\limits_{j=0}^{\infty}\left(\sum\limits_{k=j+1}^\infty \norm{\hat\bB_{k}}_\infty\right)^{q}$, and
\begin{equation*}
     {J_{N,T}^*}:=\sup\limits_{y\in\mathbb{R}}\abs{\P^*\left(\norm{\frac{1}{\sqrt{T}}\sum_{t=1}^T\bx_t^*}_{\infty}\leq y\right)-\P^*\left(\norm{ \bz}_\infty\leq y\right)},
\end{equation*}
where $\bz\sim N(\bzero,\bSigma)$.
\begin{enumerate}
\item Under \Cref{ass:subgaussian}.\ref{ass:subgaussian1}, on $\mathcal{P}\bigcap\mathcal{Q}\bigcap\mathcal{T}_1\bigcap\mathcal{U}_1$,
\begin{equation*}\begin{split}
    {J_{N,T}^*}&\leq C\left\lbrace\log(N)\log(T)\psi_N^2\left[d_{N}\sqrt{\phi_{N,T}}+\frac{d_N}{\sqrt{T}}+\xi_{N,T}\psi_N\right]+\frac{\log(N)d_{N}^*\sqrt{S_2^*}}{\sqrt{T}}+\frac{1}{\log(N)}\right.\\
     &\left.\qquad+(\tilde{S}^*d_{N}^*)^2\left[\frac{\log(N)^{3/2}\log(T)}{\sqrt{T}}+\frac{\log(N)^2\log(T)^2}{T}\right]+\sqrt{\frac{\log(N)^2\log(T)\log(NT)}{T}}\right\rbrace,
\end{split}\end{equation*}
where $d_{N}^*=C\left(\sqrt{T\phi_{N,T}}+\sqrt{\log(N)}\log(T)\right)$.
    \item Under \Cref{ass:subgaussian}.\ref{ass:subgaussian2}, on $\mathcal{P}\bigcap\mathcal{Q}\bigcap\mathcal{T}_2\bigcap\mathcal{U}_2$,
   \begin{equation*}\begin{split}
    &{J_{N,T}^*}\leq C\left\lbrace\log(N)\log(T)\psi_N^2\left[d_{N}\sqrt{\phi_{N,T}}+\frac{d_N^4}{T^{3/4}}+\xi_{N,T}\psi_N\right]+(Nd_{N}^{*m}\psi_{N}^m)^{\frac{1}{m+1}}\left(\frac{\sqrt{\log(N)}}{\sqrt{T}}\right)^{\frac{m}{m+1}}\right.\\
     &+\left.(\tilde{S}^*d_{N}^*)^2\left[\frac{\log(N)^{3/2}\left(\log(T)+(\tilde{S}^*d_{N}^*)^{\frac{1}{m-1}}\right)}{\sqrt{T}}+\frac{\log(N)^2\log(T)}{T^{\frac{m-2}{m}}}\right]+\sqrt{\frac{\log(N)^2\log(T)\log(NT)}{T}}\right\rbrace,
\end{split}\end{equation*}
where $d_{N}^*= C\left(\sqrt{T\phi_{N,T}}+(NT)^{1/m}\eta_T^{-1}\right)$.
\end{enumerate}
\end{theorem}
Since $\bz$ in \Cref{thm:HDCLTforboot} is the same as in \Cref{thm:HDCLTforLP}, we can combine both theorems and a telescopic sum argument to bound the distance between distributions of $\norm{\frac{1}{\sqrt{T}}\sum_{t=1}^T\bx_t}_{\infty}$ and $\norm{\frac{1}{\sqrt{T}}\sum_{t=1}^T\bx_t^*}_{\infty}$, giving us bootstrap consistency in the following theorem.

\begin{theorem}\label{thm:bootstrapconsistency}Let \Cref{ass:covariance,ass:VARsummable,ass:VARconsistency,ass:resconsistency} hold, and define
\begin{equation*}
    {D_{N,T}}=\sup\limits_{y\in\mathbb{R}}\abs{\P\left(\norm{\frac{1}{\sqrt{T}}\sum_{t=1}^T\bx_{t}}_\infty\leq y\right)-\P^*\left(\norm{\frac{1}{\sqrt{T}}\sum_{t=1}^T\bx_t^*}_\infty\leq y\right)}
\end{equation*}
The following hold with probability converging to 1 as $N,T\to\infty$. 

Under \Cref{ass:subgaussian}.\ref{ass:subgaussian1},
\begin{equation*}\begin{split}
     {D_{N,T}}&\leq  C\left\lbrace\psi_{N}^2\left[\frac{\ell_{N}^3}{\sqrt{T}}+\ell_N\ell_T\left(\ell_N\sqrt{\phi_{N,T}}+\frac{\ell_N}{\sqrt{T}}+\xi_{N,T}\psi_N\right)\right.\right.\\
     &\qquad\qquad\qquad\qquad\qquad+\left.\left.\left(\sqrt{T\phi_{N,T}}+\ell_T\sqrt{\ell_N}\right)^2\left(\frac{\ell_N^{3/2}\ell_T}{\sqrt{T}}+\frac{\ell_N^2\ell_T^2}{T}\right)\right]+\frac{1}{\ell_N}\right\rbrace,
\end{split}\end{equation*}
where $\ell_T=\log(T)$, $\ell_{N}=\log(N)$. 

Under \Cref{ass:subgaussian}.\ref{ass:subgaussian2},
\begin{equation*}\begin{split}
  &{D_{N,T}}
    \leq C\eta_T^{-1}\Biggl\{\frac{\psi_N^4 N^{4/m}\ell_N^2\ell_T}{T^{\frac{m-2}{m}}}+\psi_N^{\frac{2m}{m-2}}\frac{\left(N^2\ell_N^{\frac{3m-8}{2}}\ell_T\ell_{NT}\right)^{\frac{1}{m-2}}}{\sqrt{T}}\\
    +&\psi_N^2\Biggl[\ell_N\ell_T\left(\frac{N^{4/m}}{T^{3/4}}+\xi_{N,T}\psi_N\right)+\left(\sqrt{T\phi_{N,T}}+(NT)^{1/m}\right)^2\frac{\ell_N^{3/2}\left(\ell_T+\psi_N^{\frac{1}{m-1}}\left(\sqrt{T\phi_{N,T}}+(NT)^{1/m}\right)^{\frac{1}{m-1}}\right)}{\sqrt{T}}\Biggr]\\
    &\qquad\qquad\qquad\qquad\qquad+\left(\psi_N\frac{\sqrt{\ell_N}}{\sqrt{T}}\right)^{\frac{m}{m+1}}\left[N^{\frac{2}{m+1}}+N^{\frac{1}{m+1}}\left(\sqrt{T\phi_{N,T}}+(NT)^{1/m}\right)^{\frac{m}{m+1}}\right]\Biggr\},
\end{split}\end{equation*}
where $\ell_{NT}=\log(NT)$.
\end{theorem}
\section{Bootstrap Consistency for VAR Estimation by the Lasso}\label{sec:lassoexample}
The application of our proposed bootstrap method requires that the lasso satisfies \Cref{ass:VARconsistency,ass:resconsistency} with sequences $\psi_{N}$, $\xi_{N,T}$, and $\phi_{N,T}$ such that the bound in \Cref{thm:bootstrapconsistency} converges to 0. In this section, we show that this is the case under both options of \Cref{ass:subgaussian}, and under both weak and exact row-wise sparsity of the underlying VAR.

As described in \Cref{sec:bootexplained} we propose to estimate the VAR equation-by-equation, using the lasso estimators in \Cref{eq:problem}.
Our goal is therefore to find bounds on $\max\limits_{j}\norm{\hat\bbeta_j-\bbeta_j}_1$ and $\max\limits_{j}\frac{1}{T}\norm{\hat\bepsilon_{j}-\bepsilon_j}_2^2=\max\limits_{j}\frac{1}{T}\sum\limits_{t=1}^T\left[(\hat\bbeta_j-\bbeta_j)^\prime\cX_t\right]^2$.
For this purpose, we will be using error bounds in Corollary 1 of our previous work in \cite{adamek2021lasso}, though similar error bounds have been derived in different contexts by other authors; see e.g.~\cite{bickel2009simultaneous}, \cite{KockCallot2015}, \cite{MedeirosMendes16}, 
and \cite{masini2019regularized}.
Next, we will elaborate on the assumptions under which these error bounds hold.

For Assumption 1 of \cite{adamek2021lasso}, we have $\E\bx_t=0\implies \E\cX_t=0$ by the structure of \Cref{eq:DGPVAR}, and $\E\bx_t\epsilon_{j,t}=0,~\forall j$, by independence of the errors. We then need to assume that $\max\limits_{j,t}\E\abs{x_{j,t}}^m\leq C$ in addition to \Cref{ass:subgaussian}.\ref{ass:subgaussian2} in this paper to ensure the first part of the assumption is satisfied. This high-level assumption on moments of $x_{j,t}$ can also be shown to hold under more primitive conditions, such as a moment condition on linear combinations of the errors, $\max\limits_{\norm{\bu}_2\leq 1, t}\E\abs{\bu^\prime\bepsilon_{t}}^m\leq C$, and a new summability condition on the rows of $\bB_k$, $\max\limits_{j}\sum\limits_{k=0}^\infty\norm{\bb_{j,k}}_2^m\leq C$:
\begin{equation*}
    \max\limits_{j,t}\norm{x_{j,t}}_{L_m}
    \leq \sum\limits_{k=0}^{\infty}\max\limits_{j,t}\norm{\bb_{j,k}\bepsilon_{t-k}}_{L_m}=\sum\limits_{k=0}^{\infty}\norm{\bb_{j,k}}_2\norm{\frac{\bb_{j,k}}{\norm{\bb_{j,k}}_2}\bepsilon_{t-k}}_{L_m}=\norm{\bu^\prime\bepsilon_{t-k}}_{L_m}\sum\limits_{k=0}^{\infty}\norm{\bb_{j,k}}_2.
\end{equation*}
Note that $m$ in this paper corresponds to $2\bar{m}$ in \cite{adamek2021lasso}. Under an additional assumption that $\psi_N\leq C$,\footnote{This additional assumption is in line with e.g.~\cite{kock2024data} who require this in their Assumption 2.(2) to obtain error bounds on the lasso.} \Cref{ass:VARsummable} ensures that the NED assumption is satisfied uniformly across equations and as $N$ grows.  
The VMA coefficients decay at an exponential rate, 
therefore satisfying any polynomial decay rate on the NED sequence, and the assumption is satisfied for any arbitrarily large $d$.
Assumption 2 of \cite{adamek2021lasso} requires that the rows of $\mA$ are weakly sparse, in the sense that $\norm{\bbeta_j}_r^r=\norm{[\mA]_{j,\cdot}}_r^r\leq s_{r,j}$ for some $0\leq r<1$. 
Assumption 3 of \cite{adamek2021lasso} requires that the covariance matrix of the regressors satisfies a form of compatibility condition; for simplicity, we can assume that $\Lambda_{\min}\left(\frac{1}{T}\sum\limits_{t=1}^T\E\cX_{t}\cX_{t}^\prime\right)$ is bounded away from zero, which is sufficient to satisfy the condition simultaneously for all equations. For an example of conditions when this is satisfied, see Equation 6 of \cite{masini2019regularized}.  
Under these conditions, we have by Corollary 1 of \cite{adamek2021lasso} that 
\begin{equation*}\begin{split}
   &\frac{1}{T}\norm{\hat\bepsilon_{j}-\bepsilon_j}_2^2\leq  C\lambda_j^{2-r}s_{r,j},\quad \norm{\hat{\boldsymbol{\beta}_j}-{\boldsymbol{\beta}}_j}_1
\leq C\lambda_j^{1-r}s_{r,j},
\end{split}\end{equation*}
with probability converging to 1 under appropriate restrictions on the $\lambda_j$, detailed in Theorem 1 of \cite{adamek2021lasso}. Note that these restrictions are a function of the dependence (NED size $d$) and sparsity ($s_{r,j}$) within each equation, so in order to satisfy \Cref{ass:VARconsistency,ass:resconsistency}, these properties should hold uniformly across equations.

To further simplify this result, we can use the asymptotic setup of Example C.1 of \cite{adamek2021lasso} where $N$, $\lambda_j$, and $s_{r,j}$ grow at a polynomial rate of T. While that example provides the full details on the tradeoff between $r$, the number of moments, and the growth rates of $s_{r,j}$ and $N$ relative to $T$, here we fix $r=1/2$ and $s_{r,j}\sim T^{1/8},~\forall j$  for illustrative purposes.
\begin{corollary}[Finite absolute moments]\label{cor:finitemoments}
Let Assumptions \ref{ass:covariance}, \ref{ass:subgaussian}.\ref{ass:subgaussian2}, and \ref{ass:VARsummable}-\ref{ass:resconsistency} hold. Furthermore, assume $\max\limits_{j,t}\E\abs{x_{j,t}}^m\leq C$, 
$\max\limits_{j}\sum\limits_{k=1}^{KN}\abs{\left[\mA\right]_{j,k}}^{1/2}\leq CT^{1/8}$,
and $\Lambda_{\min}\left(\frac{1}{T}\sum\limits_{t=1}^T\E\cX_{t}\cX_{t}^\prime\right)\geq 1/C$. Let  $K\leq C$,
$N\sim T^a$ for $a>0$, $\psi_N\leq C$, 
and $\lambda_j\sim T^{-\ell}$ for all $j$, with $\ell<\frac{3}{4}-\frac{4a+1}{m}$. The lasso then satisfies \Cref{ass:resconsistency,ass:VARconsistency} with $\xi_{N,T}=\eta_T^{-1}T^{\left(\frac{4a+1}{2m}-\frac{1}{4}\right)}$ and $\phi_{N,T}=\eta_T^{-1}T^{\left(\frac{12a+3}{2m}-1\right)}$.

When  
$m>\sqrt{36a^2+18a+5/2}+6a+1$, ${D_{N,T}}\to0$ with probability converging to 1 as $N,T\to \infty$.
\end{corollary}

While \Cref{cor:finitemoments} shows an example of conditions for bootstrap consistency using the finite absolute moments in \Cref{ass:subgaussian}.\ref{ass:subgaussian2}, the stronger assumption of sub-gaussian moments in \Cref{ass:subgaussian}.\ref{ass:subgaussian1} allows for faster growth of $N$ relative to $T$. In this scenario, we can consider the error bounds in Theorem 2 of \cite{KockCallot2015},
\begin{equation*}
    \frac{1}{T}\norm{\hat\bepsilon_{j}-\bepsilon_j}_2^2\leq  C\lambda_j^{2}s_{0,j}/\kappa_j,\quad \norm{\hat{\boldsymbol{\beta}_j}-{\boldsymbol{\beta}}_j}_1
\leq C\lambda_j s_{0,j}/\kappa_j,
\end{equation*}
with $\lambda_j=C\ell_{T}^{5/2}\ell_{N}^{2}\ell_{K}\ell_{N^2K}^{1/2}\sigma_T^2/\sqrt{T}$. Note that $\sigma_T^2$ denotes the largest variance among all $\epsilon_{j,t}$ and $x_{j,t}$, so we once again make the high level assumption that $\max\limits_{j,t}\E x_{j,t}^2\leq C$.  
To obtain these bounds, we need the additional assumption that the errors are Gaussian, so $\bepsilon_t\iid N(\bzero,\bSigma_{\epsilon})$, which implies \Cref{ass:subgaussian}.\ref{ass:subgaussian1}. Additionally, they consider the case of exact sparsity, with $\sum\limits_{k=1}^{KN}\ind{\abs{\left[\mA\right]_{j,k}}>0}\leq s_{0,j}$. Finally, $\kappa_{j}$ play a similar role to the compatibility constant in Assumption 2 of \cite{adamek2021lasso}, and are bounded away from 0 when $\Lambda_{\min}\left(\frac{1}{T}\sum\limits_{t=1}^T\E\cX_{t}\cX_{t}^\prime\right)\geq 1/C$, see the discussion on page 7 of \cite{KockCallot2015} for details. Regarding the growth rates of $N$ and $s_{0,j}$, we take a similar example to Theorem 3 of \cite{KockCallot2015}, with $N\sim e^{(T^a)}$ and $s_{0,j}\leq C T^b$.
\begin{corollary}[Sub-gaussian moments]\label{cor:subgaussianmoments}
Let Assumptions \ref{ass:covariance}, 
and \ref{ass:VARsummable}-\ref{ass:resconsistency} hold. Furthermore, assume $\max\limits_{j,t}\E\abs{x_{j,t}}^2\leq C$, 
$\max\limits_{j}\sum\limits_{k=1}^{KN}\ind{\abs{\left[\mA\right]_{j,k}}>0}\leq CT^b$ for some $b>0$,
and $\Lambda_{\min}\left(\frac{1}{T}\sum\limits_{t=1}^T\E\cX_{t}\cX_{t}^\prime\right)\geq 1/C$. Let $K\leq C$,
$N\sim e^{(T^a)}$ for $a>0$, $\psi_{N}\leq C$,
and $\lambda_j\sim\ell_T^{5/2}T^{(5a-1)/2}$. The lasso then satisfies \Cref{ass:resconsistency,ass:VARconsistency} with $\xi_{N,T}=C\ell^{5/2}T^{\frac{5a+2b-1}{2}}$ and $\phi_{N,T}=C\ell_T^5T^{5a+b-1}$.

When  
$13a+2b<1$,
${D_{N,T}}\to0$ with probability converging to 1 as $N,T\to \infty$.
\end{corollary}
\section{Simulations}\label{sec:simulations}
To evaluate the finite sample performance of our proposed method, our simulation study covers a variety of DGPs on which we compare size and power with other bootstrap methods typically used in a high-dimensional time series setting.\footnote{A replication package for these simulations is available at \href{https://github.com/RobertAdamek/sparseVARrepro}{github.com/RobertAdamek/sparseVARrepro}.}

\subsection{Setup}
We implement our proposed VAR multiplier bootstrap with two different ways of selecting the lasso penalty. First, we estimate the VAR with the penalty chosen by the Bayesian information criterion jointly over all equations (VAR-BIC). Second, we use the theoretically founded data-driven method of \cite{kock2024data} (VAR-TF).
For both methods the number of lags $K$ is chosen as the informative upper bound in Section 5 of \cite{HecqMargaritellaSmeekes2023}, as mentioned in Remark \ref{remark:Klags}. For details, see Algorithm E.1 in \cite{sparseVARsupplement}. Additionally, we leave the diagonal elements of the VAR coefficient matrices unpenalized in the lasso estimation. We believe this is good common practice with lasso VAR estimation, because a series' own lags are often more important than those of other series for explaining the dynamic properties. This approach is similar to the ``Own-Other'' hierarchical penalties in \cite{Wilms2020Hlag} or the Minnesota prior in Bayesian VAR estimation. To guarantee stability of the estimated VAR, we apply the finite sample correction in Remark \ref{remark:nonstat}.

As a benchmark, we also show results for the `oracle' method, which does no VAR estimation, and generates bootstrap samples using the true VAR coefficients (\emph{VAR-oracle}).  

In addition to the VAR-based bootstrap, we consider two block-based bootstrap methods: the block wild/multiplier bootstrap ({BWB) based on e.g.~\cite{Shao2011} or \cite{ZhangCheng2014bootstrapping}, and the moving block bootstrap based on 
%e.g.~\cite{PALM2011} or \cite{Smeekes2015} 
e.g.~\cite{Kunsch1989MBB}
(MBB). For both block-based bootstraps, we use a block length using the automatic bandwidth estimator for the Bartlett kernel in \cite{andrews1991heteroskedasticity}, see e.g.~page 974 in \cite{Goncalves2005}.

We study four DGPs used by other work in this field. Specifically, we take inspiration from \cite{Barigozzi2024FNETS,KockCallot2015,Krampe19}. 
%\cite{}, \cite{}, \cite{}. 
In all DGPs, we consider every combination of $T\in\left\lbrace 50,100,200,500\right\rbrace$, and $N\in\left\lbrace 20,40,100,200 \right\rbrace$. To estimate size, we generate the data with population mean $0$ for each variable. The nominal level is $\alpha=0.05$, and for better readability, all size plots are truncated at a rejection rate of 0.5. For power, we add a nonzero constant $\mu$ to a proportion $p$ of variables, such that the first $Np$ variables have mean $\mu$ and the remaining $N(1-p)$ variables have mean $0$. We consider $p=0.5$ for all DGPs, and choose $\mu$ separately for each DGP according to an initial calibration exercise, such that the power is relatively low (around 25\%) for $N=20$, $T=50$. In 
DGP1, we also investigate the effects on power of increasing $p$ to $0.9$, and doubling $\mu$.   

\subsection{DGP1: Diagonal VAR(1)}
This DGP is based on Experiment A of \cite{KockCallot2015}:
\begin{equation}\label{eq:VAR1sim}
    \bx_t=\bA\bx_{t-1}+\bepsilon_t,~\bepsilon_t \iid N(\bzero, \bSigma_\epsilon),~t=1,\dots,T,
\end{equation}
where $\bA=\diag(0.5,\dots,0.5)$ and $\bSigma_\epsilon=\diag(0.01,\dots,0.01)$. This DGP satisfies \Cref{ass:covariance} with $\Lambda_{\min}\left(\bSigma\right)=\max\limits_{1\leq j\leq N}\sigma_j^2=0.04$ for all $N$, \Cref{ass:subgaussian}.\ref{ass:subgaussian1} with Gaussian errors and \Cref{ass:VARsummable} with $\theta=0.5$, $\psi_N=1$. This DGP is the ``best-case'' setup for our proposed method because the lasso generally performs well in sparse models, and all the true non-zero parameters in this DGP are left unpenalized. 

\begin{figure}[ht]
\centering
\includegraphics[width=\linewidth]{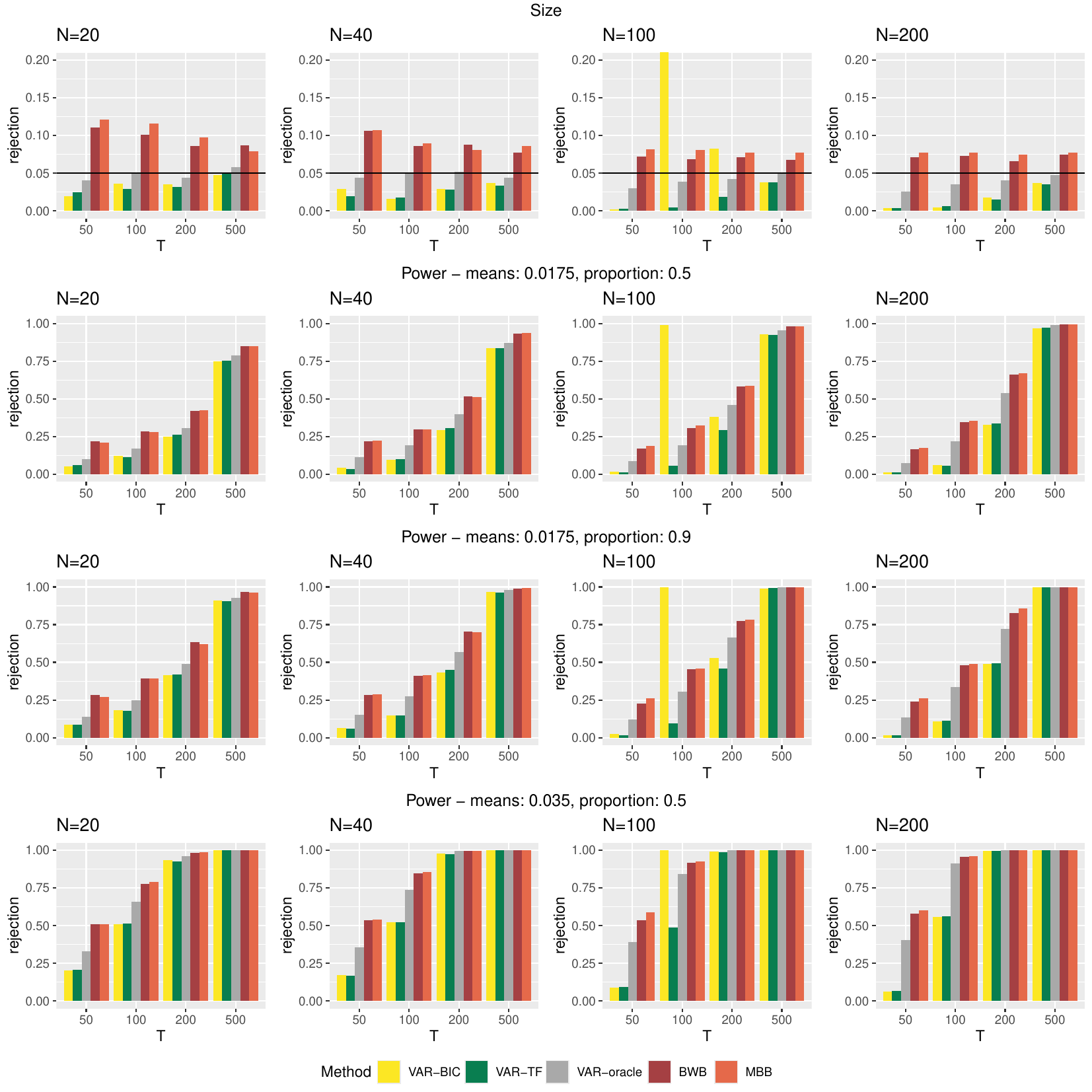}
\caption{DGP1: Diagonal VAR(1), size and power.}\label{fig:dgp2_size_power}
\end{figure}

Regarding the size in the top row of \Cref{fig:dgp2_size_power}, we generally see the VAR-based methods achieve correct, slightly conservative size. With the exception of $N=100,~T=100$, VAR-BIC and VAR-TF perform very similarly, being slightly more conservative than the oracle method. They are generally more conservative at larger $N$, but improve and reach close to nominal size as $T$ increases. At $N=100,~T=100$, BIC tends to select a very low value of the tuning parameter, often reaching the lower edge of the grid. This results in models with almost no regularization, excessive variance, and poor performance of VAR-BIC.\footnote{Note that we truncate the size plots at 0.2 rejection rate. In cases where the VAR-BIC fails, its rejection rate reaches as high as 0.45-0.9.} This phenomenon is also observed in later DGPs, so this seems to be a somewhat pervasive issue with BIC. Both block-based bootstrap methods have comparable performance, reaching size between 5 and 15\%. This large size is most pronounced at low $N$, though we see improvement with growing $T$. At $N=200$, both methods exceed 5\% only slightly, with the BWB outperforming the MBB.

Power is given in the bottom three rows of \Cref{fig:dgp2_size_power}. We see similar patterns across all three settings: For all methods, power grows 
considerably with $T$, and slightly with $N$, and reaches close to 100\% at $N=200,~T=500$. The VAR-based methods have slightly lower power than the oracle method, and the block-based methods beat the oracle. This is not necessarily an indictment against the VAR-based methods, as the block-based methods do not achieve size control. The abnormal behavior of the BIC is also reflected in the power, reaching 100\% at $N=100,~T=100$. Comparing between the three settings, we see that increasing the nonzero proportion from $p=0.5$ to $p=0.9$ increases the power only slightly, by around 5-15 percentage points. Doubling the mean from $\mu=0.0175$ to $\mu=0.035$ had a much larger impact, more than doubling the power in most cases. This is not a surprising pattern, given that the test statistic is based on the maximum of means.

\subsection{DGP2: Block-diagonal VAR(1)}\label{subsec:KKP}
DGP2 is based on Example 1 of \cite{Krampe19}. It follows \Cref{eq:VAR1sim} with $\bA$ and $\bSigma_\epsilon$ having a block-diagonal structure. The blocks are $20\times20$ in both cases; their precise definition\footnote{We use the $\xi=0.6$ version of this DGP.} can be found in Appendix D of \cite{Krampe19}, and we provide a visual overview of the pattern within blocks in \Cref{fig:krampe_sim} in Appendix C.\footnote{Note that we do not shuffle the indices of variables like \cite{Krampe19}.} This DGP satisfies \Cref{ass:covariance} with $\Lambda_{\min}\left(\bSigma\right) \approx 0.0782$ and $\max\limits_{1\leq j\leq N}\sigma_j^2 \approx 38.322$ for all $N$ (in multiples of 20), \Cref{ass:subgaussian}.\ref{ass:subgaussian1} with Gaussian errors and \Cref{ass:VARsummable} with $\theta=0.8$, $\psi_N\approx 3.121$. We expect our proposed method to perform well in this DGP: most of the structure within the blocks is on the unpenalized diagonal, and is quite sparse even in the last 6 rows.

\begin{figure}[t]
\centering
\includegraphics[width=\linewidth]{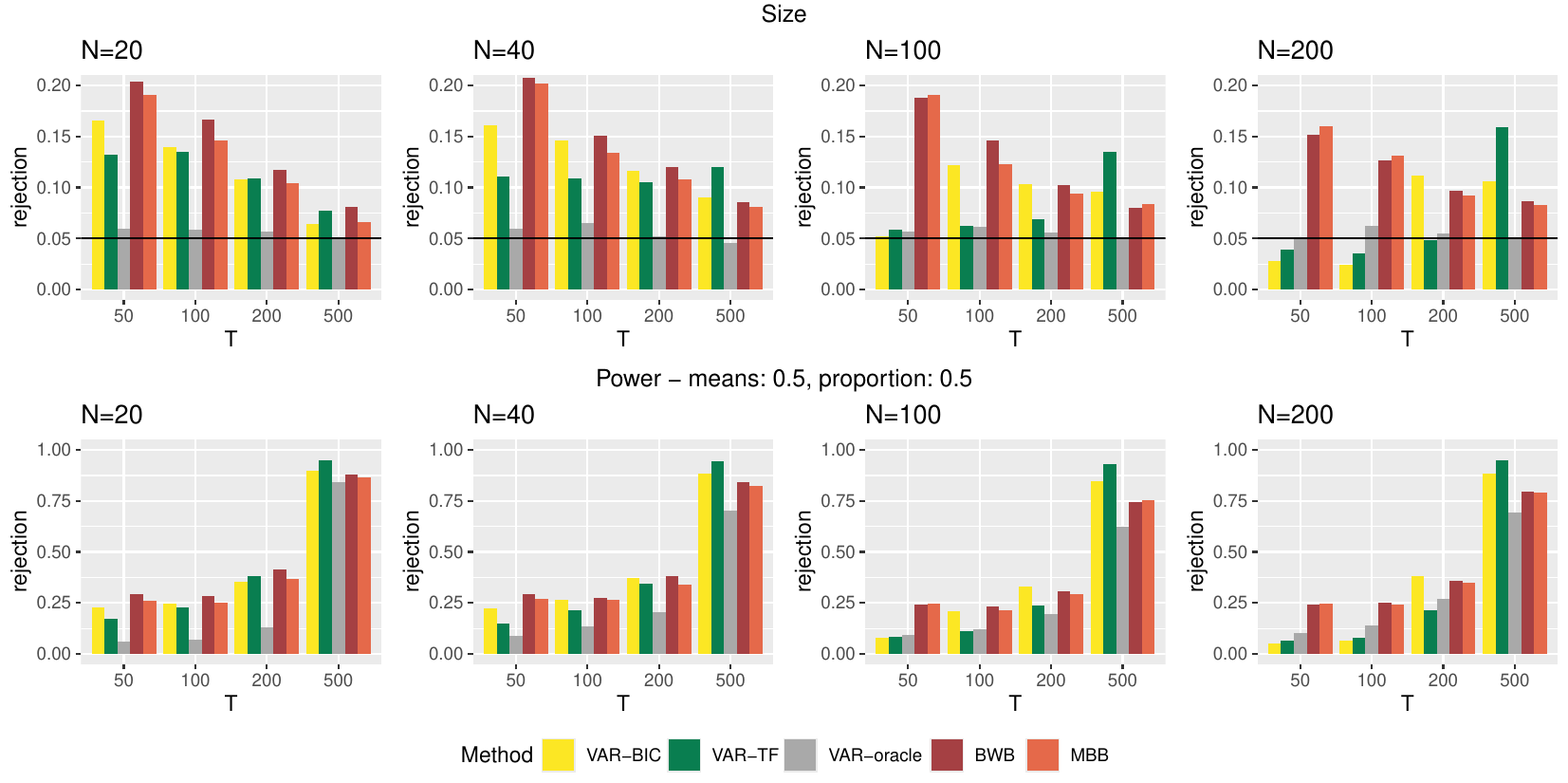}
\caption{Block-diagonal VAR(1), size and power.}\label{fig:dgp1_size_power}
\end{figure}

In terms of size (top row of \Cref{fig:dgp1_size_power}), all methods other than the oracle are generally oversized. Between the VAR-based methods, the VAR-TF has better size than VAR-BIC except at $T=500$ and $N=100,200$, where VAR-TF performs the worst with around $15\%$ size. Except the $T=500$ case, VAR-TF has the best size, with performance close to the oracle at $N=100,200$. For low $T$, VAR-BIC's performance changes significantly over different $N$, with size around $15\%$ at $N=20$, but well below $5\%$ at $N=200$.

Given that the oracle method has the correct size, the relatively poor performance of the VAR-based methods is largely due to estimation. Estimation is challenging in this DGP because of high persistency with $\rho(\mA)=0.8$, as VAR estimates can be heavily biased in such cases, even when using least squares estimation. A classic solution to this issue in low-dimensional settings is the double bootstrap of \cite{Killian1998doublebootstrap}; it is an interesting avenue of future research to investigate whether the results would improve using a similar approach 
in our setting. The block-based methods both have similar performance, with size around $15-20\%$ at $T=50$, and reaching $5-10\%$ at $T=500$. The high persistence of this DGP also hampers the block-based methods, since they need long blocks to accurately capture the dependence.

Regarding the power results displayed in the bottom row of \Cref{fig:dgp1_size_power}, we see large improvements with growing $T$, and changes over $N$ are broadly in line with the changes in rejection rates seen in the size plots.

\subsection{DGP3: Weakly sparse VAR(1)}
This DGP is based on Experiment D of \cite{KockCallot2015}. It follows \Cref{eq:VAR1sim} with $\bA$ having a Toeplitz structure and exponentially decaying off-diagonals, $a_{ij}=(-1)^{\abs{i-j}} \rho^{\abs{i-j}+1}$, $\rho=0.3$.
$\bSigma_\epsilon$ is the same as in 
DGP1. For \Cref{ass:covariance}, $\bSigma$ changes as $N$ grows,
but its properties stabilize at $\Lambda_{\min}\left(\bSigma\right) \approx 0.0234$ and $\max\limits_{1\leq j\leq N}\sigma_j^2 \approx 0.0142$. \Cref{ass:VARsummable} is satisfied with $\theta=0.6$ and $\psi_N=1$. While this DGP is not sparse in the exact sense, it is weakly sparse with elements far from the diagonal taking values very close to zero. The lasso will inevitably set most parameters equal to zero, but we do not 
expect this to have a large impact on performance, since the effect of these near-zero parameters on the dynamic properties of the process is negligible.

\begin{figure}[ht]
\centering
\includegraphics[width=\linewidth]{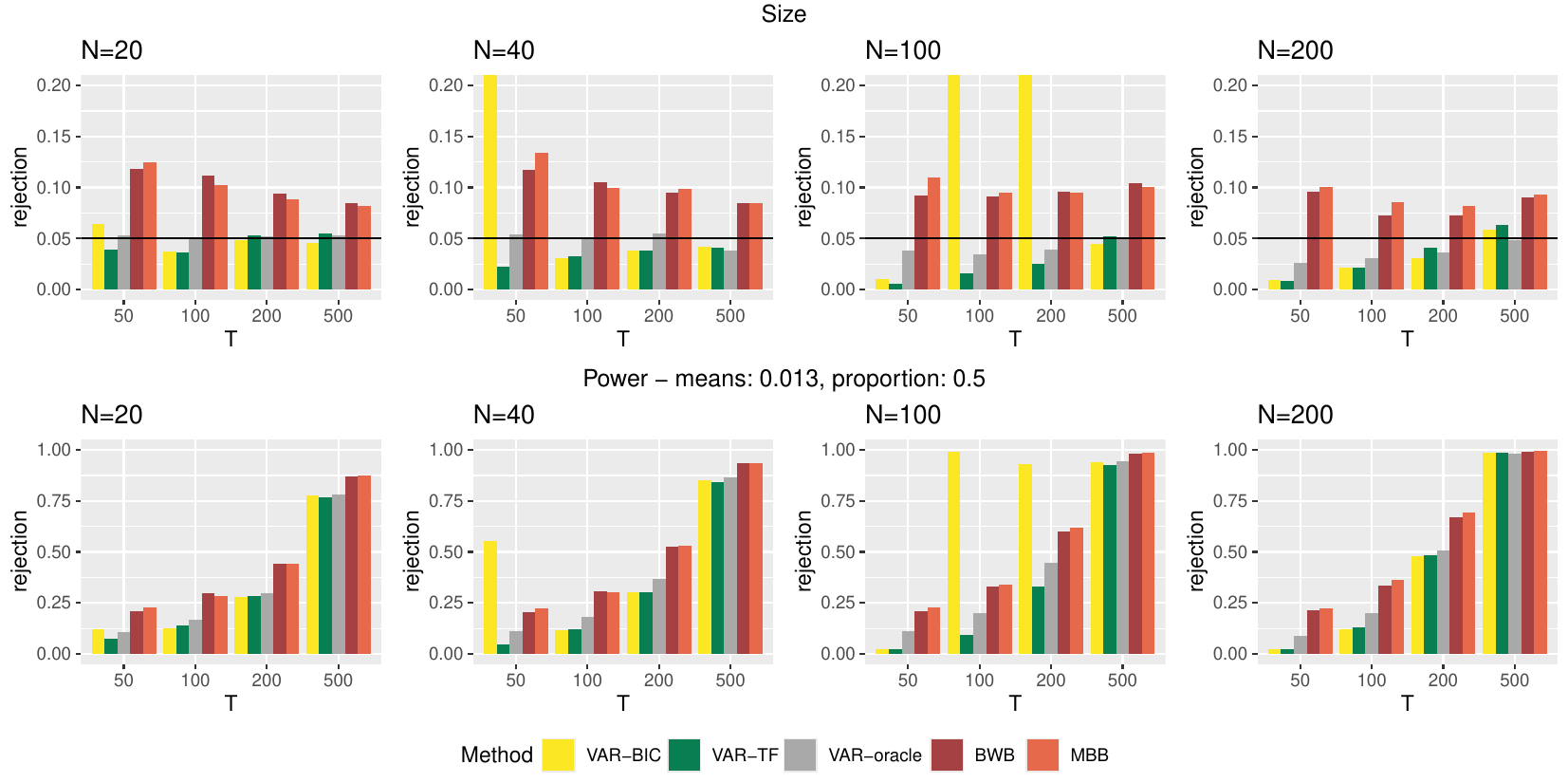}
\caption{Weakly sparse VAR(1), size and power.}\label{fig:dgp9_size_power}
\end{figure}

We see a similar pattern in the size (top row of \Cref{fig:dgp9_size_power}) as for 
DGP1: the VAR-based methods perform similarly, being slightly conservative, except a few cases where VAR-BIC fails. The block-based methods are oversized again, with size around $10\%$.

For power (bottom row of \Cref{fig:dgp9_size_power}) the pattern is also similar to 
DGP1: power generally increases greatly with $T$, and slightly with $N$, and the relative power of different methods is in line with the differences in size.

\subsection{DGP4: Factor model with sparse idiosyncratic component}
This DGP is based on the simulation setup (E1)+(C2) in Appendix E.1 of \cite{Barigozzi2024FNETS}:
\begin{equation*}\begin{split}
    \bx_t&=\bchi_t+\bxi_t,~t=1,\dots,T\\
    \chi_{i,t}&=w_i\sum\limits_{\ell=1}^2 \blambda_{i,\ell}^\prime\bbf_{t-\ell+1},~i=1,\dots,N,\\
    \underset{2\times 1}{\bbf_t}&=\bD\bbf_{t-1}+\bu_t,~\bu_t\iid N(\bzero, \bI),\\
    \bxi_t&=\bA\bxi_{t-1} + \bepsilon_t,~\bepsilon_t\iid N(\bzero,\bI),
\end{split}\end{equation*}
where the entries of $\blambda_{i,\ell}\in\mathds{R}^2$ are generated as i.i.d.~standard Gaussian, $\bD=\bD_0\cdot 0.7 /\Lambda_{\max}(\bD_0)$, where $\bD\in\mathds{R}^{2\times 2}$ has off-diagonal elements generated i.i.d.~from $U[0,0.3]$ and diagonal elements generated from $U[0.5,0.8]$. The $w_i$ are such that the sample estimate of $\text{Var}(\chi_{i,t})/\text{Var}(\xi_{i,t})=1,~\forall i$. To generate $\bA$, first $\bA_0$ is generated, with its entries drawn i.i.d.~from $Bernoulli(1/N)\cdot 0.275$. Then, if $\Lambda_{\max}(\bA_0)\leq 0.9$, $\bA=\bA_0$; otherwise $\bA=\bA_0\cdot 0.9/\Lambda_{\max}(\bA_0)$. This DGP does not fit the VAR structure in \Cref{eq:DGPVAR}, and Assumptions \ref{ass:covariance}-\ref{ass:VARsummable} do not hold. The process is stationary, but if a VAR representation exists, it is likely not sparse due to the factor structure. We expect our proposed method to perform more poorly relative to the block-based bootstrap methods, since it is an adverse setting for the lasso. Note that since the DGP is not a VAR model, the oracle method is not implemented for this DGP.

\begin{figure}[ht]
\centering
\includegraphics[width=\linewidth]{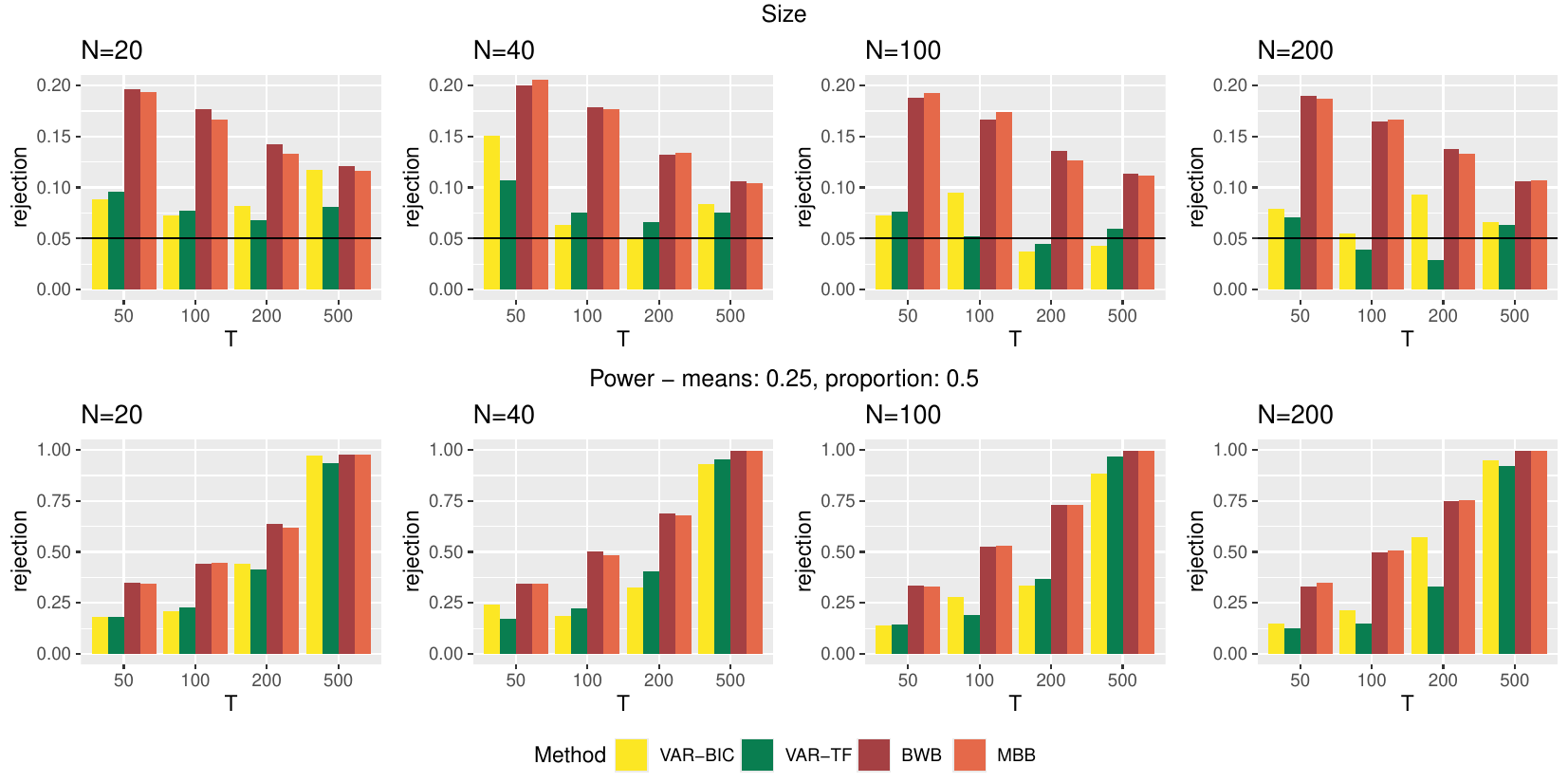}
\caption{Factor model, size and power.}\label{fig:dgp0_size_power}
\end{figure}

Contrary to our expectations, size results in the top row of \Cref{fig:dgp0_size_power}, demonstrate good performance of the VAR-based methods, especially compared to the block-based methods. They are slightly oversized at around $10\%$ for $T=50$, but are close to nominal for larger $T$. On the other hand, the block-based methods are oversized across the board, with size at $20\%$ at $T=50$ and only decreasing to $10\%$ at $T=500$. Power in the bottom row of \Cref{fig:dgp0_size_power} shows improvements with increasing $T$ as for the other DGPs, and not much change with $N$. The relative powers of the different methods is broadly in line with the size differences.

\section{Empirical Application}\label{sec:empirical}
We demonstrate our proposed method on a macroeconomic application. We use data from the Penn World Table version 11.0 \citep{FeenstraInklaarTimmer2015PWT}, available for download at \url{www.ggdc.net/pwt}, and consider the mean growth rate of real GDP\footnote{In particular, we use the variable $\text{RGDP}^{\text{NA}}$, which is Real DGP using national-accounts growth rates.} for $N=158$ countries, spanning the years 1970--2023. For the full list of countries we include and their country codes, see Table \ref{tab:country_list} in Appendix C.3.\footnote{A replication package for this empirical application is available at \href{https://github.com/RobertAdamek/sparseVARrepro}{github.com/RobertAdamek/sparseVARrepro}.}

We perform a one-sided test for the hypothesis
\begin{equation*}
    H_0:\mu_{j}\leq 2\%\text{ for all } j=1,\dots,N \text{ vs. }H_1: \mu_j>2\%\text{ for at least one }j,
\end{equation*}
and compute the associated $p$-values for individual countries using the stepdown procedure described in Section 4.5 of \cite{chernozhukov2022high}. Let $t_1\geq\dots\geq t_N$ be the \textit{ordered} $t$-statistics associated with the original time series
with $t_j=\left(\bar{\bx}_j-\mu_0\right)/\hat{\sigma}_j$
and $t_j^{*b}$ the corresponding $t$-statistics of the bootstrap data where %I.e. $t_j=\left(\bar{\bx}_j-\mu_0\right)/\hat{\sigma}_j$ and 
$t_j^{*b}=\bar{\bx}_j^{*b}/\hat{\sigma}_j^{*b}$ for $j = 1,\dots, N$. We then compute $p$-values recursively as $p_0=0$, $p_j=\max\left[\frac{1}{B+1}\left(\sum\limits_{b=1}^B\mathds{1}\left\lbrace\underset{s\in[j,N]}{\max} t_s^{*b}\geq t_j\right\rbrace+1\right),p_{j-1}\right]$, for $j =1,\dots,N$

In line with Section \ref{sec:simulations}, we estimate these $p$-values by our proposed method VAR-TF, and compare them with the block bootstrap methods BWB and MBB. To demonstrate the importance of correctly treating the time dependence in the data, we additionally include $p$-values estimated by the simple Gaussian multiplier bootstrap (MB) and empirical bootstrap (EB), which are equivalent to BWB and MBB respectively when setting the block length to 1. Finally, we include a VAR extension of the EB, labeled %We additionally include the method 
``VAR-EB" which uses the same VAR estimation as VAR-TF, but the bootstrap errors are generated by drawing with replacement from the residuals, rather than using multipliers as in steps 6 and 7 of Algorithm \ref{alg:boot}.  As such, we consider three groups of bootstrap methods: VAR-based, block-based, and time-naive; and within in each group one is a wild/multiplier bootstrap, and the other is an empirical bootstrap.

\begin{figure}[ht]
\centering
\includegraphics[width=\linewidth]{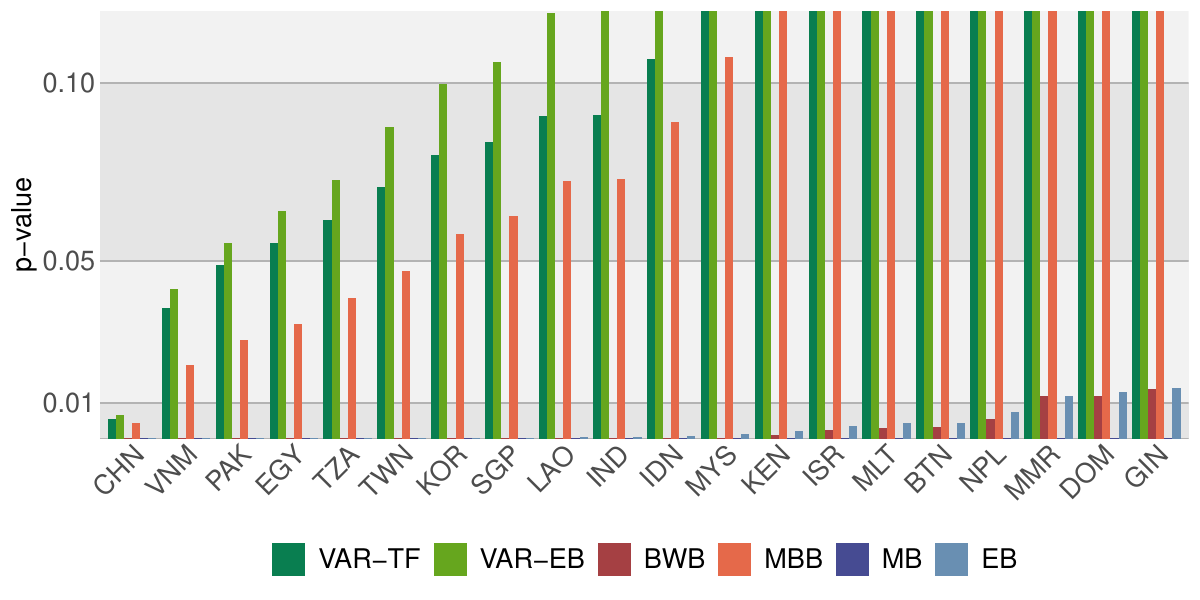}
\caption{Individual $p$-values for the 20 ``most significant" countries across the six bootstrap methods.}\label{fig:application_p_values}
\end{figure}

Figure \ref{fig:application_p_values} shows the $p$-values for the 20 ``most-significant" countries in our sample. The VAR-based bootstrap methods are generally more conservative than the block-based methods, and the time-naive methods are much less conservative. The former is corroborated by our simulation results in Section \ref{sec:simulations}, and the latter is to be expected.

The close agreement between the two VAR-based methods indicates that both capture the underlying time series dynamics adequately.
We also observe a common pattern in which the empirical bootstrap methods tend to be more conservative than their wild/multiplier counterparts. However, while this difference is relatively small for the VAR-based methods, it becomes more pronounced for both the block-based and time-naive methods. This suggests that the choice of bootstrap method can materially affect inference in those settings, leading to different conclusions regarding rejections across countries. % which would lead to drastically different conclusions about rejections in different countries. %Given that we would not expect large differences based on the error resampling method, we see the relative similarity of the two VAR-based methods as an indication that they more adequately capture the underlying time series dynamics.

%VAR-TF and MBB generally give much larger p-values than the MB and EB benchmarks, as we would expect in a time series setting. We also find that VAR-TF is slightly more conservative than MBB, which is broadly in line with our simulation results in Section \ref{sec:simulations}. 
%Interestingly, despite BWB and MBB performing quite similarly in simulations, BWB appears to have much lower p-values in this example. Similarly, we also see much lower p-values for MB compared to EB.
\section{Conclusion}\label{sec:conclusion}
In this paper, we introduce a VAR multiplier bootstrap procedure which approximates the distribution of scaled high-dimensional means, using the lasso to estimate the VAR. We motivate the usefulness of this procedure as a  tool for inference in high-dimensional time series, allowing for non-conservative simultaneous testing of a large set of hypotheses. We show that the bootstrap is consistent under two different moment assumptions on the errors: sub-gaussian moments, and a finite number of absolute moments. Under the former, $N$ can grow at an exponential rate of $T$. Under the latter, $N$ can only grow at a polynomial rate of $T$, with the growth rate of $N$ limited by the number of absolute moments available. 

We provide guidance for estimating the VAR bootstrap model by the lasso as a running example. We show that the lasso satisfies appropriate error bounds for consistency of the bootstrap distribution, under the assumption that the underlying VAR process is (row-wise) sparse. In our examples, we derive explicit limits on the growth rate of $N$ relative to $T$ thereby allowing for exact and weak sparsity of the VAR.

To establish the consistency of the VAR multiplier bootstrap, we derive a Gaussian approximation for the maximum mean of a linear process, which may be of independent interest. Our results can be applied to more complex statistics than simple means, and we believe that extending this method to inference for linear model coefficients is an interesting avenue for future research. Our simulation results show generally good performance of the lasso-VAR-based bootstrap, with the exception of highly persistent DGPs. We believe that another interesting extension would be a bias-corrected version of the bootstrap to improve performance in highly persistent DGPs.
\bibliographystyle{chicago}
\bibliography{bibliography} 

\numberwithin{lemma}{section}
\numberwithin{equation}{section}

\begin{appendices}
\section{Preliminary Lemmas}
\begin{lemma}\label{lma:subgaussian}\hspace{0mm}
\begin{enumerate}
        \item\label{lma:subgaussian1} Under \Cref{ass:subgaussian}.\ref{ass:subgaussian1}, $\max\limits_{t}\norm{\max\limits_{j}\abs{\epsilon_{j,t}}}_{\psi_2}\leq d_{N}$ with $d_{N}=C\sqrt{\log(N)}\geq 1$.
        \item\label{lma:subgaussian2} Under \Cref{ass:subgaussian}.\ref{ass:subgaussian2}, $\max\limits_{t}\norm{\max\limits_{j}\abs{\epsilon_{j,t}}}_{L_m}\leq d_{N},$ with $d_{N}=C N^{1/m}\eta_T^{-1}\geq 1$, where $\eta_T^{-1}\geq 1$.
    \end{enumerate}
\end{lemma}
\begin{lemma}\label{lma:ChernozhukovHDCLT} Let \Cref{ass:covariance} hold, and define 
\begin{equation*}
    {M_{N,T}}:=\sup\limits_{y\in\mathbb{R}}\abs{\P\left(\norm{\frac{1}{\sqrt{T}}\sum_{t=1}^T\cB(1)\bepsilon_t}_\infty\leq y\right)-\P\left(\norm{\bz}_{\infty}\leq y\right)},
\end{equation*}
where $\bz\sim N(\bzero,\bSigma)$, $\bSigma=\cB(1)\bSigma_{\bepsilon}\cB(1)^\prime$. 
\begin{enumerate}
    \item Under \Cref{ass:subgaussian}.\ref{ass:subgaussian1}
\begin{equation*}
    {M_{N,T}}\leq C\left( \frac{b_T\log(N)^{3/2}\log(T)}{\sqrt{T}}+\frac{b_T\log(N)^2}{\sqrt{T}}\right),
\end{equation*}
where $b_T=\tilde{S}^2 d_{N}^2$.
\item Under \Cref{ass:subgaussian}.\ref{ass:subgaussian2}
\begin{equation*}
    {M_{N,T}}\leq  C\left(\frac{ b_T(\log N)^{3/2}\log (T)}{\sqrt{T}}+\frac{b_T^2\log(N)^2\log (T)}{T^{1-2/m}}+\left[\frac{b_T^m \log(N)^{3m/2-4}\log(T)\log (NT)}{T^{m/2-1}}\right]^{\frac{1}{m-2}}\right),
\end{equation*}
where $b_T=
\tilde{S}^2 d_{N}^2$. 
\end{enumerate}
\end{lemma} 

\begin{lemma}\label{lma:leftover} Define $\tilde{\cB}(L)=\sum\limits_{j=0}^{\infty}\sum\limits_{k=j+1}^\infty\bB_k$. 
\begin{enumerate}
    \item Under \Cref{ass:subgaussian}.\ref{ass:subgaussian1}, for any $y>0$
 \begin{equation*}
     \P\left(\norm{\frac{1}{\sqrt{T}}\tilde{\cB}(L)\bepsilon_{T}}_{\infty}>y\right) \leq 2N\exp\left(-C\frac{y^2 T}{d_{N}^2 S_2}\right).
 \end{equation*}
 \item Under \Cref{ass:subgaussian}.\ref{ass:subgaussian2}, for any $y>0$
 \begin{equation*}
     \P\left(\norm{\frac{1}{\sqrt{T}}\tilde{\cB}(L)\bepsilon_{T}}_{\infty}>y\right) \leq \frac{N d_{N}^{m} S_1^m}{\left(y\sqrt{T}\right)^m}.
 \end{equation*}
\end{enumerate}
\end{lemma}

\begin{lemma}\label{lma:trueVMAsummable}
Under \Cref{ass:VARsummable},
for any constant $1\leq q<\infty$, 
\begin{enumerate}
    \item $\tilde{S}=\sum\limits_{j=0}^{\infty} \norm{\bB_{j}}_\infty\leq C_1\psi_{N}$.
    \item $\sum\limits_{j=0}^{\infty}\norm{\bB_j}_{\infty}^q\leq C_1^q\psi_{N}^q$.
    \item $S_q=\sum\limits_{j=0}^{\infty}\left(\sum\limits_{k=j+1}^\infty \norm{\bB_{k}}_\infty\right)^{q}\leq C_3^q\psi_{N}^q$.
\end{enumerate}
Additionally, under \Cref{ass:VARconsistency}, on $\mathcal{P}$,
\begin{enumerate}
\item[4.] $\rho(\hat{\mA})<1$.
\item[5.] $\tilde{S}^*=\sum\limits_{j=0}^\infty \norm{\hat\bB_j}_{\infty}\leq C_5\psi_{N}$.
    \item[6.] $\sum\limits_{j=0}^\infty \norm{\hat\bB_j-\bB_j}_{\infty}^q\leq C_6^q\xi_{N,T}^q\psi_{N}^{2q}$.
     \item[7.] $S_q^*=\sum\limits_{j=0}^{\infty}\left(\sum\limits_{k=j+1}^\infty \norm{\hat\bB_{k}}_\infty\right)^{q}\leq C_7^q\psi_{N}^q$.

\end{enumerate}
\end{lemma}
\begin{lemma}\label{lma:epsilonConsistent}
    Define the set 
     \begin{equation*}
        \mathcal{R}_{1}:=\left\lbrace\max\limits_{1\leq j\leq N}\abs{\frac{1}{T}\sum\limits_{t=1}^T\epsilon_{j,t}^2}\leq Cd_{N}^2\right\rbrace.
    \end{equation*}
    Under \Cref{ass:subgaussian}.\ref{ass:subgaussian1},
    $\lim\limits_{N,T\to\infty}\P(\mathcal{R}_{1})=1$. Furthermore, 
     define the set
    \begin{equation*}
         \mathcal{R}_{2}:=\left\lbrace\max\limits_{1\leq j\leq N}\abs{\frac{1}{T}\sum\limits_{t=1}^T\epsilon_{j,t}^2}\leq Cd_{N}^2\right\rbrace.
     \end{equation*}
     Under \Cref{ass:subgaussian}.\ref{ass:subgaussian2},
     $\lim\limits_{N,T\to\infty}\P(\mathcal{R}_{2})=1$.
\end{lemma}

\begin{lemma}\label{lma:hatstonohats2}
On either $\mathcal{Q}\bigcap\mathcal{R}_1$ or $\mathcal{Q}\bigcap\mathcal{R}_2$,
\begin{equation*}
    \norm{\frac{1}{T}\sum\limits_{t=1}^T\hat\bepsilon_{t}\hat\bepsilon_{t}^\prime-\frac{1}{T}\sum\limits_{t=1}^T\bepsilon_{t}\bepsilon_{t}^\prime}_{\max}\leq C\left(\phi_{N,T}+d_{N} \sqrt{\phi_{N,T}}\right).
\end{equation*}
\end{lemma}

\begin{lemma}\label{lma:nohatstoexp}
Define the set
\begin{equation*}
    \mathcal{S}_1:=\left\lbrace\norm{\frac{1}{T}\sum\limits_{t=1}^T\bepsilon_{t}\bepsilon_{t}^\prime-\frac{1}{T}\sum\limits_{t=1}^T\E\bepsilon_{t}\bepsilon_{t}^\prime}_{\max}\leq \frac{d_N}{\sqrt{T}}\right\rbrace.
\end{equation*}
Under \Cref{ass:subgaussian}.\ref{ass:subgaussian1}, $\lim\limits_{N,T}\P\left(S_1\right)=1$. Furthermore, define the set 
\begin{equation*}
    \mathcal{S}_2:=\left\lbrace\norm{\frac{1}{T}\sum\limits_{t=1}^T\bepsilon_{t}\bepsilon_{t}^\prime-\frac{1}{T}\sum\limits_{t=1}^T\E\bepsilon_{t}\bepsilon_{t}^\prime}_{\max}\leq \frac{d_N^4}{T^{3/4}}\right\rbrace.
\end{equation*}
for some sequence $\eta_T\to0$.
Under \Cref{ass:subgaussian}.\ref{ass:subgaussian2}, $\lim\limits_{N,T\to\infty}\P\left(S_2\right)=1$.

\end{lemma}

\begin{lemma}\label{lma:epsilonbounded}
  Define the set  
  \begin{equation*}
      \mathcal{U}_1:=\left\lbrace\max\limits_{j,t}\abs{\epsilon_{j,t}}\leq d_{N}\log(T)\right\rbrace.
  \end{equation*} 
  Under \Cref{ass:subgaussian}.\ref{ass:subgaussian1}, $\lim\limits_{N,T\to\infty}\P\left(\mathcal{U}_1\right)=1$. 
  Furthermore, define the set 
  \begin{equation*}
      \mathcal{U}_2:=\left\lbrace\max\limits_{j,t}\abs{\epsilon_{j,t}}\leq d_{N}T^{1/m}\right\rbrace.
  \end{equation*}
  Under \Cref{ass:subgaussian}.\ref{ass:subgaussian2}, $\lim\limits_{N,T\to\infty}\P\left(\mathcal{U}_2\right)=1$. 
\end{lemma}

\begin{lemma}\label{lma:bootsubgaussian} \hspace{0mm}
\begin{enumerate}
    \item\label{lma:bootsubgaussian1} On $\mathcal{U}_1\bigcap\mathcal{Q}$, $\max\limits_{t}\norm{\max\limits_{j}\epsilon_{j,t}^*}_{\psi_2}^*\leq d_{N}^*$, with $d_{N}^*=C\left(\sqrt{T\phi_{N,T}}+d_{N}\log(T)\right)$,
    \item\label{lma:bootsubgaussian2} On $\mathcal{U}_2\bigcap\mathcal{Q}$,  $\max\limits_{t}\norm{\max\limits_{j}\epsilon_{j,t}^*}_{L_m}^*\leq d_{N}^*,$ with $d_{N}^*= C\left(\sqrt{T\phi_{N,T}}+d_{N}T^{1/m}\right)$.
\end{enumerate}
\end{lemma}

\begin{lemma}\label{lma:bootCLT} Let \Cref{ass:covariance} hold, and
 define 
 \begin{equation*}
     {M_{N,T}^*}:=\sup\limits_{y\in\mathds{R}}\abs{\P^*\left(\norm{\frac{1}{\sqrt{T}}\sum\limits_{t=1}^T\cB(1)^*\bepsilon_{t}^*}_\infty\right)-\P^*\left(\norm{\bz}_{\infty}\leq y\right)},
 \end{equation*}
 where $\bz\sim N(\bzero,\bSigma)$.  On $\mathcal{T}_1\bigcap\mathcal{U}_1\bigcap\mathcal{Q}$
 \begin{equation*}\begin{split}
    &{M_{N,T}^*}\leq C\left\lbrace\log(N)\log(T)\psi_N^2\left[d_{N}\sqrt{\phi_{N,T}}+\frac{d_N}{\sqrt{T}}+\xi_{N,T}\psi_N\right]\right.\\
     &\left.+(\tilde{S}^*d_{N}^*)^2\left[\frac{\log(N)^{3/2}\log(T)}{\sqrt{T}}+\frac{\log(N)^2\log(T)^2}{T}\right]+\sqrt{\frac{\log(N)^2\log(T)\log(NT)}{T}}\right\rbrace.
\end{split}\end{equation*}
On $\mathcal{T}_2\bigcap\mathcal{U}_2\bigcap\mathcal{Q}$
\begin{equation*}\begin{split}
     &{M_{N,T}^*}\leq C\left\lbrace\log(N)\log(T)\psi_N^2\left[d_{N}\sqrt{\phi_{N,T}}+\frac{d_N^4}{T^{3/4}}+\xi_{N,T}\psi_N\right]\right.\\
     &+\left.(\tilde{S}^*d_{N}^*)^2\left[\frac{\log(N)^{3/2}\left(\log(T)+(\tilde{S}^*d_{N}^*)^{\frac{1}{m-1}}\right)}{\sqrt{T}}+\frac{\log(N)^2\log(T)}{T^{\frac{m-2}{m}}}\right]+\sqrt{\frac{\log(N)^2\log(T)\log(NT)}{T}}\right\rbrace.
\end{split}\end{equation*}

\end{lemma}

\section{Proofs}
\begin{proof}[\hypertarget{p:L0}{\textbf{Proof of \Cref{lma:subgaussian}}}]
Following Lemma 2.2.2 of \cite{vandervaart1996weak},\footnote{We take $\psi(x)=e^{x^2}-1$ (see the explanation of their page 97), and note that $\sqrt{\log(1+N)}\leq C\sqrt{\log{N}}$ when $N>1$.}
\begin{equation*}
    \max\limits_{t}\norm{\max\limits_{j}\abs{\epsilon_{j,t}}}_{\psi_2}\leq C\sqrt{\log(N)} \max\limits_{j,t}\norm{\epsilon_{j,t}}_{\psi_2},
\end{equation*}
and by the statement on page 96 of \cite{vandervaart1996weak},
\begin{equation*}
    \max\limits_{t}\norm{\max\limits_{j}\abs{\epsilon_{j,t}}}_{L_m}\leq N^{1/m} \max\limits_{j,t}\norm{\epsilon_{j,t}}_{L_m}\leq  N^{1/m} \max\limits_{j,t}\norm{\epsilon_{j,t}}_{L_m} \eta_T^{-1}.\qedhere
\end{equation*}
\end{proof}

\begin{proof}[\hypertarget{p:L1}{\textbf{Proof of \Cref{lma:ChernozhukovHDCLT}}}]
Note that $\frac{1}{\sqrt{T}}\sum_{t=1}^T\cB(1)\bepsilon_t$ is a scaled sum of iid random variables, and the proof will proceed by applying the Gaussian approximation in Corollary 2.1 of \cite{chernozhukov2020nearly}. In particular, we will use either the second or third clause of this corollary, depending on whether we use \Cref{lma:subgaussian}.\ref{lma:subgaussian1} or \Cref{lma:subgaussian}.\ref{lma:subgaussian2}.  

First, using \Cref{lma:subgaussian}.\ref{lma:subgaussian1} we use the second clause, which needs their conditions (E.2) and (M). For (E.2), we have by \Cref{lma:subgaussian}.\ref{lma:subgaussian1} that
\begin{equation*}
     \norm{\frac{x_{j,t}}{\sigma_j}}_{\psi_2}=\norm{\frac{\cB(1)_{j}\bepsilon_t}{\sigma_j}}_{\psi_2}\leq \norm{\frac{\norm{\cB(1)_j}_1\max\limits_{j}\abs{\epsilon_{j,t}}}{\sigma_j}}_{\psi_2}\leq   \frac{\norm{\cB(1)_{j}}_{1}}{\abs{\sigma_j}} \norm{\max\limits_{j}\abs{\epsilon_{j,t}}}_{\psi_2}\leq C \tilde{S} d_{N},
\end{equation*}
where $\cB(1)_j$ denotes the $j$th row of $\cB(1)$. The last inequality comes from bounding
$\sigma_j^2\geq\Lambda_{\min}(\bSigma)\geq 1/C$ by \Cref{ass:covariance}, and

\begin{equation*}
   \norm{\cB(1)_{j}}_{1}=\norm{\sum\limits_{j=0}^\infty \bb_{j,k}}_{1}\leq \sum\limits_{j=0}^\infty \norm{\bb_{j,k}}_{1}\leq \sum\limits_{j=0}^\infty \norm{ \bB_{k}}_{\infty}=\tilde{S},
\end{equation*} 
 where $\bb_{j,k}$ is the $j$th row of $\bB_k$.
For (M), 
\begin{equation*}
   \E\abs{\frac{x_{j,t}}{\sigma_j}}^4=\norm{\frac{\cB(1)_{j}\bepsilon_t}{\sigma_j}}_{L_4}^4\leq C \norm{\frac{\cB(1)_{j}\bepsilon_t}{\sigma_j}}_{\psi_2}^4\leq  C\tilde{S}^4 d_{N}^4,
\end{equation*}
by equation (2.15) in \cite{vershynin2019high}. To satisfy the second clause of Corollary 2.1 in \cite{chernozhukov2020nearly}, we then need a sequence $b_T$ such that $C\tilde{S}d_N\leq b_T$ and $C\tilde{S}^4 d_{N}^4\leq b_T^2$. Note that $\tilde{S}\geq 1$ since $\bB_0=\bI$, and $d_{N}\geq 1$ by assumption, so these inequalities are satisfied when $b_T\sim \tilde{S}^2 d_{N}^2$. It therefore follows that 
\begin{equation*}
    {M_{N,T}}\leq  C\left(\frac{ b_T(\log N)^{3/2}\log T}{\sqrt{T}\Lambda_{\min}\left(\tilde\bSigma\right)}+\frac{b_T(\log N)^2}{\sqrt{T}\sqrt{\Lambda_{\min}\left(\tilde\bSigma\right)}}\right),
\end{equation*}
where $\tilde\bSigma$ is the correlation matrix of $\bx_t$. To show that $\Lambda_{\min}\left(\tilde\bSigma\right)$ is bounded away from 0, write $\tilde\bSigma=\bD\bSigma\bD$, where $\bD=\diag(1/\sigma_{1},\dots,1/\sigma_{N})$. Since $\bD$ and $\bSigma$ are symmetric and positive definite by \Cref{ass:covariance}, we have $\Lambda_{\min}\left(\tilde\bSigma\right)\geq\Lambda_{\min}\left(\bD\right)^2\Lambda_{\min}\left(\bSigma\right)$.
The eigenvalues of a diagonal matrix are just its diagonal entries, which are bounded away from 0 since the variances $\sigma_j$ are bounded, and $\Lambda_{\min}\left(\bSigma\right)$ is bounded away from 0; both by \Cref{ass:covariance}. The result of the first statement then follows.

Second, using \Cref{lma:subgaussian}.\ref{lma:subgaussian2}, we use the third clause of Corollary 2.1 in \cite{chernozhukov2020nearly}, which needs their conditions (E.3) and (M). For (E.3),
\begin{equation*}\begin{split}
    \norm{\max\limits_{1\leq j\leq N}\abs{\frac{x_{j,t}}{\sigma_{j}}}}_{L_m}&\leq \max\limits_{j}\abs{1/\sigma_j}\norm{\max\limits_{j}x_{j,t}}_{L_m}\\
    &\leq C\norm{\max\limits_{j}\cB(1)_{j}\bepsilon_t}_{L_m}\leq C\norm{\cB(1)}_{\infty}\norm{\max\limits_{j}\abs{\epsilon_{j,t}}}_{L_m}\leq C\tilde{S}d_{N}.
\end{split}\end{equation*}
For (M), 
\begin{equation*}\begin{split}
   \E\abs{\frac{x_{j,t}}{\sigma_j}}^4=\norm{\frac{\cB(1)_{j}\bepsilon_t}{\sigma_j}}_{L_4}^4\leq  C\tilde{S}^4 d_{N}^4.
\end{split}\end{equation*}
Similarly to before, we need the sequence $b_T$ to satisfy 
$\tilde{S}d_{N}\leq b_T$, and $\tilde{S}^4 d_{N}^4\leq b_T^2$, which is satisfied when taking $b_T\sim 
\tilde{S}^2 d_{N}^2$. Therefore
\begin{equation*}
\begin{split}
    {M_{N,T}}&\leq  C\left(\frac{ b_T(\log N)^{3/2}\log T}{\sqrt{T}\Lambda_{\min}\left(\tilde\bSigma\right)}+\frac{b_T^2(\log N)^2\log T}{T^{1-2/m}\Lambda_{\min}\left(\tilde\bSigma\right)}
    +\left[\frac{b_T^m (\log N)^{3m/2-4}(\log T)\log (NT)}{T^{m/2-1}\Lambda_{\min}\left(\tilde\bSigma\right)^{m/2}}\right]^{\frac{1}{m-2}}\right),
\end{split}
\end{equation*}
and the result of the second statement follows.
\end{proof}
\begin{proof}[\hypertarget{p:L2}{\textbf{Proof of \Cref{lma:leftover}}}]
\begin{equation}\label{eq:firststep}\begin{split}
    \P\left(\norm{\frac{1}{\sqrt{T}}\tilde{\cB}(L)\bepsilon_{T}}_{\infty}>y\right)=&\P\left(\max\limits_{1\leq p\leq N}\frac{1}{\sqrt{T}}\abs{\left[\tilde{\cB}(L)\right]_{p,\cdot}\bepsilon_{T}}>y\right)\\
    =&\P\left(\max\limits_{1\leq p\leq N}\frac{1}{\sqrt{T}}\abs{\left[\sum\limits_{j=0}^\infty\left(\sum\limits_{k=j+1}^\infty \bB_{k}\right)L^j\right]_{p,\cdot}\bepsilon_{T}}>y\right)\\
    =&\P\left(\max\limits_{1\leq p\leq N}\frac{1}{\sqrt{T}}\abs{\sum\limits_{j=0}^\infty\left(\sum\limits_{k=j+1}^\infty \bb_{p,k}\right)\bepsilon_{T-j}}>y\right),
\end{split}\end{equation}
where $\bb_{p,k}$ is the $p$th row of $\bB_k$. 

By \Cref{lma:subgaussian}.\ref{lma:subgaussian1}, we proceed from \Cref{eq:firststep} with the union bound and Hoeffding's inequality (see Theorem 2.6.2 in \cite{vershynin2019high})
\begin{equation*}\begin{split}
    &\P\left(\max\limits_{1\leq p\leq N}\frac{1}{\sqrt{T}}\abs{\sum\limits_{j=1}^\infty\left(\sum\limits_{k=j}^\infty \bb_{p,k}\right)\bepsilon_{T+1-j}}>y\right)\leq \sum\limits_{p=1}^N\P\left(\abs{\sum\limits_{j=1}^\infty\left(\sum\limits_{k=j}^\infty \bb_{p,k}\right)\bepsilon_{T+1-j}}>y\sqrt{T}\right)\\
    &\leq \sum\limits_{p=1}^N 2\exp\left(-C\frac{\left[y\sqrt{T}\right]^2}{\sum\limits_{j=1}^\infty\norm{\left(\sum\limits_{k=j}^\infty \bb_{p,k}\right)\bepsilon_{T+1-j}}_{\psi_2}^2}\right).
\end{split}\end{equation*}
Using \Cref{lma:subgaussian}.\ref{lma:subgaussian1} and arguments similar to those in the proof of \Cref{lma:ChernozhukovHDCLT}, we can bound
\begin{equation*}
    \norm{\left(\sum\limits_{k=j}^\infty \bb_{p,k}\right)\bepsilon_{T+1-j}}_{\psi_2}\leq  d_{N}\sum\limits_{k=j}^{\infty}\norm{\bB_k}_{\infty},
\end{equation*}
and therefore
\begin{equation*}\begin{split}
    &\P\left(\max\limits_{1\leq p\leq N}\frac{1}{\sqrt{T}}\abs{\sum\limits_{j=1}^\infty\left(\sum\limits_{k=j}^\infty \bb_{p,k}\right)\bepsilon_{T+1-j}}>y\right)\leq 2N\exp\left(-C\frac{y^2 T}{d_{N}^2 S_2}\right),
\end{split}\end{equation*}
so the first statement follows.
For the second statement, by \Cref{lma:subgaussian}.\ref{lma:subgaussian2}, we proceed from \Cref{eq:firststep} with the union bound and Markov's inequality

\begin{equation}\label{eq:markov}\begin{split}
    &\P\left(\max\limits_{1\leq p\leq N}\frac{1}{\sqrt{T}}\abs{\sum\limits_{j=0}^\infty\left(\sum\limits_{k=j+1}^\infty \bb_{p,k}\right)\bepsilon_{T-j}}>y\right)\\
    &\leq \sum\limits_{p=1}^N\P\left(\abs{\sum\limits_{j=0}^\infty\left(\sum\limits_{k=j+1}^\infty \bb_{p,k}\right)\bepsilon_{T-j}}>y\sqrt{T}\right)
    \leq \sum\limits_{p=1}^N\frac{\E\left[\abs{\sum\limits_{j=0}^\infty\left(\sum\limits_{k=j+1}^\infty \bb_{p,k}\right)\bepsilon_{T-j}}^{m}\right]}{\left(y\sqrt{T}\right)^m}.
\end{split}\end{equation}
For the numerator, we continue with Minkownski's inequality  and \Cref{lma:subgaussian}.\ref{lma:bootsubgaussian2}
\begin{equation*}\begin{split}
&\left(\E\left[\abs{\sum\limits_{j=0}^\infty\left(\sum\limits_{k=j+1}^\infty \bb_{p,k}\right)\bepsilon_{T-j}}^{m}\right]\right)^{1/m}\leq\sum\limits_{j=0}^\infty\left(\E\left[\abs{\left(\sum\limits_{k=j+1}^\infty \bb_{p,k}\right)\bepsilon_{T-j}}^{m}\right]\right)^{1/m}\\
&\leq \sum\limits_{j=0}^\infty\left(\norm{\sum\limits_{k=j+1}^\infty \bb_{p,k}}_{1}^m\E\left[\max\limits_{p}\abs{\bepsilon_{p,T-j}}^{m}\right]\right)^{1/m}\leq\max\limits_{t}\norm{\max\limits_{p}\abs{\bepsilon_{p,t}}}_{L_m}\sum\limits_{j=0}^\infty\left(\norm{\sum\limits_{k=j+1}^\infty \bb_{p,k}}_{1}^m\right)^{1/m}\\
&\leq d_N\sum\limits_{j=0}^\infty\left(\sum\limits_{k=j+1}^\infty \norm{\bb_{p,k}}_{1}\right)\leq d_N\sum\limits_{j=0}^\infty\left(\sum\limits_{k=j+1}^\infty \norm{\bB_{k}}_{\infty}\right)=d_N S_1.
\end{split}\end{equation*}
Continuing from \Cref{eq:markov}, we therefore obtain
\begin{equation*}
     \sum\limits_{p=1}^N\frac{\E\left[\abs{\sum\limits_{j=0}^\infty\left(\sum\limits_{k=j+1}^\infty \bb_{p,k}\right)\bepsilon_{T-j}}^{m}\right]}{\left(y\sqrt{T}\right)^m}\leq  \sum\limits_{p=1}^N \frac{d_{N}^m S_1^m}{\left(y\sqrt{T}\right)^m}=\frac{Nd_{N}^m S_1^m}{\left(y\sqrt{T}\right)^m}.
     \qedhere
\end{equation*}
\end{proof}
\begin{proof}[\hypertarget{p:T1}{\textbf{Proof of \Cref{thm:HDCLTforLP}}}]
We first write the Beveridge-Nelson decomposition of the process
\begin{equation*}
    \bx_t=\cB(L)\bepsilon_t=\cB(1)\bepsilon_t-(1-L)\tilde{\cB}(L)\bepsilon_t,
\end{equation*}
where $\tilde{\cB}(L)=\sum_{j=0}^\infty\tilde{\bB}_j L^j, \tilde{\bB}_j=\sum_{k=j+1}^\infty\bB_k$, such that
\begin{equation*}
    \frac{1}{\sqrt{T}}\sum_{t=1}^T\bx_t=\frac{1}{\sqrt{T}}\sum_{t=1}^T\cB(1)\bepsilon_t-\frac{1}{\sqrt{T}}\tilde{\cB}(L)\bepsilon_{T}+\frac{1}{\sqrt{T}}\tilde{\cB}(L)\bepsilon_{0}.
\end{equation*}
Note that by assumption $\bepsilon_{t}=\bzero$ for $t<1$, so $\frac{1}{\sqrt{T}}\tilde{\cB}(L)\bepsilon_{0}=\bzero$.
Define 
\begin{equation*}\begin{split}
    x_{T}^{(\max)}=\norm{\frac{1}{\sqrt{T}}\sum_{t=1}^T\bx_t}_\infty,\qquad \epsilon_T^{(\max)}=\norm{\frac{1}{\sqrt{T}}\sum_{t=1}^T\cB(1)\bepsilon_t}_\infty, \qquad z_T^{(\max)}=\norm{\bz}_\infty,
\end{split}\end{equation*}
\begin{equation*}\begin{split}
    &F_{1,T}(y):=\P\left(x_{T}^{(\max)}\leq y\right)\quad
    F_{2,T}(y):=\P\left(\epsilon_T^{(\max)}\leq y\right)\\
    &G_{T}(y):=\P\left(z_T^{(\max)}\leq y\right)\quad 
    r_T:=x_{T}^{(\max)}-\epsilon_T^{(\max)}
\end{split}\end{equation*}
Then
\begin{equation*}\begin{split}
    \abs{r_T}=&\abs{ \norm{\frac{1}{\sqrt{T}}\sum_{t=1}^T\bx_t}_\infty-\norm{\frac{1}{\sqrt{T}}\sum_{t=1}^T\cB(1)\bepsilon_t}_\infty}\\
    \leq&\norm{\frac{1}{\sqrt{T}}\sum_{t=1}^T\bx_t-\frac{1}{\sqrt{T}}\sum_{t=1}^T\cB(1)\bepsilon_t}_\infty=\norm{\frac{1}{\sqrt{T}}\tilde{\cB}(L)\bepsilon_{T}}_\infty=R_T.
\end{split}\end{equation*}
By 
\Cref{lma:leftover} we have
$\P(\abs{r_T}>\eta_{T,1})\leq\P(R_T>\eta_{T,1})\leq 2N\exp\left(-C\frac{\eta_{T,1}^2 T}{d_{N}^2 S_2}\right)=:\eta_{T,2}$.
Continue with 
\begin{align*}
&\abs{F_{1,T}(y) - G_T (y)} \\
&\leq \abs{\P\left(\left.\epsilon_T^{(\max)}+ r_T \leq y \right| \abs{r_T} \leq \eta_{T,1} \right) \P\left(\abs{r_T} \leq \eta_{T,1}\right) - \P\left(z_T^{(\max)}\leq y\right)}\\
&\quad\qquad + \P\left(\left.x_{T}^{(\max)} \leq y \right| \abs{r_T} > \eta_{T,1}\right) \P\left(\abs{r_T} > \eta_{T,1}\right)\\
&\quad \leq \abs{\P\left(\epsilon_T^{(\max)} \leq y + \eta_{T,1} \right) - \P\left(z_T^{(\max)} \leq y\right)} + \eta_{T,2}\\
&\quad \leq \underbrace{\abs{\P\left(\epsilon_T^{(\max)}  \leq y + \eta_{T,1} \right) - \P(z_T^{(\max)}  \leq y + \eta_{T,1})}}_{A_{T,1}(y+\eta_{T,1})}\\
&\quad \quad + \underbrace{\abs{\P\left(z_T^{(\max)}  \leq y + \eta_{T,1} \right) - \P(z_T^{(\max)}  \leq y)}}_{A_{T,2}(y)}  + \eta_{T,2}.
\end{align*}
Note that $\sup\limits_{y\in\mathbb{R}}A_{T,1}(y+\eta_{T,1})={M_{N,T}}$ which can be bounded by \Cref{lma:ChernozhukovHDCLT}, 
and  $\sup\limits_{y\in\mathbb{R}}A_{T,2}(y)$ can be bounded by Lemma A.1 in \cite{CCK17}, which states that for centered Gaussian vectors $\bz\in \mathbb{R}^N$ with variances uniformly bounded away from 0 (as is the case here by \Cref{ass:covariance}), for all $\by\in\mathbb{R}^N$ and $a>0$
\begin{equation*}
    \P\left(\bz\leq\by+a\right)-\P\left(\bz\leq\by\right)\leq C a\sqrt{\log(N)}.
\end{equation*}
Note that this applies to $\norm{\bz}_{\infty}$ as well, since
\begin{equation*}
    \P\left(\norm{\bz}_{\infty}\leq y+a\right)-\P\left(\norm{\bz}_{\infty}\leq y\right)=2\left[\P\left(\bz\leq \by+a\right)-\P\left(\bz\leq\by\right)\right],
\end{equation*}
when $\by$ has each element equal to $y$, and if the bound holds for all $\by\in\mathbb{R}^N$, it also holds for the supremum over $y\in\mathbb{R}$. We therefore have the bound
\begin{equation*}
    \sup\limits_{y\in\mathbb{R}}\abs{F_{1,T}(y) - G_T (y)} \leq {M_{N,T}}+ C_1\left[\eta_{T,1}\sqrt{\log{N}}+N\exp\left(-C_2\frac{\eta_{T,1}^2 T}{d_{N}^2 S_2}\right)\right].
\end{equation*}
In order for this expression to converge, we need to choose $\eta_{T,1}$ converging to 0 fast enough such that $\eta_{T,1}\sqrt{\log(N)}\to0$, but slow enough such that $N\exp\left(-C_2\frac{\eta_{T,1}^2 T}{d_{N}^2 S_2}\right)\to0$. One such choice is $\eta_{T,1}=\sqrt{\log(N\log(N))\frac{d_{N}^2S_2}{C_2 T}}$ (assuming $N>1$), which lets us bound 
\begin{equation*}\begin{split}
    C_1\left[\eta_{T,1}\sqrt{\log{N}}+N\exp\left(-C_2\frac{\eta_{T,1}^2 T}{d_{N}^2 S_2}\right)\right]&\leq C\left[\frac{d_{N}\sqrt{S_2}}{\sqrt{T}}\sqrt{\log(N)\log(N\log(N))}+\frac{1}{\log(N)}\right]\\
    &\leq C\left[\frac{\log(N)d_{N}\sqrt{S_2}}{\sqrt{T}}+\frac{1}{\log(N)}\right],
\end{split}\end{equation*}
and the result of the first statement follows.

For the second statement, by \Cref{lma:subgaussian}.\ref{lma:subgaussian2}, we may follow the same steps as above, taking $\eta_{T,2}:=2\frac{N d_{N}^{m} S_1^m}{\left(\eta_{T,1}\sqrt{T}\right)^m}$ by the second clause of \Cref{lma:leftover}. We then have the bound
\begin{equation*}
    \sup\limits_{y\in\mathbb{R}}\abs{F_{1,T}(y) - G_T (y)} \leq {M_{N,T}}+ C_1\left[\eta_{T,1}\sqrt{\log{N}}+\frac{N d_{N}^{m} S_1^m}{\left(\eta_{T,1}\sqrt{T}\right)^m}\right].
\end{equation*}
In this case, we can solve for the optimal rate of convergence for $\eta_{T,1}$, which has both terms converging at the same rate, $\eta_{T,1}= \left(\frac{N d_{N}^{m} S_1^m}{\sqrt{T}^m\sqrt{\log(N)}}\right)^{\frac{1}{m+1}}$. We then have 
\begin{equation*}
    \eta_{T,1}\sqrt{\log{N}}=\frac{N d_{N}^{m} S_1^m}{\left(\eta_{T,1}\sqrt{T}\right)^m}=\left(Nd_{N}^mS_1^m \right)^{\frac{1}{m+1}}\left(\frac{\sqrt{\log(N)}}{\sqrt{T}}\right)^{\frac{m}{m+1}},
\end{equation*}
and the result of the second statement follows.
\end{proof}

\begin{proof}[\hypertarget{p:L5}{\textbf{Proof of \Cref{lma:trueVMAsummable}}}]
Under Assumption \ref{ass:VARsummable}, using Gelfand's formula,
\begin{equation}\label{VAR_stationary}
    \rho(\mA)=\lim\limits_{j\to\infty}\norm{\mA^j}_\infty^{1/j}\overset{\text{Ass.}\ref{ass:VARsummable}}{\leq} \lim\limits_{j\to\infty}(\psi_N\theta^j)^{1/j}=\theta \lim\limits_{j\to\infty}\psi_N^{1/j} =\theta<1.
\end{equation}
The process is therefore invertible, and we have
$\bB_k=\bJ\mA^k\bJ^\prime$, where $\underset{N\times KN}{\bJ}=\left(\bI,\bzero,\dots,\bzero\right)$:
\begin{equation}\label{VMA_to_VAR}
    \norm{\bB_k}_{\infty}\overset{(\ref{VAR_stationary})}{=}\norm{\bJ\mA^k\bJ^\prime}_{\infty}\leq \norm{\bJ}_{\infty}\norm{\mA^k}_{\infty}\norm{\bJ^\prime}_{\infty}=\norm{\mA^k}_{\infty}.
\end{equation}
\begin{equation}\label{VMA_summable}
    \tilde{S}=\sum\limits_{j=0}^{\infty} \norm{\bB_{j}}_\infty \overset{(\ref{VMA_to_VAR})}{\leq} \sum\limits_{j=0}^{\infty} \norm{\mA^j}_\infty \overset{\text{Ass.}\ref{ass:VARsummable}}{\leq} \sum\limits_{j=0}^{\infty} \psi_N \theta^j=\psi_N\sum\limits_{j=0}^{\infty} \theta^j=\frac{1}{1-\theta}\psi_N.
\end{equation}
We therefore have point 1.~with
$C_{1}=\frac{1}{1-\theta}$.
By properties of (vector) $p$-norms, $\norm{\ba}_q\leq\norm{\ba}_1$ for $q\geq 1$, which implies:
\begin{equation}\label{cross_terms}
    \sum\limits_{i}\abs{a_i}^q  =\norm{\ba}_m^q\leq \norm{\ba}_1^q=\left(\sum\limits_{i}\abs{a_i}\right)^q.
\end{equation}
From point 1.~we then directly have point 2.:
\begin{equation*}
    \sum\limits_{j=0}^{\infty} \norm{\bB_{j}}_\infty^q\overset{(\ref{cross_terms})}{\leq} \left(\sum\limits_{j=0}^{\infty} \norm{\bB_{j}}_\infty\right)^q{\leq}\left({C_{1}\psi_N}\right)^q=C_{1}^q\psi_N^q.
\end{equation*}
As an intermediate result, we have:
\begin{equation}\label{VMA_j_summable_power}
    \sum\limits_{j=0}^{\infty} j^q\norm{\bB_{j}}_\infty^q\overset{(\ref{VMA_to_VAR})}{\leq} \sum\limits_{j=0}^{\infty} j^q\norm{\mA^{j}}_\infty^q \overset{(\ref{cross_terms})}{\leq} \left(\sum\limits_{j=0}^{\infty} j\norm{\mA^{j}}_\infty\right)^q\overset{\text{Ass.}\ref{ass:VARsummable}}{\leq}\psi_N^q\left(\sum\limits_{j=0}^{\infty} j\theta^j\right)^q=\psi_N^q\left(\frac{\theta}{(1-\theta)^2}\right)^q.
\end{equation}
For point 3.:
\begin{equation*}\begin{split}
&\sum\limits_{j=0}^{\infty}\left(\sum\limits_{k=j+1}^{\infty}\norm{\bB_k}_\infty\right)^q\overset{(\ref{cross_terms}),(\ref{VMA_to_VAR})}{\leq} \left(\sum\limits_{j=0}^{\infty}\sum\limits_{k=j+1}^{\infty}\norm{\mA^k}_\infty \right)^q \overset{\text{Ass.}\ref{ass:VARsummable}}{\leq}\psi_N^q \left(\sum\limits_{j=0}^{\infty}\sum\limits_{k=j+1}^{\infty}\theta^k \right)^q \\
&=\psi_N^q \left(\sum\limits_{j=0}^{\infty}\left[\sum\limits_{k=0}^{\infty}\theta^k-\sum\limits_{k=0}^{j}\theta^k \right]\right)^q=\psi_N^q \left(\sum\limits_{j=0}^{\infty}\left[\frac{1}{1-\theta}-\frac{1-\theta^{j+1}}{1-\theta}\right]\right)^q\\
&=\psi_N^q \left(\sum\limits_{j=0}^{\infty}\frac{\theta^{j+1}}{1-\theta}\right)^q=\psi_N^q\left(\frac{\theta}{(1-\theta)^{2}}\right)^q,
\end{split}\end{equation*}
with $C_3=\frac{\theta}{(1-\theta)^{2}}$.
As an intermediate result, we have
\begin{equation}\label{VAR_power_diff_sum}
    \sum\limits_{j=0}^{k}\norm{\hat{\mA}^j-\mA^j}_\infty\leq \frac{1}{1-\theta}\xi_{N,T}\psi_N\sum\limits_{j=0}^k\norm{\hat{\mA}^j}_\infty,\text{ for any }j\geq 0.
\end{equation}

To see this, by the proof of Lemma 11 in \cite{Krampe19}, we have:
\begin{equation}\label{krampe}
     \hat\mA^j-\mA^j=\sum\limits_{s=0}^{j-1}\hat\mA^s(\hat\mA-\mA)\mA^{j-1-s}.
\end{equation}
\begin{equation*}
\begin{split}
    &\sum\limits_{j=0}^{k}\norm{\hat{\mA}^j-\mA^j}_\infty=\sum\limits_{j=1}^{k}\norm{\hat{\mA}^j-\mA^j}_\infty\overset{(\ref{krampe})}{=}\sum\limits_{j=1}^{k}\norm{\sum\limits_{s=0}^{j-1}\hat\mA^s(\hat\mA-\mA)\mA^{j-1-s}}_\infty\\
    &{\leq}\sum\limits_{j=1}^{k}\sum\limits_{s=0}^{j-1}\norm{\hat\mA^s(\hat\mA-\mA)\mA^{j-1-s}}_\infty\overset{}{\leq}\sum\limits_{j=1}^{k}\sum\limits_{s=0}^{j-1}\norm{\hat{\mA}^s}_\infty\norm{\hat\mA-\mA}_\infty\norm{\mA^{j-1-s}}_\infty\\
    &\overset{\text{Ass.}\ref{ass:VARsummable},\mathcal{P}}{\leq}\xi_{N,T}\psi_N\sum\limits_{j=1}^{k}\sum\limits_{s=0}^{j-1}\norm{\hat{\mA}^s}_\infty\theta^{j-1-s}=\xi_{N,T}\psi_N\sum\limits_{j=1}^{k}\sum\limits_{s=0}^{k-1}\norm{\hat{\mA}^s}_\infty\theta^{j-1-s}\mathds{1}_{\{s\leq j-1\}}\\
    &=\xi_{N,T}\psi_N\sum\limits_{s=0}^{k-1}\norm{\hat{\mA}^s}_\infty\sum\limits_{j=1}^{k}\theta^{j-1-s}\mathds{1}_{\{s\leq j-1\}}=\xi_{N,T}\psi_N\sum\limits_{s=0}^{k-1}\norm{\hat{\mA}^s}_\infty\sum\limits_{j=s+1}^{k}\theta^{j-1-s}\\
    &=\xi_{N,T}\psi_N\sum\limits_{s=0}^{k-1}\norm{\hat{\mA}^s}_\infty\sum\limits_{j=0}^{k-1-s}\theta^{j}\overset{}{\leq}  \xi_{N,T}\psi_N\sum\limits_{s=0}^{k}\norm{\hat{\mA}^s}_\infty\sum\limits_{j=0}^{\infty}\theta^{j}=\frac{1}{1-\theta}\xi_{N,T}\psi_N\sum\limits_{s=0}^{k}\norm{\hat{\mA}^s}_\infty.
    \end{split}
\end{equation*}
We can then show that the VAR coefficient powers are summable:
\begin{equation*}\begin{split}
    &\sum\limits_{j=0}^{k}\norm{\hat{\mA}^{j}}_\infty{\leq}\sum\limits_{j=0}^{k}\norm{\hat{\mA}^j-\mA^j}_\infty+\sum\limits_{j=0}^{k}\norm{\mA^{j}}_\infty\overset{(\ref{VMA_summable})}{\leq}\sum\limits_{j=0}^{k}\norm{\hat{\mA}^j-\mA^j}_\infty+\frac{1}{1-\theta}\psi_{N}\\
    &\overset{(\ref{VAR_power_diff_sum})}{\leq} \frac{1}{1-\theta}\xi_{N,T}\psi_N\sum\limits_{j=0}^k\norm{\hat{\mA}^j}_\infty+\frac{1}{1-\theta}\psi_{N}.
    \end{split}
\end{equation*}
Under Assumption \ref{ass:VARconsistency}, we have 
\begin{equation*}
     1-\frac{1}{1-\theta}\xi_{N,T}\psi_N\geq 1-\frac{1}{(1-\theta)^2}\xi_{N,T}\psi_N \overset{\text{Ass.}\ref{ass:VARconsistency}}{\geq} 1-\frac{1}{(1-\theta)^2}\bar{C}(1-\theta)^2= 1-\bar{C},
\end{equation*}
and because $0<\bar{C}<1$, 
\begin{equation}\label{sufficiently_large1}
    \left(1-\frac{1}{1-\theta}\xi_{N,T}\psi_N\right)^{-1}\leq \frac{1}{1-\bar{C}}.
\end{equation}
Factorizing the sum:
\begin{equation*}
\begin{split}
    &\sum\limits_{j=0}^{k}\norm{\hat{\mA}^{j}}_\infty\leq \frac{1}{1-\theta}\xi_{N,T}\psi_N\sum\limits_{j=0}^k\norm{\hat{\mA}^j}_\infty+\frac{1}{1-\theta}\psi_{N}\\
    &\sum\limits_{j=0}^{k}\norm{\hat{\mA}^{j}}_\infty (1-\frac{1}{1-\theta}\xi_{N,T}\psi_N)\leq \frac{1}{1-\theta}\psi_{N}\\
    &\sum\limits_{j=0}^{k}\norm{\hat{\mA}^{j}}_\infty \leq \left(1-\frac{1}{1-\theta}\xi_{N,T}\psi_N\right)^{-1}\frac{1}{1-\theta}\psi_{N}\overset{(\ref{sufficiently_large1})}{\leq} \frac{1}{1-\bar{C}}\frac{1}{1-\theta}\psi_{N}.
    \end{split}
\end{equation*}
Then, 
\begin{equation}\label{boot_VMA_summable}
\sum\limits_{j=0}^{\infty}\norm{\hat{\mA}^{j}}_\infty=\lim\limits_{k\to\infty}\left\lbrace\sum\limits_{j=0}^{k}\norm{\hat{\mA}^{j}}_\infty\right\rbrace\leq \lim\limits_{k\to\infty}\left\lbrace \frac{1}{1-\bar{C}}\frac{1}{1-\theta}\psi_{N}\right\rbrace=\frac{1}{1-\bar{C}}\frac{1}{1-\theta}\psi_{N}.
\end{equation}
As another intermediate result,
\begin{equation}\label{VAR_power_diff_j_sum}
   \sum\limits_{j=0}^{k}j\norm{\hat{\mA}^j-\mA^j}_\infty\leq \frac{1}{(1-\theta)^2}\xi_{N,T}\psi_N\left(\sum\limits_{j=0}^{k}j\norm{\hat{\mA}^j}_\infty+\frac{1}{1-\bar{C}}\frac{1}{1-\theta}\psi_{N}\right).
\end{equation}
To see this, we will use the following inequality:
\begin{equation}\label{quadratic}
    j+s+1\leq (j+1)(s+1)\text{ for all }j,s\geq0.
\end{equation}
\begin{equation*}\begin{split}
    &\sum\limits_{j=0}^{k}j\norm{\hat{\mA}^j-\mA^j}_\infty\overset{(\ref{krampe})}{=}\sum\limits_{j=0}^{k}j\norm{\sum\limits_{s=0}^{j-1}\hat\mA^s(\hat\mA-\mA)\mA^{j-1-s}}_\infty\overset{\text{Ass.}\ref{ass:VARsummable}, \mathcal{P}}{\leq}\xi_{N,T}\psi_N\sum\limits_{j=0}^{k}\sum\limits_{s=0}^{j-1}j\norm{\hat{\mA}^s}_\infty\theta^{j-1-s}\\
    &=\xi_{N,T}\psi_N\sum\limits_{j=0}^{k}\sum\limits_{s=0}^{k-1}j\norm{\hat{\mA}^s}_\infty\theta^{j-1-s}\ind{s\leq j-1}=\xi_{N,T}\psi_N\sum\limits_{s=0}^{k-1}\sum\limits_{j=s+1}^{k}j\norm{\hat{\mA}^s}_\infty\theta^{j-1-s}\\
    &=\xi_{N,T}\psi_N\sum\limits_{s=0}^{k-1}\sum\limits_{j=0}^{k-1-s}(j+s+1)\norm{\hat{\mA}^s}_\infty\theta^{j}\overset{(\ref{quadratic})}{\leq}\xi_{N,T}\psi_N\left(\sum\limits_{s=0}^{k-1}(s+1)\norm{\hat{\mA}^s}_\infty\right)\left(\sum\limits_{j=0}^{k-1-s}(j+1)\theta^{j}\right)\\
    &\leq \xi_{N,T}\psi_N\left(\sum\limits_{s=0}^{k}(s+1)\norm{\hat{\mA}^s}_\infty\right)\left(\sum\limits_{j=0}^{\infty}j\theta^{j}+\theta^j \right)=\xi_{N,T}\psi_N\left(\sum\limits_{s=0}^{k}(s+1)\norm{\hat{\mA}^s}_\infty\right)\left(\frac{\theta}{(1-\theta)^2}+\frac{1}{1-\theta} \right)\\
    &=\frac{1}{(1-\theta)^2}\xi_{N,T}\psi_N\left(\sum\limits_{s=0}^{k}s\norm{\hat{\mA}^s}_\infty+\norm{\hat{\mA}^s}_\infty\right)\overset{(\ref{boot_VMA_summable})}{\leq} \frac{1}{(1-\theta)^2}\xi_{N,T}\psi_N\left(\sum\limits_{s=0}^{k}s\norm{\hat{\mA}^s}_\infty+\frac{1}{1-\bar{C}}\frac{1}{1-\theta}\psi_N\right).
\end{split}\end{equation*}
We then have the following summability result:
\begin{equation}\label{boot_VMA_j_summable}
\sum\limits_{j=0}^{\infty}j^q\norm{\hat{\mA}^j}_{\infty}^q\leq \left(\frac{1}{1-\bar{C}}\left[\frac{\bar{C}}{1-\bar{C}}\frac{1}{1-\theta}+\frac{\theta}{(1-\theta)^2}\right]\right)^q\psi_N^q.
\end{equation}
To see this:
\begin{equation*}
\sum\limits_{j=0}^{\infty}j^q\norm{\hat{\mA}^j}_{\infty}^q\overset{(\ref{cross_terms})}{\leq}\left(\sum\limits_{j=0}^{\infty}j\norm{\hat{\mA}^j}_{\infty}\right)^q=\left(\lim\limits_{k\to\infty}\left\lbrace\sum\limits_{j=0}^{k}j\norm{\hat{\mA}^j}_{\infty}\right\rbrace\right)^q
\end{equation*}
\begin{equation*}\begin{split}
    &\sum\limits_{j=0}^{k}j\norm{\hat{\mA}^j}_{\infty}{\leq} \sum\limits_{j=0}^{k}j\norm{\hat{\mA}^j-\mA^j}_{\infty}+\sum\limits_{j=0}^{k}j\norm{\mA^j}_{\infty}\\
    &\overset{(\ref{VAR_power_diff_j_sum}),(\ref{VMA_j_summable_power})}{\leq}\frac{1}{(1-\theta)^2}\xi_{N,T}\psi_N\left(\sum\limits_{j=0}^{k}j\norm{\hat{\mA}^j}_\infty+\frac{1}{1-\bar{C}}\frac{1}{1-\theta}\psi_{N}\right)+\frac{\theta}{(1-\theta)^2}\psi_N.
\end{split}\end{equation*}
Factorizing the sum:
\begin{equation*}\begin{split}
    &\sum\limits_{j=0}^{k}j\norm{\hat{\mA}^j}_{\infty}\leq \frac{1}{(1-\theta)^2}\xi_{N,T}\psi_N\left(\sum\limits_{j=0}^{k}j\norm{\hat{\mA}^j}_\infty+\frac{1}{1-\bar{C}}\frac{1}{1-\theta}\psi_{N}\right)+\frac{\theta}{(1-\theta)^2}\psi_N\\
    &\sum\limits_{j=0}^{k}j\norm{\hat{\mA}^j}_{\infty}\left(1-\frac{1}{(1-\theta)^2}\xi_{N,T}\psi_N\right)\leq \frac{1}{1-\bar{C}}\frac{1}{(1-\theta)^3}\xi_{N,T}\psi_N^2+\frac{\theta}{(1-\theta)^2}\psi_{N}\\
    &\sum\limits_{j=0}^{k}j\norm{\hat{\mA}^j}_{\infty}\overset{(\ref{sufficiently_large1})}{\leq}  \frac{1}{1-\bar{C}}\left[\frac{1}{1-\bar{C}}\frac{1}{(1-\theta)^3}\xi_{N,T}\psi_N^2+\frac{\theta}{(1-\theta)^2}\psi_{N}\right].
\end{split}\end{equation*}

\begin{equation*}\begin{split}
\left(\lim\limits_{k\to\infty}\left\lbrace\sum\limits_{j=0}^{k}j\norm{\hat{\mA}^j}_{\infty}\right\rbrace\right)^q&{\leq} \left(\frac{1}{1-\bar{C}}\left[\frac{1}{1-\bar{C}}\frac{1}{(1-\theta)^3}\xi_{N,T}\psi_N^2+\frac{\theta}{(1-\theta)^2}\psi_{N}\right]\right)^q\\
&\overset{\text{Ass.}\ref{ass:VARconsistency}}{\leq} \left(\frac{1}{1-\bar{C}}\left[\frac{\bar{C}}{1-\bar{C}}\frac{1}{1-\theta}+\frac{\theta}{(1-\theta)^2}\right]\right)^q\psi_N^q.
\end{split}\end{equation*}
We also have summability for the differences of powers: 
\begin{equation}\label{VAR_power_diff_m_sum}
   \sum\limits_{j=0}^{\infty}\norm{\hat{\mA}^j-\mA^j}_\infty^q\leq \left(\frac{1}{1-\bar{C}}\frac{1}{(1-\theta)^2}\right)^q\xi_{N,T}^q\psi_N^{2q}
\end{equation}
To see this:
\begin{equation*}
    \sum\limits_{j=0}^{\infty}\norm{\hat{\mA}^j-\mA^j}_\infty^q\overset{(\ref{cross_terms})}{\leq}\left(\sum\limits_{j=0}^{\infty}\norm{\hat{\mA}^j-\mA^j}_\infty\right)^q=\left(\lim\limits_{k\to\infty}\sum\limits_{j=0}^{k}\norm{\hat{\mA}^j-\mA^j}_\infty\right)^q
\end{equation*}
\begin{equation*}\begin{split}
   &\sum\limits_{j=0}^{k}\norm{\hat{\mA}^j-\mA^j}_\infty \overset{(\ref{VAR_power_diff_sum})}{\leq} \frac{1}{1-\theta}\xi_{N,T}\psi_N\sum\limits_{j=0}^k\norm{\hat{\mA}^j}_\infty\overset{(\ref{boot_VMA_summable})}{\leq}\frac{1}{1-\bar{C}}\frac{1}{(1-\theta)^2}\xi_{N,T}\psi_N^2.
\end{split}\end{equation*}
For point 4., first note that by equation (3.247) of \cite{gentle2007matrix}, for any square matrix $\bA$, $\lim\limits_{k\to\infty} \bA^k=\bzero \text{ if and only if } \rho(\bA)<1$. 
We also have (for any square matrix $\bA$),
\begin{equation*}
    \lim\limits_{k\to\infty}\bA^k=\bzero \iff \lim\limits_{k\to\infty}\norm{\bA^k}_\infty=\bzero.
\end{equation*}
``$\implies$'' follows from the Continuous mapping theorem: $\lim\limits_{k\to\infty}\norm{\bA^k}_\infty=\norm{\lim\limits_{k\to\infty}\bA^k}_\infty=\norm{\bzero}_\infty=0$, because $\norm{\cdot}_{\infty}$ is a  continuous function in the entries of $\bA$ (the limit is with respect to the power of $\bA$, not the dimension of $\bA$, so this argument should work). ``$\impliedby$'' follows from:
\begin{equation*}\begin{split}
    0\leq\abs{[\bA^k]_{i,j}}&\leq \norm{\bA^k}_\infty\\
    0\leq\lim\limits_{k\to\infty}\abs{[\bA^k]_{i,j}}&\leq \lim\limits_{k\to\infty}\norm{\bA^k}_\infty=0,
\end{split}\end{equation*}
which implies that $\lim\limits_{k\to\infty}\abs{[\bA^k]_{i,j}}=0$ for all $i,j$, i.e. $\lim\limits_{k\to\infty}\bA^k=\bzero$. To summarize: For any square matrix $\bA$
\begin{equation}\label{gentle_equiv}
    \rho(\bA)<1\iff \lim\limits_{k\to\infty}\norm{\bA^k}_\infty=\bzero.
\end{equation}
Applying this to $\hat\mA$:
\begin{equation*}\begin{split}
    0\leq\norm{\hat\mA^k}_\infty&{\leq}\norm{\hat\mA^k-\mA^k}_\infty+\norm{\mA^k}_\infty\\
    0\leq\lim\limits_{k\to\infty}\norm{\hat\mA^k}_\infty&{\leq}\lim\limits_{k\to\infty}\norm{\hat\mA^k-\mA^k}_\infty+\lim\limits_{k\to\infty}\norm{\mA^k}_\infty
\end{split}\end{equation*}
From (\ref{VMA_summable}) we have $\sum\limits_{k=0}^\infty\norm{\mA^k}_\infty\leq \frac{1}{1-\theta}\psi_{N}$ and from (\ref{VAR_power_diff_m_sum}), $\sum\limits_{k=0}^{\infty}\norm{\hat{\mA}^k-\mA^k}_\infty\leq \frac{1}{1-\bar{C}}\frac{1}{(1-\theta)^2}\xi_{N,T}\psi_N^{2}$. This means that $\norm{\mA^k}_\infty$ and $\norm{\hat{\mA}^k-\mA^k}_\infty$ are absolutely summable sequences, which implies they both converge to $0$.
\begin{equation*}\begin{split}
    0\leq\lim\limits_{k\to\infty}\norm{\hat\mA^k}_\infty&{\leq}\lim\limits_{k\to\infty}\norm{\hat\mA^k-\mA^k}_\infty+\lim\limits_{k\to\infty}\norm{\mA^k}_\infty=0\\
    0\leq\lim\limits_{k\to\infty}\norm{\hat\mA^k}_\infty&\leq0 \overset{(\ref{gentle_equiv})}{\iff} \rho(\hat\mA)<1.
\end{split}\end{equation*}
Given point 4., the estimated VAR is invertible, and we have
\begin{equation}\label{boot_VMA_to_VAR}
    \norm{\hat{\bB}_k}_{\infty}=\norm{\bJ\hat\mA^k\bJ^\prime}_{\infty}\leq \norm{\bJ}_{\infty}\norm{\hat{\mA}^k}_{\infty}\norm{\bJ^\prime}_{\infty}=\norm{\hat{\mA}^k}_{\infty},
\end{equation}
and
\begin{equation}\label{VMA_to_VAR_diff}
    \norm{\hat{\bB}_k-\bB_k}_{\infty}=\norm{\bJ\left(\hat{\mA}^k-\mA^k\right)\bJ^\prime}_{\infty}\leq \norm{\bJ}_{\infty}\norm{\hat{\mA}^k-\mA^k}_{\infty}\norm{\bJ^\prime}_{\infty}=\norm{\hat{\mA}^k-\mA^k}_{\infty}.
\end{equation}
For point 5., 
\begin{equation*}
    \tilde{S}^*=\sum\limits_{j=0}^{\infty}\norm{\hat{\bB}_{j}}_\infty\overset{(\ref{boot_VMA_to_VAR})}{\leq}\sum\limits_{j=0}^{\infty}\norm{\hat{\mA}^{j}}_\infty\overset{(\ref{boot_VMA_summable})}{\leq}\frac{1}{1-\bar{C}}\frac{1}{1-\theta}\psi_{N},
\end{equation*}
with $C_5=\frac{1}{1-\bar{C}}\frac{1}{1-\theta}$.
For point 6.,
\begin{equation*}
    \sum\limits_{j=0}^{\infty}\norm{\hat{\bB}_j-\bB_j}_\infty^q\overset{(\ref{VMA_to_VAR_diff})}{\leq} \sum\limits_{j=0}^{\infty}\norm{\hat{\mA}^j-\mA^j}_\infty^q \overset{(\ref{VAR_power_diff_m_sum})}{\leq}  \left(\frac{1}{1-\bar{C}}\frac{1}{(1-\theta)^2}\right)^q\xi_{N,T}^q\psi_N^{2q},
\end{equation*}
with $C_6=\frac{1}{1-\bar{C}}\frac{1}{(1-\theta)^2}$. For point 7.,
\begin{equation*}\begin{split}
    &\sum\limits_{j=0}^{\infty}\left(\sum\limits_{k=j+1}^{\infty}\norm{\hat{\bB}_k}_\infty\right)^q\overset{(\ref{boot_VMA_to_VAR})}{\leq}\sum\limits_{j=0}^{\infty}\left(\sum\limits_{k=j+1}^{\infty}\norm{\hat{\mA}^k}_\infty\right)^q\overset{(\ref{cross_terms})}{\leq}\left(\sum\limits_{j=0}^{\infty}\sum\limits_{k=j+1}^{\infty}\norm{\hat{\mA}^k}_\infty\right)^q\\
    &=\left(\sum\limits_{k=1}^{\infty}\sum\limits_{j=0}^{\infty}\norm{\hat{\mA}^k}_\infty\ind{k\geq j+1}\right)^q=\left(\sum\limits_{k=1}^{\infty}\sum\limits_{j=0}^{k-1}\norm{\hat{\mA}^k}_\infty\right)^q=\left(\sum\limits_{k=1}^{\infty}k\norm{\hat{\mA}^k}_\infty\right)^q\\
    &\overset{(\ref{boot_VMA_j_summable})}{\leq} \left(\frac{1}{1-\bar{C}}\left[\frac{\bar{C}}{1-\bar{C}}\frac{1}{1-\theta}+\frac{\theta}{(1-\theta)^2}\right]\psi_N\right)^q,
\end{split}\end{equation*}
with $C_{7}=\frac{1}{1-\bar{C}}\left[\frac{\bar{C}}{1-\bar{C}}\frac{1}{1-\theta}+\frac{\theta}{(1-\theta)^2}\right]$.
\end{proof}

\begin{proof}[\hypertarget{p:L6}{\textbf{Proof of \Cref{lma:epsilonConsistent}}}] 
By Markov's inequality and \Cref{lma:subgaussian}.\ref{lma:subgaussian1},
which implies \\$\E\exp(\max\limits_{j}\epsilon_{j,t}^2/d_{N}^2)\leq 2$, we have that
\begin{equation*}\begin{split}
    \P\left(\max\limits_{j}\sum\limits_{t=1}^T\epsilon_{j,t}^2>Ty\right)&=\P\left(\exp\left(\max\limits_{j}\sum\limits_{t=1}^T\epsilon_{j,t}^2/d_{N}^2\right)>\exp\left(Ty/d_{N}^2\right)\right)\\
    &\leq \frac{\E\exp\left(\max\limits_{j}\sum\limits_{t=1}^T\epsilon_{j,t}^2/d_{N}^2\right)}{\exp \left(Ty/d_{N}^2\right)}
    \leq \frac{\prod\limits_{t=1}^T\E\exp\left(\max\limits_{j}\epsilon_{j,t}^2/d_{N}^2\right)}{\exp \left(Ty/d_{N}^2\right)}\leq \frac{2^T}{\exp \left(Ty/d_{N}^2\right)}. 
\end{split}\end{equation*}
Therefore
\begin{equation*}
    \P\left(\max\limits_{j}\frac{1}{T}\sum\limits_{t=1}^T\epsilon_{j,t}^2\leq y\right)\geq 1- \frac{2^T}{\exp \left(Ty/d_{N}^2\right)},
\end{equation*}
and we need to choose $y$ such that this converges to 1. In particular, we take 
$y=Cd_{N}^2$, and the first statement follows. 

For the second statement, we use the union bound, Markov's and Minkowski's inequalities, and \Cref{ass:subgaussian}.\ref{ass:subgaussian2} 
\begin{equation*}\begin{split}
    \P\left(\max\limits_{j}\sum\limits_{t=1}^T\epsilon_{j,t}^2>Ty\right)&\leq \sum\limits_{j=1}^N\P\left(\sum\limits_{t=1}^T\epsilon_{j,t}^2>Ty\right)\leq \sum\limits_{j=1}^N\frac{\E\abs{\sum\limits_{t=1}^T\epsilon_{j,t}^2}^{m/2}}{(Ty)^{m/2}}\leq \sum\limits_{j=1}^N\frac{\left(\sum\limits_{t=1}^T\left[\E\abs{\epsilon_{j,t}}^{m}\right]^{2/m}\right)^{m/2}}{(Ty)^{m/2}}\\
    &\leq \frac{NT^{m/2}\max\limits_{j,t}\norm{\epsilon_{j,t}}_{L_m}^m}{(Ty)^{m/2}}\leq \frac{CNT^{m/2}}{(Ty)^{m/2}}. 
\end{split}\end{equation*}
Therefore
\begin{equation*}
    \P\left(\max\limits_{j}\frac{1}{T}\sum\limits_{t=1}^T\epsilon_{j,t}^2\leq y\right)\geq 1-\frac{CNT^{m/2}}{(Ty)^{m/2}},
\end{equation*}
which converges to 1 when $y= d_N^2$: $\frac{CNT^{m/2}}{(Ty)^{m/2}}=C\eta_T^m\to0$.

\end{proof}

\begin{proof}[\hypertarget{p:L7}{\textbf{Proof of \Cref{lma:hatstonohats2}}}]
We have that
    \begin{equation*}\begin{split}
        &\norm{\frac{1}{T}\sum\limits_{t=1}^T\hat\bepsilon_{t}\hat\bepsilon_{t}^\prime-\frac{1}{T}\sum\limits_{t=1}^T\bepsilon_{t}\bepsilon_{t}^\prime}_{\max}
        =\norm{\frac{1}{T}\sum\limits_{t=1}^T\left[\left(\hat\bepsilon_{t}-\bepsilon_{t}\right)\left(\hat\bepsilon_{t}^\prime-\bepsilon_{t}^\prime\right)+\left(\hat\bepsilon_{t}-\bepsilon_{t}\right)\bepsilon_{t}^\prime+\bepsilon_{t}\left(\hat\bepsilon_{t}^\prime-\bepsilon_{t}^\prime\right)\right]}_{\max}\\
        &\leq\norm{\frac{1}{T}\sum\limits_{t=1}^T\left(\hat\bepsilon_{t}-\bepsilon_{t}\right)\left(\hat\bepsilon_{t}^\prime-\bepsilon_{t}^\prime\right)}_{\max}+2\norm{\frac{1}{T}\sum\limits_{t=1}^T\left(\hat\bepsilon_{t}-\bepsilon_{t}\right)\bepsilon_{t}^\prime }_{\max}.
    \end{split}\end{equation*}
    By the Cauchy-Schwarz inequality,
     \begin{equation*}\begin{split}
       \norm{\frac{1}{T}\sum\limits_{t=1}^T\left(\hat\bepsilon_{t}-\bepsilon_{t}\right)\left(\hat\bepsilon_{t}^\prime-\bepsilon_{t}^\prime\right)}_{\max}&=\max\limits_{r,s}\abs{\frac{1}{T}\sum\limits_{t=1}^T\left(\hat\epsilon_{r,t}-\epsilon_{r,t}\right)\left(\hat\epsilon_{s,t}-\epsilon_{s,t}\right)}\\
       &\leq \max\limits_{r,s}\left\lbrace\frac{1}{T}\left(\sum\limits_{t=1}^T\abs{\hat\epsilon_{r,t}-\epsilon_{r,t}}^2\right)^{1/2}\left(\sum\limits_{t=1}^T\abs{\hat\epsilon_{s,t}-\epsilon_{s,t}}^2\right)^{1/2}\right\rbrace\\
       &=\frac{1}{T}\max\limits_{r}\left(\sum\limits_{t=1}^T\abs{\hat\epsilon_{r,t}-\epsilon_{r,t}}^2\right)=\frac{1}{T}\max\limits_{r}\norm{\hat\bepsilon_{r}-\bepsilon_{r}}_{2}^2\overset{\mathcal{Q}}{\leq}\phi_{N,T}.
    \end{split}\end{equation*}
    Then
     \begin{equation*}\begin{split}
         \norm{\frac{1}{T}\sum\limits_{t=1}^T\left(\hat\bepsilon_{t}-\bepsilon_{t}\right)\bepsilon_{t}^\prime }_{\max}&=\max\limits_{s,r}\abs{\frac{1}{T}\sum\limits_{t=1}^T\left(\hat\epsilon_{r,t}-\epsilon_{r,t}\right)\epsilon_{s,t}}\\
         &\leq \max\limits_{s,r}\abs{\frac{1}{T}\left(\sum\limits_{t=1}^T\abs{\hat\epsilon_{r,t}-\epsilon_{r,t}}^2\right)^{1/2}\left(\sum\limits_{t=1}^T\abs{\epsilon_{s,t}}^2\right)^{1/2}}\\
         &\leq \max\limits_{r}\abs{\left(\sum\limits_{t=1}^T\abs{\hat\epsilon_{r,t}-\epsilon_{r,t}}^2\right)^{1/2}}\max\limits_{s}\abs{\frac{1}{T}\left(\sum\limits_{t=1}^T\abs{\epsilon_{s,t}}^2\right)^{1/2}}\\
         &=\frac{1}{\sqrt{T}} \max\limits_{r}\norm{\hat\bepsilon_{r}-\bepsilon_{r}}_{2}\max\limits_{s}\abs{\frac{1}{T}\sum\limits_{t=1}^T\epsilon_{s,t}^2}^{1/2}
         \overset{\mathcal{Q},\mathcal{R}_{1}}{\leq} Cd_{N}\sqrt{\phi_{N,T}} 
    \end{split}\end{equation*}
    and the first statement follows. The second statement follows by identical steps except the last, where we use the set $\mathcal{R}_2$ to bound $ \max\limits_{r}\abs{\frac{1}{T}\sum\limits_{t=1}^T\epsilon_{r,t}^2}^{1/2}\leq Cd_{N}$.
\end{proof}

\begin{proof}[\hypertarget{p:L8}{\textbf{Proof of \Cref{lma:nohatstoexp}}}]
By the union bound
\begin{equation*}
    \P\left(\norm{\frac{1}{T}\sum\limits_{t=1}^T\bepsilon_{t}\bepsilon_{t}^\prime-\frac{1}{T}\sum\limits_{t=1}^T\E\bepsilon_{t}\bepsilon_{t}^\prime}_{\max}\leq y\right)\geq 1-\sum\limits_{1\leq s,r \leq N}\P\left(\abs{\frac{1}{T}\sum\limits_{t=1}^T\left[\epsilon_{r,t}\epsilon_{s,t}-\E\epsilon_{r,t}\epsilon_{s,t}\right]}>y\right).
\end{equation*}
Note that by Lemma 2.7.7 and  Exercise 2.7.10  of \cite{vershynin2019high} we have that under \Cref{ass:subgaussian}.\ref{ass:subgaussian1}, $\epsilon_{r,t}\epsilon_{s,t}$ is sub-exponential with $\norm{\epsilon_{r,t}\epsilon_{s,t}-\E\epsilon_{r,t}\epsilon_{s,t}}_{\psi_{1}}\leq C\norm{\epsilon_{r,t}\epsilon_{s,t}}_{\psi_{1}}\leq \norm{\epsilon_{r,t}}_{\psi_2}\norm{\epsilon_{s,t}}_{\psi_2}\leq C$. Furthermore, by Theorem 2.8.1  of \cite{vershynin2019high}, we have Bernstein's inequality
\begin{equation*}
\begin{split}
    &\P\left(\abs{\sum\limits_{t=1}^T\left(\epsilon_{r,t}\epsilon_{s,t}-\E\epsilon_{r,t}\epsilon_{s,t}\right)}>Ty\right)\\
    &\leq 2\exp\left(-C\min\left\lbrace\frac{T^2y^2}{\sum\limits_{t=1}^T\norm{\epsilon_{r,t}\epsilon_{s,t}-\E\epsilon_{r,t}\epsilon_{s,t}}_{\psi_1}^2},\frac{Ty}{\max\limits_{t}\norm{\epsilon_{r,t}\epsilon_{s,t}-\E\epsilon_{r,t}\epsilon_{s,t}}_{\psi_1}}\right\rbrace\right)
\end{split}
\end{equation*}
We separately bound the terms in the minimum, $\sum\limits_{t=1}^T\norm{\epsilon_{r,t}\epsilon_{s,t}-\E\epsilon_{r,t}\epsilon_{s,t}}_{\psi_1}^2\leq C T$, and\\ $\max\limits_{t}\norm{\epsilon_{r,t}\epsilon_{s,t}-\E\epsilon_{r,t}\epsilon_{s,t}}_{\psi_1}\leq C$, so this simplifies to 
\begin{equation*}
    \P\left(\abs{\sum\limits_{t=1}^T\left(\epsilon_{r,t}\epsilon_{s,t}-\E\epsilon_{r,t}\epsilon_{s,t}\right)}>Ty\right)\leq 2\exp\left(-C\min\left\lbrace Ty^2,Ty\right\rbrace\right).
\end{equation*}
since we will choose $y\to0$, the first term is smaller, and we obtain the bound $2\exp\left(-CTy^2\right)$, and 
\begin{equation*}
    \P\left(\norm{\frac{1}{T}\sum\limits_{t=1}^T\bepsilon_{t}\bepsilon_{t}^\prime-\frac{1}{T}\sum\limits_{t=1}^T\E\bepsilon_{t}\bepsilon_{t}^\prime}_{\max}\leq y\right)\geq 1-C_1N^2\exp\left(-C_2 Ty^2\right).
\end{equation*}
We then find $y$ by bounding $C_1N^2\exp\left(-C_2Ty^2\right)\leq N^{-1}\implies y\geq C\frac{\sqrt{\log(N)}}{\sqrt{T}}$, and the first result follows by taking $y\sim\frac{d_N}{\sqrt{T}}$.

For the second result, by Markov's, Marcinkiewicz–Zygmund (twice) and Minkowski's inequalities 
\begin{equation*}
\begin{split}
    &\P\left(\abs{\frac{1}{T}\sum\limits_{t=1}^T\left[\epsilon_{r,t}\epsilon_{s,t}-\E\epsilon_{r,t}\epsilon_{s,t}\right]}>y\right)\leq \frac{\E\left[\abs{\sum\limits_{t=1}^T\left[\epsilon_{r,t}\epsilon_{s,t}-\E\epsilon_{r,t}\epsilon_{s,t}\right]}^{m/2}\right]}{T^{m/2}y^{m/2}}\\
    &\leq C\frac{\E\left[\left(\sum\limits_{t=1}^T\abs{\epsilon_{r,t}\epsilon_{s,t}-\E\epsilon_{r,t}\epsilon_{s,t}}^{4}\right)^{m/8}\right]}{T^{m/2}y^{m/2}}=C\frac{\norm{\sum\limits_{t=1}^T\abs{\epsilon_{r,t}\epsilon_{s,t}-\E\epsilon_{r,t}\epsilon_{s,t}}^{4}}_{L_{m/8}}^{m/8}}{T^{m/2}y^{m/2}}\\
    &\leq C\frac{\left(\sum\limits_{t=1}^T\norm{\abs{\epsilon_{r,t}\epsilon_{s,t}-\E\epsilon_{r,t}\epsilon_{s,t}}^{4}}_{L_{m/8}}\right)^{m/8}}{T^{m/2}y^{m/2}}=C\frac{\left(\sum\limits_{t=1}^T\norm{\epsilon_{r,t}\epsilon_{s,t}-\E\epsilon_{r,t}\epsilon_{s,t}}_{L_{m/2}}^{4}\right)^{m/8}}{T^{m/2}y^{m/2}}
\end{split}
\end{equation*}
By triangle, Jensen's, and Cauchy-Schwarz inequalities, and \Cref{ass:subgaussian}.\ref{ass:subgaussian2}
\begin{equation*}
    \norm{\epsilon_{r,t}\epsilon_{s,t}-\E\epsilon_{r,t}\epsilon_{s,t}}_{L_{m/2}}\leq C\norm{\epsilon_{r,t}\epsilon_{s,t}}_{L_{m/2}}\leq C\norm{\epsilon_{r,t}}_{L_{m}}\norm{\epsilon_{s,t}}_{L_{m}}\leq C,
\end{equation*}
so
\begin{equation*}
    \P\left(\norm{\frac{1}{T}\sum\limits_{t=1}^T\bepsilon_{t}\bepsilon_{t}^\prime-\frac{1}{T}\sum\limits_{t=1}^T\E\bepsilon_{t}\bepsilon_{t}^\prime}_{\max}\leq y\right)\geq 1-CN^2\frac{T^{m/8} }{T^{m/2}y^{m/2}}=1-C N^2 T^{-3m/8} y^{-m/2}. 
\end{equation*}
This probability then converges to 1 when $y\sim \frac{N^{4/m}}{T^{3/4}}\eta_T^{-1}$, so the second result follows when taking $y\sim \frac{d_N^4}{T^{3/4}}$.
\end{proof}

\begin{proof}[\hypertarget{p:T2}{\textbf{Proof of \Cref{thm:CovarianceCloseness}}}]
For $N\times N$ matrices $\bA,\bB, \bC$, 
\begin{equation*}\begin{split}
    \norm{\bA\bB\bC^{\prime}}_{\max}&=\max\limits_{1\leq r,s\leq N}\norm{\ba_r\bB\bc_s^\prime}=\max\limits_{r,s}\norm{\sum\limits_{1\leq i,j\leq N}\ba_{r,i}b_{i,j}\bc_{s,j}^\prime}\leq \max\limits_{i,j}\abs{b_{i,j}}\max\limits_{r,s}\left\lbrace\sum\limits_{i,j}\abs{a_{r,i}}\abs{c_{s,j}}\right\rbrace\\
    &=\max\limits_{i,j}\abs{b_{i,j}}\max\limits_{r,s}\left\lbrace\norm{\ba_{r}}_{1}\norm{\bc_{s}}_1\right\rbrace\leq \norm{\bB}_{\max}\norm{\bA}_{\infty}\norm{\bC}_\infty.
\end{split}\end{equation*}
Using telescoping sums, sub-additivity of the $\norm{\cdot}_{\max}$ norm, and the result above, we can rewrite 
\begin{equation*}\begin{split}
     &\norm{\hat\bSigma-\bSigma}_{\max}= \norm{\hat\cB(1)\hat\bSigma_{\epsilon}\hat\cB(1)^\prime-\cB(1)\bSigma_{\epsilon}\cB(1)^\prime}_{\max}\\
     &\leq \norm{\Delta\hat\bSigma_{\epsilon}}_{\max}\norm{\Delta\hat\cB(1)}_{\infty}^2+ \norm{\bSigma_{\epsilon}}_{\max}\norm{\Delta\hat\cB(1)}_{\infty}^2+ \norm{\Delta\hat\bSigma_{\epsilon}}_{\max}\norm{\cB(1)}_{\infty}^2\\
     &+2 \norm{\Delta\hat\bSigma_{\epsilon}}_{\max}\norm{\Delta\hat\cB(1)}_{\infty}\norm{\cB(1)}_{\infty}+2 \norm{\bSigma_{\epsilon}}_{\max}\norm{\Delta\hat\cB(1)}_{\infty}\norm{\cB(1)}_{\infty},
\end{split}\end{equation*}
where $\Delta\hat\bSigma_{\epsilon}=\hat\bSigma_{\epsilon}-\bSigma_{\epsilon}$ and $\Delta\hat\cB(1)=\hat\cB(1)-\cB(1)$. There are therefore 4 distinct expressions we need to bound. 
On $\mathcal{Q}\bigcap\mathcal{R}_1\bigcap\mathcal{S}_1$, by \Cref{lma:hatstonohats2}
\begin{equation*}\begin{split}
     \norm{\Delta\hat\bSigma_{\epsilon}}_{\max}&\leq  \norm{\frac{1}{T}\sum\limits_{t=1}^T\hat\bepsilon_{t}\hat\bepsilon_{t}^\prime-\frac{1}{T}\sum\limits_{t=1}^T\bepsilon_{t}\bepsilon_{t}^\prime}_{\max}+\norm{\frac{1}{T}\sum\limits_{t=1}^T\bepsilon_{t}\bepsilon_{t}^\prime-\frac{1}{T}\sum\limits_{t=1}^T\E\bepsilon_{t}\bepsilon_{t}^\prime}_{\max}\\
     &\leq C\left(\phi_{N,T}+d_{N}\sqrt{\phi_{N,T}}+\frac{d_N}{\sqrt{T}}\right).
\end{split}\end{equation*}
On $\mathcal{Q}\bigcap\mathcal{R}_2\bigcap\mathcal{S}_2$
\begin{equation*}\begin{split}
     \norm{\Delta\hat\bSigma_{\epsilon}}_{\max}
     \leq C\left(\phi_{N,T}+d_{N}\sqrt{\phi_{N,T}}+\frac{d_N^4}{T^{3/4}}\right).
\end{split}\end{equation*}
By \Cref{lma:trueVMAsummable}.6, on $\mathcal{P}$, 
\begin{equation*}
    \norm{\Delta\hat\cB(1)}_{\infty}\leq \sum\limits_{k=0}^\infty\norm{\hat\bB_k-\bB_k}_{\infty}\leq C\xi_{N,T}\psi_N^2.
\end{equation*}
By Cauchy-Schwarz and \Cref{ass:subgaussian}
\begin{equation*}
     \norm{\bSigma_{\epsilon}}_{\max}=\max\limits_{r,s}\abs{\frac{1}{T}\sum\limits_{t=1}^T\E\epsilon_{r,t}\epsilon_{s,t}}\leq \max\limits_{r,t}\norm{\epsilon_{r,t}}_{L_2}^2
     \leq C.
\end{equation*}
Note that the above argument works also under \Cref{ass:subgaussian}.\ref{ass:subgaussian2}: By Equation (2.15) in \cite{vershynin2019high} 
\begin{equation*}
    \max\limits_{r,t}\norm{\epsilon_{r,t}}_{L_2}^2\leq C m \max\limits_{r,t}\norm{\epsilon_{r,t}}_{\psi_2}^2\leq C.
\end{equation*}

Under \Cref{ass:VARsummable}, by \Cref{lma:trueVMAsummable}.1
\begin{equation*}
    \norm{\cB(1)}_{\infty}\leq \sum\limits_{k=0}^\infty\norm{\bB_{k}}_{\infty}=\tilde S\leq C\psi_{N}.
\end{equation*}
Plugging these in, we find
\begin{equation*}\begin{split}
     \norm{\hat\bSigma-\bSigma}_{\max}&\leq C_1\norm{\Delta\hat\bSigma_{\epsilon}}_{\max}\psi_N^2+C_2\xi_{N,T}\psi_N^3.\\
\end{split}\end{equation*}
Plugging in the respective bounds on $\norm{\Delta\hat\bSigma_{\epsilon}}_{\max}$, we obtain the bounds in $\mathcal{T}_1$ and $\mathcal{T}_2$.
\end{proof}
\begin{proof}[\hypertarget{p:L9}{\textbf{Proof of \Cref{lma:epsilonbounded}}}]
By the union bound and equation (2.14) in \cite{vershynin2019high}, and using \Cref{lma:subgaussian}.\ref{lma:subgaussian1}
\begin{equation*}\begin{split}
    \P\left(\max\limits_{j,t}\abs{\epsilon_{j,t}}\leq y\right)&=1-\sum\limits_{t=1}^T\P\left(\max\limits_{j}\abs{\epsilon_{j,t}}>y\right)\geq 1-\sum\limits_{t=1}^T2\exp\left(-Cy^2/\norm{\max\limits_{j}\abs{\epsilon_{j,t}}}_{\psi_2}^2\right)\\&\geq 1-2T\exp\left(\frac{-Cy^2}{d_{N}^2}\right).
\end{split}\end{equation*}
This probability converges to 1 when taking $y=d_{N}\log(T)$, showing the first statement. By union bound, Markov's inequality and the arguments in the proof of \Cref{lma:subgaussian}.\ref{lma:subgaussian2},
\begin{equation*}\begin{split}
    \P\left(\max\limits_{j,t}\abs{\epsilon_{j,t}}\leq y\right)&\geq1-\sum\limits_{t=1}^T\P\left(\max\limits_{j}\abs{\epsilon_{j,t}}>y\right)\geq 1-\sum\limits_{t=1}^T\frac{\E\left[\max\limits_{j}\abs{\epsilon_{j,t}}^m\right]}{y^m}\\
    &\geq 1-T\frac{\max\limits_{t}\norm{\max\limits_{j}\abs{\epsilon_{j,t}}}^m_{L_m}}{y^m}\geq 1-TN y^{-m}.
\end{split}\end{equation*}
This probability converges to 1 when $y=d_{N} T^{1/m}$, showing the second statement.
\end{proof}

\begin{proof}[\hypertarget{p:L10}{\textbf{Proof of \Cref{lma:bootsubgaussian}}}]
By submultiplicativity of the Orlicz norm,
\begin{equation*}
    \max\limits_{t}\norm{\max\limits_{j}\epsilon_{j,t}^*}_{\psi_2}^*=\max\limits_{t}\norm{\max\limits_{j}\hat\epsilon_{j,t}\gamma_t}_{\psi_2}^*\leq \max\limits_{t}\norm{\max\limits_{j}\hat\epsilon_{j,t}}_{\psi_2}^*\max\limits_{t}\norm{\gamma_t}_{\psi_2}^*.
\end{equation*}
Since $\gamma_t$ is by construction independent of $\bX$ and identically Gaussian distributed, we have by Example 2.5.8 in \cite{vershynin2019high} $\max\limits_{t}\norm{\gamma_t}_{\psi_2}^*=\max\limits_{t}\norm{\gamma_t}_{\psi_2}\leq C$. 
\begin{equation*}\begin{split}
     \max\limits_{t}\norm{\max\limits_{j}\hat\epsilon_{j,t}}_{\psi_2}^*&= \max\limits_{t}\inf\left\lbrace \lambda>0:\E^*\exp\left(\abs{\max\limits_{j}\hat\epsilon_{j,t}}^2/\lambda^2\right)\leq 2\right\rbrace\\
     &=\max\limits_{t}\inf\left\lbrace \lambda>0:\exp\left(\abs{\max\limits_{j}\hat\epsilon_{j,t}}^2/\lambda^2\right)\leq 2\right\rbrace\\
     &\leq\max\limits_{t}\inf\left\lbrace \lambda>0:\exp\left(\max\limits_{j}\abs{\hat\epsilon_{j,t}}^2/\lambda^2\right)\leq 2\right\rbrace\\
      &=\max\limits_{t}\inf\left\lbrace \lambda>0:\max\limits_{j}\abs{\hat\epsilon_{j,t}}\leq \sqrt{\log(2)}\lambda\right\rbrace.
\end{split}\end{equation*}
Therefore, up to a $\sqrt{\log(2)}$ constant, any bound on $\max\limits_{j,t}\abs{\hat\epsilon_{j,t}}$ is also a bound on $\max\limits_{t}\norm{\max\limits_{j}\hat\epsilon_{j,t}}_{\psi_2}^*$. By triangle inequality, $\max\limits_{j,t}\abs{\hat\epsilon_{j,t}}\leq   \max\limits_{j,t}\abs{\hat\epsilon_{j,t}-\epsilon_{j,t}}+  \max\limits_{j,t}\abs{\epsilon_{j,t}}$, and we further bound the individual terms using $\mathcal{Q}$
\begin{equation*}
\max\limits_{j,t}\abs{\hat\epsilon_{j,t}-\epsilon_{j,t}}\leq \max\limits_{j}\sqrt{\sum\limits_{t=1}^T\abs{\hat\epsilon_{j,t}-\epsilon_{j,t}}^2}=\sqrt{T}\max\limits_{j}\sqrt{\frac{1}{T}\norm{\hat\bepsilon_j-\bepsilon_j}_2^2}\leq \sqrt{T\phi_{N,T}}.
\end{equation*}
Then, on $\mathcal{U}_1$, $\max\limits_{j,t}\abs{\epsilon_{j,t}}\leq d_{N}\log(T)$, and the first statement follows.

For the second statement, since $\gamma_t$ is again i.i.d.~Gaussian, we have $\max\limits_{t}\norm{\gamma_t}_{L_m}\leq C$ for all $0<m<\infty$, so 
\begin{equation*}\begin{split}
    \max\limits_{t}\norm{\max\limits_{j}\epsilon_{j,t}^*}_{L_m}^*&=  \max\limits_{t}\left(\E^{*}\max\limits_{j}\abs{\epsilon_{j,t}^*}^m\right)^{1/m}=\max\limits_{t}\left(\E^{*}\max\limits_{j}\abs{\hat\epsilon_{j,t}\gamma_t}^m\right)^{1/m}\\
    &=\max\limits_{t}\left(\max\limits_{j}\abs{\hat\epsilon_{j,t}}^m\E\abs{\gamma_t}^m\right)^{1/m}\leq C\max\limits_{j,t}\abs{\hat\epsilon_{j,t}}.
\end{split}\end{equation*}
We use the same arguments for bounding this term as for the first statement, using that on $\mathcal{U}_2$, $\max\limits_{j,t}\abs{\epsilon_{j,t}}\leq d_{N}T^{1/m}$, and the second statement is obtained.
\end{proof}

\begin{proof}[\hypertarget{p:L11}{\textbf{Proof of \Cref{lma:bootCLT}}}]
By Theorem 2.2 in \cite{chernozhukov2020nearly}, for all $\lambda>0$
\begin{equation*}
    {M_{N,T}^*}\leq C\left\lbrace\log(T)\left(\Delta_0+\sqrt{\Delta_1\log(N)}+\frac{(\mathcal{M}\log(N))^2}{T\Lambda_{\min}(\tilde\bSigma)}\right)+\sqrt{\frac{\Lambda_1M(\lambda)}{T\Lambda_{\min}^2(\tilde\bSigma)}}+\frac{\lambda \log(N)^{3/2}}{\sqrt{T\Lambda_{\min}(\tilde\bSigma)}}\right\rbrace,
\end{equation*}
where $\tilde\bSigma$ is the correlation matrix of $\bx_t$,
\begin{equation*}
    \Delta_0=\frac{\log(N)}{\Lambda_{\min}(\tilde\bSigma)}\norm{\bSigma-\bSigma^*}_{\max},
\end{equation*}
and
\begin{equation*}\begin{split}
    \bSigma^*&=\E^*\left[\left(\frac{1}{\sqrt{T}}\sum\limits_{t=1}^T\cB(1)^*\bepsilon_{t}^*\right)\left(\frac{1}{\sqrt{T}}\sum\limits_{t=1}^T\cB(1)^*\bepsilon_{t}^*\right)^\prime\right]=\cB(1)^*\left(\frac{1}{T}\sum\limits_{s,t}\E^*\bepsilon_{s}^*\bepsilon_{t}^{*\prime}\right)\cB(1)^{*\prime}\\
    &=\cB(1)^*\left(\frac{1}{T}\sum\limits_{t}\E^*\bepsilon_{t}^*\bepsilon_{t}^{*\prime}\right)\cB(1)^{*\prime}=\cB(1)^*\left(\frac{1}{T}\sum\limits_{t}\bepsilon_{t}^*\E(\gamma_{t}^2)\bepsilon_{t}^{*\prime}\right)\cB(1)^{*\prime}=\hat\cB(1)\hat\bSigma_{\epsilon}\hat\cB(1)^\prime,
\end{split}\end{equation*}
since conditionally on $\bX$, $\bepsilon_{s}^*$ and $\bepsilon_{t}^*$ are independent for $s\neq t$. Furthermore,
\begin{equation*}
    \Delta_1=\frac{(\log N)^2}{T^2\Lambda_{\min}^2(\tilde\bSigma)}\max\limits_{j}\sum\limits_{t=1}^T\E^*\abs{\cB(1)^*_j\bepsilon_{t}^*}^4,
\end{equation*}
\begin{equation*}
    \mathcal{M}=\left(\E^*\left[\max\limits_{j,t}\abs{\cB(1)^*_j\bepsilon_{t}^*}^4\right]\right)^{1/4},
\end{equation*}
\begin{equation*}
    \Lambda_1=(\log(N))^2\log(T)\log(NT),
\end{equation*}
and
\begin{equation*}
   M(\lambda)=\max\limits_{t}\E^*\left[\norm{\cB(1)^*\bepsilon_{t}^*}_{\infty}\ind{\norm{\cB(1)^*\bepsilon_{t}^*}_{\infty}>\lambda}\right].
\end{equation*}
We now derive bounds for each of these expressions. By similar arguments to those in the proof of \Cref{lma:ChernozhukovHDCLT}, by \Cref{ass:covariance}, $\Lambda_{\min}(\tilde\bSigma)\geq 1/C$, and on $\mathcal{T}_1$ or $\mathcal{T}_2$, we have respectively
\begin{equation*}
\Delta_0\leq C\log(N)\psi_N^2\left[ \phi_{N,T}+d_{N}\sqrt{\phi_{N,T}}+\frac{d_N}{\sqrt{T}}+\xi_{N,T}\psi_N\right],
\end{equation*}
or 
\begin{equation*}
\Delta_0\leq C\log(N)\psi_N^2\left[ \phi_{N,T}+d_{N}\sqrt{\phi_{N,T}}+\frac{d_N^4}{T^{3/4}}+\xi_{N,T}\psi_N\right].
\end{equation*}
For $\Delta_1$
\begin{equation*}
    \frac{(\log N)^2}{T^2\Lambda_{\min}^2(\tilde\bSigma)}\max\limits_{j}\sum\limits_{t=1}^T\E^*\abs{\cB(1)^*_j\bepsilon_{t}^*}^4\leq C\frac{\log(N)^2\tilde S^{*4}\norm{\max\limits_{j}\abs{\epsilon_{j,t}^*}}^{*4}_{L_4}}{T},
\end{equation*}
so on $\mathcal{U}_1\bigcap\mathcal{Q}$ or $\mathcal{U}_2\bigcap\mathcal{Q}$, we have by \Cref{lma:bootsubgaussian} 
\begin{equation*}
    \Delta_1\leq C\frac{\log(N)^2\tilde S^{*4}d_{N}^{*4}}{T}.
\end{equation*}
Note that $d_{N}^*$ is different depending on which clause of \Cref{lma:bootsubgaussian} we use. 
For $\mathcal{M}$ we have 
\begin{equation*}
    \left(\E^*\left[\max\limits_{j,t}\abs{\cB(1)^*_j\bepsilon_{t}^*}^4\right]\right)^{1/4}\leq \tilde{S^*}\norm{\max\limits_{j,t}\abs{\epsilon_{j,t}^*}}_{L_4}^*\leq \tilde{S^*}\norm{\max\limits_{j,t}\abs{\epsilon_{j,t}^*}}_{L_m}^*,
\end{equation*}
so on $\mathcal{U}_1\bigcap\mathcal{Q}$ or $\mathcal{U}_2\bigcap\mathcal{Q}$, we have respectively 
\begin{equation*}
 \mathcal{M}\leq \tilde{S^*}\sqrt{\log(T)}d_{N}^{*}   \text{ or } \mathcal{M}\leq \tilde{S^*}T^{1/m}d_{N}^{*}.
\end{equation*}
For $M(\lambda)$, we have by Cauchy-Schwarz
\begin{equation*}\begin{split}
    &\max\limits_{t}\E^*\left[\norm{\cB(1)^*\bepsilon_{t}^*}_{\infty}\ind{\norm{\cB(1)^*\bepsilon_{t}^*}_{\infty}>\lambda}\right]\leq \max\limits_{t}\left\lbrace\norm{\norm{\cB(1)^*\bepsilon_{t}^*}_{\infty}}^*_{L_2}\left(\P^*(\norm{\cB(1)^*\bepsilon_{t}^*}_{\infty}>\lambda)\right)^{1/2}\right\rbrace\\
    &\leq \tilde{S}^*\max\limits_{t}\norm{\max\limits_{j}\abs{\epsilon_{j,t}}}_{L_2}^*\max\limits_{t}\left(\P^*(\norm{\cB(1)^*\bepsilon_{t}^*}_{\infty}>\lambda)\right)^{1/2}.
\end{split}\end{equation*}
On $\mathcal{U}_1\bigcap\mathcal{Q}$, by equation (2.14) in \cite{vershynin2019high}, 
\begin{equation*}
    \P^*(\norm{\cB(1)^*\bepsilon_{t}^*}_{\infty}>\lambda)\leq 2\exp\left(-C\frac{\lambda^2}{d_{N}^{*2}\tilde{S}^{*2}}\right), 
\end{equation*}
and we may let $\lambda=Cd_{N}^*\tilde{S}^*\sqrt{\log(d_{N}^*\tilde{S}^*)}$ such that $M(\lambda)\leq C$. On $\mathcal{U}_2\bigcap\mathcal{Q}$, we use H\"older's inequality instead of Cauchy-Schwarz, 
\begin{equation*}\begin{split}
    \max\limits_{t}\E^*\left[\norm{\cB(1)^*\bepsilon_{t}^*}_{\infty}\ind{\norm{\cB(1)^*\bepsilon_{t}^*}_{\infty}>\lambda}\right]\leq  \tilde{S}^*\max\limits_{t}\norm{\max\limits_{j}\abs{\epsilon_{j,t}}}_{L_m}^*\max\limits_{t}\left(\P^*(\norm{\cB(1)^*\bepsilon_{t}^*}_{\infty}>\lambda)\right)^{\frac{m-1}{m}}.
\end{split}\end{equation*}
By Markov's inequality
\begin{equation*}
    \P^*(\norm{\cB(1)^*\bepsilon_{t}^*}_{\infty}>\lambda)\leq \frac{\E^*\abs{\max\limits_{j}\abs{\epsilon_{j,t}^*}}}{\lambda/\tilde{S}^*}\leq \frac{d_{N}^*\tilde{S}^*}{\lambda}.
\end{equation*}
We then take $\lambda=C(d_{N}^*\tilde{S}^*)^{\frac{2m-1}{m-1}}$ such that $M(\lambda)\leq C$. The result then follows by plugging in the bounds on these terms, and using that $\phi_{N,T}\to0$, $d_{N}\geq1$, $\tilde{S}^*\geq 1$, $d_{N}^{*}\to\infty$ to omit asymptotically dominated terms.
\end{proof}

\begin{proof}[\hypertarget{p:T3}{\textbf{Proof of \Cref{thm:HDCLTforboot}}}] This proof largely follows the same structure as the proof of \Cref{thm:HDCLTforLP}. 
By \Cref{lma:trueVMAsummable}.4, the bootstrap process is invertible, and we write the Beveridge-Nelson decomposition of the process: 
\begin{equation*}
    \bx_t^*=\cB(L)^*\bepsilon_t^*=\cB(1)^*\bepsilon_t^*-(1-L)\tilde{\cB}^*(L)\bepsilon_t^*,\text{ where } \tilde{\cB}^*(L)=\sum_{j=0}^\infty\tilde{\bB}_j^* L^j, \tilde{\bB}_j^*=\sum_{k=j+1}^\infty\bB_k^*,
\end{equation*}
\begin{equation*}
    \frac{1}{\sqrt{T}}\sum_{t=1}^T\bx_t^*=\frac{1}{\sqrt{T}}\sum_{t=1}^T\cB(1)^*\bepsilon_t^*-\frac{1}{\sqrt{T}}\tilde{\cB}^*(L)\bepsilon_{T}^*+\frac{1}{\sqrt{T}}\tilde{\cB}^*(L)\bepsilon_{0}^*.
\end{equation*}
Since $\bepsilon_0=\bzero$, it is natural to take $\bepsilon^*_0=\bzero$ as well, giving $\frac{1}{\sqrt{T}}\tilde{\cB}^*(L)\bepsilon_{0}^*=\bzero$. Define 
\begin{equation*}\begin{split}
    x_{T}^{(\max)*}=\norm{\frac{1}{\sqrt{T}}\sum_{t=1}^T\bx_t^*}_\infty,\qquad \epsilon_T^{(\max)*}=\norm{\frac{1}{\sqrt{T}}\sum_{t=1}^T\cB(1)^*\bepsilon_t^*}_\infty, \qquad z_T^{(\max)}=\norm{\bz}_\infty,
\end{split}\end{equation*}
\begin{equation*}\begin{split}
    &F_{1,T}^*(y):=\P\left(x_{T}^{(\max)*}\leq y\right)\quad
    F_{2,T}^*(y):=\P\left(\epsilon_T^{(\max)*}\leq y\right)\\
    &G_{T}^*(y):=\P\left(z_T^{(\max)*}\leq y\right)\quad 
    r_T^*:=x_{T}^{(\max)*}-\epsilon_T^{(\max)*}
\end{split}\end{equation*}
Then
\begin{equation*}\begin{split}
    \abs{r_T^*}
    \leq&\norm{\frac{1}{\sqrt{T}}\tilde{\cB}(L)^*\bepsilon_{T}^*}_\infty=R_T^*.
\end{split}\end{equation*}
For $R_T^*$, we may simply apply \Cref{lma:leftover} to the bootstrap quantity directly, using \Cref{lma:bootsubgaussian} instead of \Cref{lma:subgaussian}: On $\mathcal{U}_1\bigcap\mathcal{Q}$, by \Cref{lma:bootsubgaussian}.\ref{lma:bootsubgaussian1}, we have 
\begin{equation*}
     \P^*\left(R_T^*>\eta_{T}\right) \leq 2N\exp\left(-C\frac{\eta_T^2 T}{d_{N}^{*2} S^*_2}\right).
 \end{equation*}
 Similarly, on $\mathcal{U}_2\bigcap\mathcal{Q}$, by \Cref{lma:bootsubgaussian}.\ref{lma:bootsubgaussian2}
 \begin{equation*}
     \P^*\left(R_T^*>\eta_{T}\right) \leq C\frac{N d_{N}^{*m} S^{*m}_1}{\left(\eta_{T}\sqrt{T}\right)^m}.
 \end{equation*}

\textit{Under \Cref{ass:subgaussian}.\ref{ass:subgaussian1}}, we can bound
\begin{equation*}\begin{split}
    \P^*(\abs{r_T^*}>\eta_{T,1})&\leq \P^*(R_T^*>\eta_{T,1})
    \leq 2N\left[\exp\left(-C\frac{\eta_{T,1}^{2} T}{d_{N}^{*2} S^*_2}\right)\right]=:\eta_{T,2}.
\end{split}\end{equation*}   
Continue with 
\begin{align*}
\abs{F_{1,T}^*(y) - G_T^*(y)} 
\quad &\leq \underbrace{\abs{\P^*\left(\epsilon_T^{(\max)^*}  \leq y + \eta_{T,1} \right) - \P^*(z_T^{(\max)^*}  \leq y + \eta_{T,1})}}_{A_{T,1}^*(y+\eta_{T,1})}\\
&+ \underbrace{\abs{\P^*\left(z_T^{(\max)^*}  \leq y + \eta_{T,1} \right) - \P^*(z_T^{(\max)^*}  \leq y)}}_{A_{T,2}^*(y)}  + \eta_{T,2}.
\end{align*}
Note that $\sup\limits_{y\in\mathbb{R}}A_{T,1}^*(y+\eta_{T,1})={M_{N,T}^*}$ which can be bounded by \Cref{lma:bootCLT}, 
and  $\sup\limits_{y\in\mathbb{R}}A_{T,2}^*(y)\leq C\eta_{T,1}\sqrt{\log(N)}$ by Lemma A.1 in \cite{CCK17}. We therefore have the bound
\begin{equation*}
    \sup\limits_{y\in\mathbb{R}}\abs{F_{1,T}^*(y) - G_T^*(y)} \leq {M_{N,T}^*}+ C_1\left[\eta_{T,1}\sqrt{\log{N}}+N\exp\left(-C_2\frac{\eta_{T,1}^{2} T}{d_{N}^{*2} S^*_2}\right)\right].
\end{equation*}
Following the same argument as in the proof of \Cref{thm:HDCLTforLP}, we choose $\eta_{T,1}=\sqrt{\log(N\log(N))\frac{d_{N}^{*2} S^*_2}{CT}}$,
which lets us bound 
\begin{equation*}\begin{split}
    &C_1\left[\eta_{T,1}\sqrt{\log{N}}+N\exp\left(-C_2\frac{\eta_{T,1}^{2} T}{d_{N}^{*2} S^*_2}\right)\right]\leq C\left[\frac{\log(N)d_{N}^*\sqrt{S_2^*}}{\sqrt{T}}+\frac{1}{\log(N)}\right]
\end{split}\end{equation*}
and the result of the first statement follows.

\textit{Under  \Cref{ass:subgaussian}.\ref{ass:subgaussian2}}, we may follow the same steps as above, taking
\begin{equation*}
    \eta_{T,2}:=C\frac{N  S_1^{*m} d_{N}^{*m}}{\left(\eta_{T,1}\sqrt{T}\right)^m}.
\end{equation*} We then have the bound
\begin{equation*}
\begin{split}
    \sup\limits_{y\in\mathbb{R}}\abs{F_{1,T}^*(y) - G_T^*(y)} &\leq {M_{N,T}^*}+ C\left[\eta_{T,1}\sqrt{\log{N}}+\frac{N  S_1^{*m} d_{N}^{*m}}{\left(\eta_{T,1}\sqrt{T}\right)^m}\right]\\&\leq {M_{N,T}^*}+C(Nd_{N}^{*m}\psi_{N}^m)^{\frac{1}{m+1}}\left(\frac{\sqrt{\log(N)}}{\sqrt{T}}\right)^{\frac{m}{m+1}},
\end{split}
\end{equation*}
and the result of the second statement follows.
\end{proof}

\begin{proof}[\hypertarget{p:T4}{\textbf{Proof of \Cref{thm:bootstrapconsistency}}}]
With a simple telescopic sum argument
\begin{small}
\begin{equation*}\begin{split}
    &\sup\limits_{y\in\mathbb{R}}\abs{\P\left(\norm{\frac{1}{\sqrt{T}}\sum_{t=1}^T\bx_{t}}_\infty\leq y\right)-\P^*\left(\norm{\frac{1}{\sqrt{T}}\sum_{t=1}^T\bx_t^*}_\infty\leq y\right)}\\
    &\leq \sup\limits_{y\in\mathbb{R}}\abs{\P\left(\norm{\frac{1}{\sqrt{T}}\sum_{t=1}^T\bx_{t}}_\infty\leq y\right)-\P\left(\norm{\bz}_\infty\leq y\right)}+ \sup\limits_{y\in\mathbb{R}}\abs{\P^*\left(\norm{\frac{1}{\sqrt{T}}\sum_{t=1}^T\bx_{t}^*}_\infty\leq y\right)-\P\left(\norm{\bz}_\infty\leq y\right)}\\
    &=\sup\limits_{y\in\mathbb{R}}\abs{\P\left(\norm{\frac{1}{\sqrt{T}}\sum_{t=1}^T\bx_{t}}_\infty\leq y\right)-\P\left(\norm{\bz}_\infty\leq y\right)}+ \sup\limits_{y\in\mathbb{R}}\abs{\P^*\left(\norm{\frac{1}{\sqrt{T}}\sum_{t=1}^T\bx_{t}^*}_\infty\leq y\right)-\P^*\left(\norm{\bz}_\infty\leq y\right)}\\
    \leq & {J_{N,T}}+{J_{N,T}^*},
\end{split}\end{equation*}
\end{small}
which are bounded by \Cref{thm:HDCLTforLP,thm:HDCLTforboot} respectively. The bounds provided by these theorems only hold under \Cref{ass:covariance,ass:subgaussian,ass:VARsummable}, on the set $\mathcal{P}\bigcap\mathcal{Q}\bigcap\mathcal{T}_i\bigcap\mathcal{U}_i$ ($i\in \{1,2\}$), depending on which moment assumption we make in \Cref{ass:subgaussian}) and for sufficiently large $N,T$. The latter is satisfied as we look consider the asymptotic case as $N,T\to\infty$ in this theorem. Consider first the set $\mathcal{T}_i$. By \Cref{thm:CovarianceCloseness}, it holds (with probability equal to 1) on the set  $\mathcal{P}\bigcap\mathcal{Q}\bigcap\mathcal{R}_i\bigcap\mathcal{S}_i$. These sets then hold with probability converging to 1 individually by \Cref{ass:VARconsistency}, \Cref{ass:resconsistency}, \Cref{lma:epsilonConsistent}, and \Cref{lma:nohatstoexp} respectively. By the union bound, we then have
\begin{equation*}
    \P\left(\mathcal{P}\bigcap\mathcal{Q}\bigcap\mathcal{R}_i\bigcap\mathcal{S}_i\right)\geq 1-\left[\P(\mathcal{P}^c)+\P(\mathcal{Q}^c)+\P(\mathcal{R}^c_i)+\P(\mathcal{S}^c_i)\right]{\to} 1,
\end{equation*}
as $N,T\to\infty$. We therefore also have $\lim\limits_{N,T\to\infty}\P(\mathcal{T}_i)=1$, unconditionally. To see why, we may alternatively phrase the result of \Cref{thm:CovarianceCloseness} as $\P(\mathcal{T}_i\vert\mathcal{P}\bigcap\mathcal{Q}\bigcap\mathcal{R}_i\bigcap\mathcal{S}_i)=1$. We may then write the unconditional probability as 
\begin{equation*}\begin{split}
    \lim\limits_{N,T\to\infty}\P(\mathcal{T}_i)=& \lim\limits_{N,T\to\infty}\underset{=1}{\underbrace{\P(\mathcal{T}_i\vert\mathcal{P}\bigcap\mathcal{Q}\bigcap\mathcal{R}_i\bigcap\mathcal{S}_i)}}\times\underset{\to 1}{\underbrace{\P(\mathcal{P}\bigcap\mathcal{Q}\bigcap\mathcal{R}_i\bigcap\mathcal{S}_i)}}\\
    +&\underset{\leq 1}{\underbrace{\P(\mathcal{T}_i\vert\mathcal{P}^c\bigcup\mathcal{Q}^C\bigcup\mathcal{R}_i^c\bigcup\mathcal{S}_i^c)}}\times\underset{\to0}{\underbrace{\P(\mathcal{P}^c\bigcup\mathcal{Q}^C\bigcup\mathcal{R}_i^c\bigcup\mathcal{S}_i^c)}}=1.
\end{split}\end{equation*}
We can apply the same logic to the bounds on ${J_{N,T}},{J_{N,T}^*}$, and ${M_{N,T}^*}$ in \Cref{lma:bootCLT} (the bound on ${M_{N,T}}$ in \Cref{lma:ChernozhukovHDCLT} holds deterministically), noting that we also have $\lim\limits_{N,T\to\infty}\P(\mathcal{U}_i)=1$ by \Cref{lma:epsilonbounded}. Then if each bound holds with probability converging to 1, the bound obtained by combining them all holds with probability converging to 1 also.

Combining the bounds on ${J_{N,T}}$ and ${J_{N,T}^*}$ under \Cref{ass:subgaussian}.\ref{ass:subgaussian1}, we obtain the bound 
\begin{equation*}\begin{split}
   &C \underset{{J_{N,T}}}{\underbrace{ \left[\frac{(\tilde{S}d_N)^2\log(N)^{3/2}\log(T)}{\sqrt{T}}+\frac{(\tilde{S}d_N)^2\log(N)^2}{\sqrt{T}}+\frac{\log(N)d_{N}\sqrt{S_2}}{\sqrt{T}}+\frac{1}{\log(N)}\right.}}\\
   &+\underset{{J_{N,T}^*}}{\underbrace{\log(N)\log(T)\psi_N^2\left[d_{N}\sqrt{\phi_{N,T}}+\frac{d_N}{\sqrt{T}}+\xi_{N,T}\psi_N\right]+\frac{\log(N)d_{N}^*\sqrt{S_2^*}}{\sqrt{T}}+\frac{1}{\log(N)}}}\\
   &+\underset{{J_{N,T}^*}}{\underbrace{\left.(\tilde{S}^*d_{N}^*)^2\left[\frac{\log(N)^{3/2}\log(T)}{\sqrt{T}}+\frac{\log(N)^2\log(T)^2}{T}\right]+\sqrt{\frac{\log(N)^2\log(T)\log(NT)}{T}}\right]}}
\end{split}\end{equation*}
We plug in the bounds $\tilde{S}\leq C\psi_N$ by \Cref{lma:trueVMAsummable}.1, $S_2\leq C\psi_N^2$ by \Cref{lma:trueVMAsummable}.3, $\tilde{S}^*\leq C\psi_N$ by \Cref{lma:trueVMAsummable}.5, $S_2^*\leq C\psi_N^2$ by \Cref{lma:trueVMAsummable}.7, $d_N=C\sqrt{\log(N)}$, $d_N^*=C\left(\sqrt{T\phi_{N,T}}+\sqrt{\log(N)}\log{T}\right)$. We then eliminate dominated terms using $\psi_N\geq 1$, $\log(T)\geq 1$ and $\log(N)\geq 1$, and use the shorthand notation $\ell_N:=\log(N)$, $\ell_T:=\log(T)$ to simplify this expression to the following:
\begin{equation*}\begin{split}
    C\left\lbrace\psi_N^2\left[\frac{\ell_N^3}{\sqrt{T}}+\ell_N\ell_T\left(\ell_N\sqrt{\phi_{N,T}}+\frac{\ell_N}{\sqrt{T}}+\xi_{N,T}\psi_N\right)+\left(\sqrt{T\phi_{N,T}+\sqrt{\ell_N}\ell_T}\right)^2\left(\frac{\ell_N^{3/2}}{\sqrt{T}}+\frac{\ell_N^2\ell_T^2}{T}\right)\right]+\frac{1}{\ell_N}\right\rbrace.
\end{split}\end{equation*}

Combining the bounds on ${J_{N,T}}$ and ${J_{N,T}^*}$ under \Cref{ass:subgaussian}.\ref{ass:subgaussian2}, we obtain the bound 
\begin{equation*}\begin{split}
   &C \underset{{J_{N,T}}}{\underbrace{ \left[\frac{(\tilde{S}d_N)^2\log(N)^{3/2}\log(T)}{\sqrt{T}}+\frac{(\tilde{S}d_N)^2\log(N)^2}{\sqrt{T}}+\frac{\log(N)d_{N}\sqrt{S_2}}{\sqrt{T}}+\frac{1}{\log(N)}\right.}}\\
   &+\underset{{J_{N,T}^*}}{\underbrace{\log(N)\log(T)\psi_N^2\left[d_{N}\sqrt{\phi_{N,T}}+\frac{d_N}{\sqrt{T}}+\xi_{N,T}\psi_N\right]+\frac{\log(N)d_{N}^*\sqrt{S_2^*}}{\sqrt{T}}+\frac{1}{\log(N)}}}\\
   &+\underset{{J_{N,T}^*}}{\underbrace{\left.(\tilde{S}^*d_{N}^*)^2\left[\frac{\log(N)^{3/2}\log(T)}{\sqrt{T}}+\frac{\log(N)^2\log(T)^2}{T}\right]+\sqrt{\frac{\log(N)^2\log(T)\log(NT)}{T}}\right]}}
\end{split}\end{equation*}
\end{proof}

\begin{proof}[\hypertarget{p:C1}{\textbf{Proof of \Cref{cor:finitemoments}}}]
 Under this choice of growth rates, we may take $\lambda_j=T^{\frac{4a+1}{m}-\frac{3}{4}}\eta_{T}^{-1}$,
 such that 
 $\max\limits_{j}\frac{1}{T}\norm{\hat\bepsilon_{j}-\bepsilon_j}_2^2\leq  C\lambda_j^{2-r}s_{r,j}=C\frac{T^{\frac{12a+3}{m}}}{T}$
 and 
 $\norm{\hat\mA-\mA}_{\infty}=\max\limits_{j}\norm{\hat{\boldsymbol{\beta}_j}-{\boldsymbol{\beta}}_j}_1 \leq C\lambda_js_{r,j}=C\frac{T^{\frac{4a+1}{2m}}}{T^{1/4}}$.
 We also have $\log(N)\sim a\log(T)\leq C\log(T)$, and similarly $\log(NT)\leq C\log(T)$. Therefore, we may take $\xi_{N,T}=\eta_T^{-1}\frac{T^{\frac{4a+1}{m}}}{T^{1/4}}$, $\phi_{N,T}=\eta_T^{-1}\frac{T^{\frac{12a+3}{m}}}{T}$. Plugging these into the bound of \Cref{thm:bootstrapconsistency} and eliminating dominated terms, we see it converges to 0 when 
 \begin{equation*}
     \eta_T^{-1}\frac{\ell_T^{3/2}T^{\frac{12a+3}{2m}+\frac{12a+3}{4m(m-1)}}}{\sqrt{T}}\to0.
 \end{equation*} 
 Note that any $\log(T)$ term is dominated by a term polynomial in $T$, so this terms converges to 0 when $m>\sqrt{36a^2+18a+5/2}+6a+1$. 
\end{proof}

\begin{proof}[\hypertarget{p:C1}{\textbf{Proof of \Cref{cor:subgaussianmoments}}}]
 Under this choice of growth rates, we have the bounds 

 $\max\limits_{j}\frac{1}{T}\norm{\hat\bepsilon_{j}-\bepsilon_j}_2^2\leq  C\lambda_j^{2}s_{0,j}/\kappa_{j}\leq C\ell_T^5 T^{5a+b-1}=\phi_{N,T}$
 and 
 $\norm{\hat\mA-\mA}_{\infty}=\max\limits_{j}\norm{\hat{\boldsymbol{\beta}_j}-{\boldsymbol{\beta}}_j}_1 \leq C\lambda_js_{0,j}/\kappa_j=C \ell_T^{5/2} T^{\frac{5a+2b-1}{2}}=\xi_{N,T}$.
 We also have $\log(N)\sim T^a$. Plugging these into the bound of \Cref{thm:bootstrapconsistency} and eliminating dominated terms, we see it converges to 0 when \begin{equation*}
     C\ell_T^{6}T^{\frac{13a+2b-1}{2}}\to0.
 \end{equation*} 
 Note that any $\log(T)$ term is dominated by a term polynomial in $T$, so this terms converges to 0 when $13a+2b<1$. 
\end{proof}

\section{}\label{app:misc}
\subsection{Algorithm for choosing the lag length}
\begin{algorithm}[h!]
\nl Choose a large maximum lag $K_{\max}$\;
\For{$K=1,\dots,K_{\max}$}{
\nl For each $j=1,\dots,N$, estimate by OLS the (univariate) autoregressive models $x_{j,t}=\sum\limits_{k=1}^K \rho_{j,k}^{(K)}x_{j,t-k}+\varepsilon_{j,t}^{(K)}$, and save the residuals $\hat{\varepsilon}_{j,t}^{(K)}$\;
\nl Let $\hat\omega^{(K)}_j=\frac{1}{T}\sum\limits_{t=1}^T(\hat{\varepsilon}_{j,t}^{(K)})^2$, and $\hat\bOmega^{(K)}=\diag(\hat\omega^{(K)}_{1}, \dots, \hat\omega^{(K)}_{N})$\;
\nl Let $IC^*(K)=\log(\det \hat\bOmega^{(K)})+C_T\frac{KN}{T}=\sum\limits_{j=1}^N\log\hat{\omega}^{(K)}_j+C_T\frac{KN}{T}$\;
}
\nl Use the lag length $K^*=\argmin\limits_{1\leq K\leq K_{\max}} IC^*(K)$.
\caption{Informative upper bound on lag length}\label{alg:lag_selection}
\end{algorithm}
In step 4, $C_T$ takes the standard values for well-known criteria: $C_T=\log(T)$ for BIC, $C_T=2$ for AIC.
\subsection{Details of Example 1}
\begin{example}\label{ex:off_diagonal_VAR}
    Consider the model in \Cref{eq:DGPVAR}, with $K=1$, $\bA_1=\frac{1}{2}\left(\bI + \bPsi\right)$, where $\bPsi_{i,j}=\ind{j=i+1}$, and $\left(\bSigma_{\epsilon}\right)_{i,j}=\ind{i=j}-\frac{1}{2}\ind{\abs{i-j}=1}$, i.e. 
    \begin{equation*}
        \bA_1=\left[\begin{matrix}
                \frac{1}{2} & \frac{1}{2} &  0 & 0 & \dots \\
                0 & \frac{1}{2} & \frac{1}{2} &  0 & \dots \\
                0 & 0 & \frac{1}{2} & \frac{1}{2} & \dots \\
                \vdots & \vdots & \vdots & \ddots & \ddots \\
            \end{matrix}\right],~~
        \bSigma_{\epsilon}=\left[\begin{matrix}
            1 & -\frac{1}{2} &  0 & 0 & \dots \\
            -\frac{1}{2} & 1 & -\frac{1}{2} &  0 & \dots \\
             0 & -\frac{1}{2} & 1 & -\frac{1}{2} & \dots \\
            \vdots & \vdots & \ddots & \ddots & \ddots \\
        \end{matrix}\right].
    \end{equation*}
    This model satisfies \Cref{ass:covariance} with $\Lambda_{\min}(\bSigma)=2$, $\max\limits_{1\leq j\leq N}\sigma_j^2=4,~\forall N\geq2$. To satisfy \Cref{ass:VARsummable}, it is necessary that $\psi_N \geq (1/C)^{N-1}$, where $C<1$ is the constant in \Cref{ass:VARsummable}.
    \begin{proof}
    We will first show the latter result, deriving an exponential lower bound on $\psi_N$. In this simple VAR(1), $\mA=\bA_1$, and its powers satisfy 
    \begin{equation*}
        \mA^k=\frac{1}{2^k}\left(\bI+\bPsi\right)^k= \frac{1}{2^k} \sum\limits_{i=0}^{k} {k\choose i} \bPsi^i,
    \end{equation*}
    due to the binomial theorem and the fact that $\bI$ and $\bPsi$ commute. Note that $\bPsi^\ell$ has entries of $1$ on the $\ell$-th upper off-diagonal, and $0$ everywhere else: $(\bPsi^\ell)_{i,j}=\ind{j= i+\ell}$. To show this, consider the following proof by induction. The statement holds for $\ell=1$ by definition, and assuming that it holds for $\ell=k$, we can show it also holds for $\ell=k+1$:
    \begin{equation*}
        (\bPsi^{k+1})_{i,j}=\left(\bPsi^{k}\bPsi\right)_{i,j}=\sum\limits_{\ell=1}^N \left(\bPsi^k\right)_{i,\ell} \bPsi_{\ell,j}=\sum\limits_{\ell=1}^N\ind{\ell=i+k} \ind{j=\ell +1}=\ind{j=i+k+1}.
    \end{equation*}
    To satisfy \Cref{ass:VARsummable}, we need that for all $N\geq 1$ that
    \begin{equation*}
    \begin{split}
        \psi_N\geq & \max\limits_{k\geq 1}\left\lbrace\norm{\mA^k}_{\infty}/\theta^k\right\rbrace \geq \max\limits_{k\geq 1}\left\lbrace\norm{\mA^k}_{\infty}/C^k\right\rbrace\geq \max\limits_{k\geq 1}\left\lbrace\sum\limits_{j=1}^{N}\abs{\left(\mA^k\right)_{1,j}}/C^k\right\rbrace \\
        =& \max\limits_{k\geq 1}\left\lbrace\sum\limits_{j=1}^{N}\abs{
        \frac{1}{2^k}\sum\limits_{i=0}^{k}{k \choose i}\ind{j=1+k}
        }/C^k\right\rbrace\geq\abs{
        \frac{1}{2^{N-1}}\sum\limits_{i=0}^{N-1}{N-1 \choose i}
        }/C^{N-1}=(1/C)^{N-1}.
        \end{split}
    \end{equation*}
    To show the former result, we will show that $\bSigma:=\cB(1)\bSigma_{\bepsilon}\cB(1)^\prime$ has entries $\bSigma_{i,j}=4\ind{i=j}+2\ind{i\neq j}$, which has the stated minimum eigenvalue and maximum diagonal entry.  To do this, we first show that $(\cB(1))_{i,j}=2\ind{j\geq i}$, i.e.~an upper triangular matrix. From the results above, we have 
    \begin{equation*}
        \cB(1)=\sum\limits_{k=0}^{\infty}\mA^k=\sum\limits_{k=0}^{\infty}\frac{1}{2^k} \sum\limits_{i=0}^{k} {k\choose i} \bPsi^i.
    \end{equation*}
    Since $\mA^k$ is a scaled sum of the upper off-diagonal matrices $\Psi^i$, it is an upper triangular Toeplitz matrix. This property is maintained under addition, so $\cB(1)$ is also upper triangular Toeplitz (if it is a convergent series). It is therefore sufficient to show that all entries in the first row of $\cB(1)$ are 2. We proceed with a proof by induction; first, we show that $(\cB(1))_{1,1}=2$. More generally for all $1\leq n\leq N$:
    \begin{equation*}
        (\cB(1))_{1,n}=\sum\limits_{k=0}^{\infty}\frac{1}{2^k} \sum\limits_{\ell=0}^{k} {k\choose \ell} (\bPsi^\ell)_{1,n} = \sum\limits_{k=0}^{\infty}\frac{1}{2^k} \sum\limits_{\ell=0}^{k} {k\choose \ell} \ind{n= 1+\ell} = \sum\limits_{k=n-1}^{\infty}\frac{1}{2^k}{k\choose n-1}.
    \end{equation*}
 By properties of geometric series,
        \begin{equation*}
        (\cB(1))_{1,1}=\sum\limits_{k=0}^{\infty}\frac{1}{2^k} =2.
    \end{equation*}
    Next, we assume that $(\cB(1))_{1,n}=2$
    for some $1\leq n\leq N-1$, and show that $(\cB(1))_{1,n+1}=2$.
    Using Pascal's identity,
    \begin{equation*}
    \begin{split}
        (\cB(1))_{1,n+1}=&\sum\limits_{k=n}^{\infty}\frac{1}{2^k}{k\choose n}=\frac{1}{2}\sum\limits_{k=n-1}^{\infty}\frac{1}{2^{k}}{k+1\choose n}=\frac{1}{2}\left[\sum\limits_{k=n-1}^{\infty}\frac{1}{2^{k}}{k\choose n-1} + \sum\limits_{k=n-1}^{\infty}\frac{1}{2^{k}}{k\choose n}\right]\\
        =&\frac{1}{2}\left[2+\sum\limits_{k=n-1}^{\infty}\frac{1}{2^{k}}{k\choose n}\right]=\frac{1}{2}\left[2+\underbrace{\frac{1}{2^{n-1}}{n-1 \choose n}}_{=0}+(\cB(1))_{1,n+1}\right].
        \end{split}
    \end{equation*}
    If $(\cB(1))_{1,n+1}$ is a convergent series, then the above equation implies
    \begin{equation*}
        (\cB(1))_{1,n+1}=\frac{1}{2}\left[2+(\cB(1))_{1,n+1}\right]\implies (\cB(1))_{1,n+1}=2.
    \end{equation*}
    To show that $(\cB(1))_{1,n+1}$ is convergent, we use the ratio test:
    \begin{equation*}
        \lim\limits_{k\to\infty}\frac{\frac{1}{2^{k+1}}{k+1\choose n}}{\frac{1}{2^k}{k\choose n}}=\frac{1}{2}\lim\limits_{k\to\infty}\frac{k+1}{k+1-n}=\frac{1}{2}<1,
    \end{equation*}
    which shows the series is absolutely convergent. To show that $\bSigma$ is as claimed,
    \begin{equation*}\begin{split}
        \bSigma_{i,j}=&\sum_{\ell=1}^N\sum_{k=1}^N (\Sigma_{\epsilon})_{\ell,k}(\cB(1))_{i,\ell}(\cB(1))_{j,k}=\sum_{\ell=1}^N\sum_{k=1}^N \left(\ind{\ell=k}-\frac{1}{2}\ind{\abs{\ell-k}=1}\right)2\ind{\ell\geq i}2\ind{k\geq j}\\
        =&\sum_{\ell=1}^N\sum_{k=1}^N \left(4\ind{\ell=k}-2\ind{\abs{\ell-k}=1}\right)(\ind{i=j}+\ind{i\neq j})\ind{\ell\geq i}\ind{k\geq j}.
        \end{split}
    \end{equation*}
    Treating this sum in four parts:
    \begin{equation*}\begin{split}
        &\sum_{\ell=1}^N\sum_{k=1}^N 4\ind{\ell=k}\ind{i=j}\ind{\ell\geq i}\ind{k\geq j}=\ind{i=j}4(N-i+1),\\
        &\sum_{\ell=1}^N\sum_{k=1}^N 4\ind{\ell=k} \ind{i\neq j})\ind{\ell\geq i}\ind{k\geq j}=\ind{i\neq j}4(N-i\vee j+1),\\
        &\sum_{\ell=1}^N\sum_{k=1}^N -2\ind{\abs{\ell-k}=1}\ind{i=j}\ind{\ell\geq i}\ind{k\geq j}=-\ind{i=j}4(N-i),\\
        &\sum_{\ell=1}^N\sum_{k=1}^N -2\ind{\abs{\ell-k}=1}\ind{i\neq j}\ind{\ell\geq i}\ind{k\geq j}=-\ind{i\neq j}(4(N-i\vee j)+2).
        \end{split}
    \end{equation*}
    Summing these terms then gives $\bSigma_{i,j}=4\ind{i=j}+2\ind{i\neq j}$.
    \end{proof}
    
\end{example}

\subsection{}

\begin{figure}[ht]
\centering
\includegraphics[width=\linewidth]{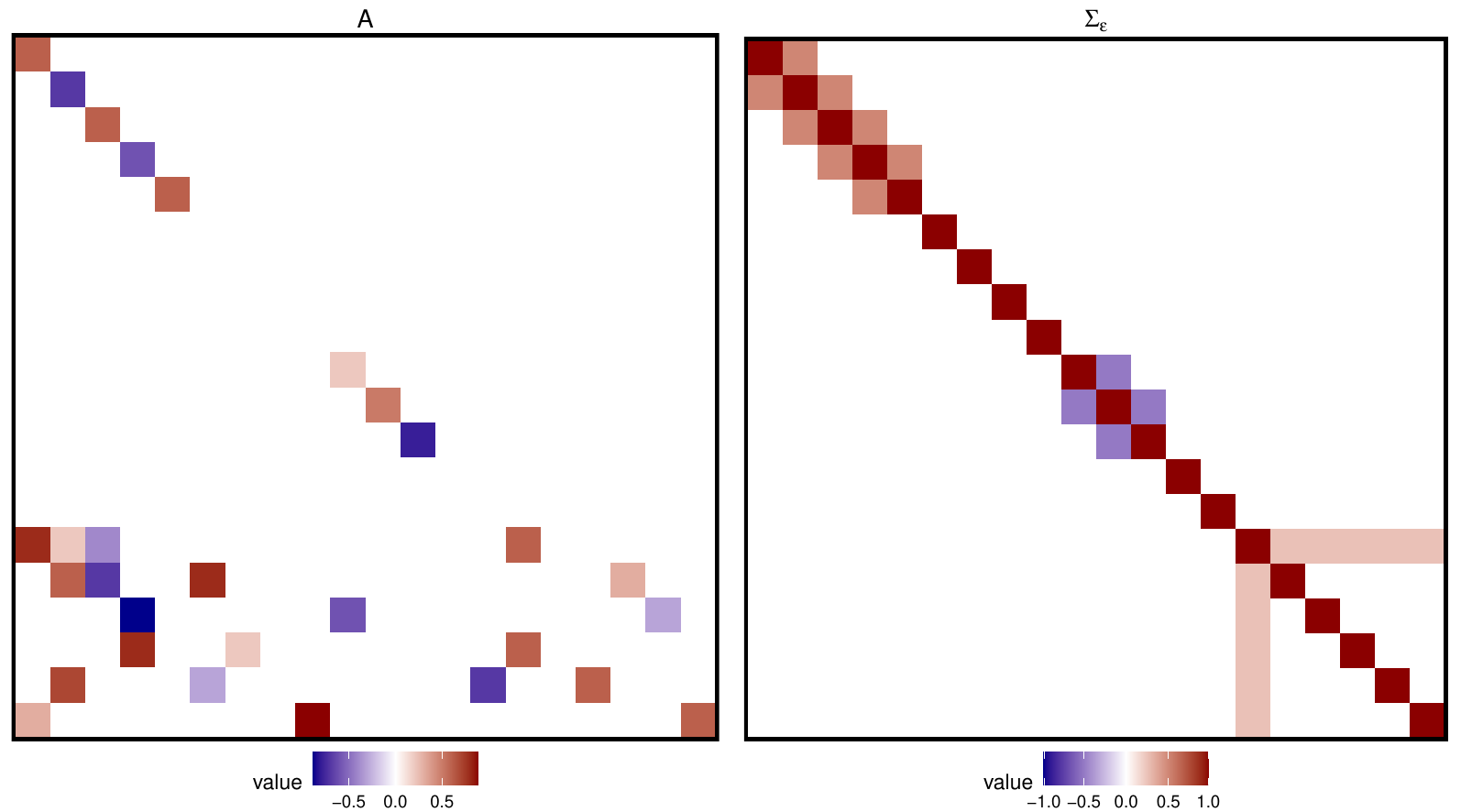}
\caption{Pattern within the blocks of $\bA$ and $\bSigma_{\epsilon}$}\label{fig:krampe_sim}
\end{figure}

\begin{table}[ht!]
\centering
\caption{Country list}\label{tab:country_list}
\resizebox{0.85\textwidth}{!}{\makebox[\textwidth]{
\begin{tabular}{ll|ll|ll}
  \hline
 Code & Name & Code & Name & Code & Name \\ 
  \hline
ABW & Aruba & GAB & Gabon & NAM & Namibia\\ 
AGO & Angola & GBR & United Kingdom & NER & Niger\\ 
AIA & Anguilla & GHA & Ghana & NGA & Nigeria\\ 
ALB & Albania & GIN & Guinea & NIC & Nicaragua\\ 
ARE & United Arab Emirates & GMB & Gambia & NLD & Netherlands\\ 
ARG & Argentina & GNB & Guinea-Bissau & NOR & Norway\\ 
ATG & Antigua and Barbuda & GNQ & Equatorial Guinea & NPL & Nepal\\ 
AUS & Australia & GRC & Greece & NZL & New Zealand\\ 
AUT & Austria & GRD & Grenada & OMN & Oman\\ 
BDI & Burundi & GTM & Guatemala & PAK & Pakistan\\ 
BEL & Belgium & GUY & Guyana & PAN & Panama\\ 
BEN & Benin & HKG & China, Hong Kong SAR & PER & Peru\\ 
BFA & Burkina Faso & HND & Honduras & PHL & Philippines\\ 
BGD & Bangladesh & HTI & Haiti & POL & Poland\\ 
BGR & Bulgaria & HUN & Hungary & PRT & Portugal\\ 
BHR & Bahrain & IDN & Indonesia & PRY & Paraguay\\ 
BHS & Bahamas & IND & India & PSE & State of Palestine\\ 
BLZ & Belize & IRL & Ireland & QAT & Qatar\\ 
BMU & Bermuda & IRN & Iran (Islamic Republic of) & ROU & Romania\\ 
BOL & Bolivia (Plurinational State of) & IRQ & Iraq & RWA & Rwanda\\ 
BRA & Brazil & ISL & Iceland & SAU & Saudi Arabia\\ 
BRB & Barbados & ISR & Israel & SDN & Sudan\\ 
BRN & Brunei Darussalam & ITA & Italy & SEN & Senegal\\ 
BTN & Bhutan & JAM & Jamaica & SGP & Singapore\\ 
BWA & Botswana & JOR & Jordan & SLE & Sierra Leone\\ 
CAF & Central African Republic & JPN & Japan & SLV & El Salvador\\ 
CAN & Canada & KEN & Kenya & SOM & Somalia\\ 
CHE & Switzerland & KHM & Cambodia & STP & Sao Tome and Principe\\ 
CHL & Chile & KNA & Saint Kitts and Nevis & SUR & Suriname\\ 
CHN & China & KOR & Republic of Korea & SWE & Sweden\\ 
CIV & Côte d'Ivoire & KWT & Kuwait & SWZ & Eswatini\\ 
CMR & Cameroon & LAO & Lao People's DR & SYC & Seychelles\\ 
COD & D.R. of the Congo & LBN & Lebanon & SYR & Syrian Arab Republic\\ 
COG & Congo & LBR & Liberia & TCA & Turks and Caicos Islands\\ 
COL & Colombia & LCA & Saint Lucia & TCD & Chad\\ 
COM & Comoros & LKA & Sri Lanka & TGO & Togo\\ 
CPV & Cabo Verde & LSO & Lesotho & THA & Thailand\\ 
CRI & Costa Rica & LUX & Luxembourg & TTO & Trinidad and Tobago\\ 
CYM & Cayman Islands & MAC & China, Macao SAR & TUN & Tunisia\\ 
CYP & Cyprus & MAR & Morocco & TUR & Türkiye\\ 
DEU & Germany & MDG & Madagascar & TWN & Taiwan\\ 
DJI & Djibouti & MDV & Maldives & TZA & U.R. of Tanzania: Mainland\\ 
DMA & Dominica & MEX & Mexico & UGA & Uganda\\ 
DNK & Denmark & MLI & Mali & URY & Uruguay\\ 
DOM & Dominican Republic & MLT & Malta & USA & United States\\ 
DZA & Algeria & MMR & Myanmar & VCT & St. Vincent and the Grenadines\\ 
ECU & Ecuador & MNG & Mongolia & VEN & Venezuela (Bolivarian Republic of)\\ 
EGY & Egypt & MOZ & Mozambique & VGB & British Virgin Islands\\ 
ESP & Spain & MRT & Mauritania & VNM & Viet Nam\\ 
ETH & Ethiopia & MSR & Montserrat & ZAF & South Africa\\ 
FIN & Finland & MUS & Mauritius & ZMB & Zambia\\ 
FJI & Fiji & MWI & Malawi & ZWE & Zimbabwe\\ 
FRA & France & MYS & Malaysia &  & \\ 
   \hline
\end{tabular}
%\end{minipage}}
}}
\end{table}
\end{appendices}
\end{document}